\documentclass[prd,tightenlines,twoside,secnumarabic,onecolumn,floatfix,nofootinbib,showpacs,superscriptaddress,11pt]{revtex4-1}

\pdfoutput=1
\usepackage[english]{babel}
\usepackage{calligra}
\usepackage{epsfig}
\usepackage{float}
\usepackage{multirow}
\usepackage{epstopdf}
\usepackage{amsmath,amssymb,bm,slashed}
\usepackage{mathtools}
\usepackage{graphicx}
\usepackage[sort&compress]{natbib}
\usepackage{xcolor}
\usepackage[normalem]{ulem}
\usepackage{hyperref}
\usepackage{cleveref}
\usepackage{subfigure} 
\definecolor{red}{rgb}{1.0, 0, 0}
\usepackage{slashed}
\usepackage{ctable}

\allowdisplaybreaks

\bibpunct{[}{]}{,}{n}{}{,}

\newcommand{\BR}{\text{BR}}

\begin{document}

\title{New Signatures of Flavor Violating Higgs Couplings}
\author{Malte Buschmann}     \email[Email: ]{buschmann@uni-mainz.de}
\author{Joachim Kopp}        \email[Email: ]{jkopp@uni-mainz.de}
\author{Jia Liu}             \email[Email: ]{liuj@uni-mainz.de}
\author{Xiao-Ping Wang}      \email[Email: ]{xiaowang@uni-mainz.de}
\affiliation{PRISMA Cluster of Excellence and Mainz Institute for Theoretical Physics,
Johannes Gutenberg University, 55099 Mainz, Germany}
\date{\today} 

\begin{abstract}
  We explore several novel LHC signatures arising from quark or lepton flavor
  violating couplings in the Higgs sector, and we constrain such couplings
  using LHC data.  Since the largest signals are possible in channels involving
  top quarks or tau leptons, we consider in particular the following flavor
  violating processes: (1) $p p \to t h h$ (top plus di-Higgs final state)
  arising from a dimension six coupling of up-type quarks to three insertions of the
  Higgs field.  We develop a search strategy for this final state and
  demonstrate that detection is possible at the high luminosity LHC if flavor
  violating top--up--Higgs couplings are not too far below the current limit.
  (2) $p p \to t H^0$, where $H^0$ is the heavy neutral CP-even Higgs boson in
  a two Higgs doublet model (2HDM). We consider the decay channels $H^0 \to t
  u, WW, ZZ, hh$ and use existing LHC data to constrain the first three of
  them. For the fourth, we adapt our search for the $thh$ final state, and we
  demonstrate that in large regions of the parameter space, it is superior to
  other searches, including searches for flavor violating top quark decays ($t
  \to h q$).  (3) $H^0 \to \tau\mu$, again in the context of a 2HDM.  This
  channel is particularly well motivated by the recent CMS excess in $h \to
  \tau\mu$, and we use the data from this search to constrain the properties of
  $H^0$.
\end{abstract}

\begin{flushright}
  MITP/16-005
\end{flushright}

\maketitle

\tableofcontents

\section{Introduction}
\label{sec:intro}

The discovery of the Higgs boson \cite{Chatrchyan:2012xdj,Aad:2012tfa},
while being a spectacular triumph for both
theoretical and experimental particle physics, is hopefully only the first step
in a new era of discoveries in the field. In particular, the hope is that precision
studies of the Higgs boson's properties will open up a pathway to physics beyond
the Standard Model (SM). Among the simplest and most promising ways in which Higgs
physics could deviate from SM predictions are \emph{flavor violating
couplings} of the Higgs boson~\cite{Bjorken:1977vt, 
  McWilliams:1980kj, 
  Babu:1999me, 
  DiazCruz:1999xe, 
  Han:2000jz, 
  Giudice:2008uua, 
  AguilarSaavedra:2009mx, 
  Goudelis:2011un, 
  Blankenburg:2012ex, 
  Kanemura:2005hr, 
  Davidson:2010xv, 
  Harnik:2012pb, 
  Davidson:2012ds, 
  Celis:2013xja, 
Gorbahn:2014sha}. 
Such couplings can arise when the vacuum
expectation value (vev) of the Higgs field $H$ is not the only source of electroweak
symmetry breaking. For example, in a two Higgs doublet model (2HDM), the vevs of two
Higgs fields contribute, and this decouples the flavor structure of the fermion mass
matrices from the flavor structure of the Yukawa couplings of the physical Higgs
bosons~\cite{
  Bjorken:1977vt, 
  DiazCruz:1999xe, 
  Han:2000jz, 
  Kanemura:2005hr, 
  Davidson:2010xv, 
  Kopp:2014rva, 
  Sierra:2014nqa, 
  Crivellin:2015mga, 
  deLima:2015pqa, 
  Omura:2015nja, 
  Dorsner:2015mja, 
  Altunkaynak:2015twa, 
  Crivellin:2015hha, 
  Altmannshofer:2015esa, 
  Botella:2015hoa, 
  Arhrib:2015maa, 
Benbrik:2015evd}. 
As a result, flavor changing processes such as $h \to \tau\mu$, $h \to \tau e$
and $t \to h q$ (where $q$ is an up quark or a charm quark) become possible.
An alternative source of misalignment between mass and Yukawa matrices can
be higher-dimensional couplings, for instance dimension six operators of the form
\begin{align}
  \mathcal{Q}_d^{ij} &\equiv \overline{Q_L^i} H d_R^j (H^\dag H) \,,
    \label{eq:Q-dij} \\
  \mathcal{Q}_u^{ij} &\equiv \overline{Q_L^i} \tilde{H} u_R^j (H^\dag H) \,,
    \label{eq:Q-uij} \\
  \mathcal{Q}_\ell^{ij} &\equiv \overline{L_L^i} H e_R^j (H^\dag H) \,.
    \label{eq:Q-eij}
\end{align}
Here, $Q_L^i$ and $L_L^i$ are as usual the left-handed quark and lepton
doublets, $u_R^j$, $d_R^j$ and $e_R^j$ are the right-handed
fermion singlets, $i, j$ are flavor indices, $H$ is the Higgs doublet, and
$\tilde{H} \equiv i \sigma^2 H^\dag$ is its charge conjugate field.

Besides the 2HDM, flavor violating Higgs couplings have been studied in the
context of warped extra dimensions~\cite{
  Blanke:2008zb, 
  Casagrande:2008hr, 
  Albrecht:2009xr, 
  Buras:2009ka, 
  Azatov:2009na}, 
supersymmetric models~\cite{
  DiazCruz:1999xe, 
  Arhrib:2012mg, 
  Arhrib:2012ax, 
  deLima:2015pqa, 
  Aloni:2015wvn}, 
models aiming to explain the flavor structure of the Standard Model~\cite{
  Dery:2013rta, 
  Campos:2014zaa, 
  Dery:2014kxa, 
  Heeck:2014qea, 
  Varzielas:2015joa, 
  Varzielas:2015sno}, 
and neutrino masses~\cite{
  DiazCruz:1999xe, 
  Arganda:2004bz, 
  Arganda:2014dta, 
  Arganda:2015naa}, 
models with vector-like fermion~\cite{
  Falkowski:2013jya}, 
leptoquark models~\cite{
  Cheung:2015yga, 
  Baek:2015mea}, 
flavored ark matter models~\cite{
  Baek:2015fma},
and composite Higgs models~\cite{
  Agashe:2009di, 
  deLima:2015pqa}. 
The connection between flavor violation and a possible new source of
CP violation has been studied in ref.~\cite{
  Kopp:2014rva}, 
and search strategies have been proposed for flavor violating Higgs couplings
in the lepton sector~\cite{
  Harnik:2012pb, 
  Davidson:2012ds} 
and in the quark sector~\cite{
  Craig:2012vj, 
  Atwood:2013ica, 
  Chen:2013qta, 
  Greljo:2014dka, 
  Wu:2014dba}. 

In fact, a recent CMS analysis~\cite{Khachatryan:2015kon} has reported a
$2\sigma$ excess in a search for $h \to \tau\mu$ decays. A
subsequent ATLAS analysis is consistent with this hint, but also with the null
hypothesis~\cite{Aad:2015gha}.  Of course, any hint for a flavor changing
neutral current (FCNC) interaction like this first has to survive scrutiny in
view of low energy precision constraints before being accepted as a potential
hint for new physics. In the case of anomalous $h\mu\tau$ couplings,
the most important constraints arise from the non-observation of flavor
violating charged lepton decays like $\tau \to \mu\gamma$ and $\tau \to
3\mu$~\cite{Blankenburg:2012ex, Harnik:2012pb}, and the CMS hint would in fact
be consistent with all constraints.  Low energy constraints are much tighter
for FCNC Higgs couplings involving only the first two generations of leptons,
thanks to the spectacular sensitivity of experiments searching for the decays
$\mu \to e\gamma$ and $\mu \to 3e$, as well as $\mu\to e$ conversion in
nuclei. Similarly, anomalous Higgs couplings not involving the top quark
are tightly constrained by searches for
anomalous rare meson decays and anomalous
contributions to neutral meson mixing.  Consequently, the only FCNC
Higgs couplings that could in principle be observable at the LHC besides $h
\mu\tau$, are $h \tau e$, $htu$, and $htc$.  (Note that the simultaneous presence
of sizeable $h\mu\tau$ and $h\mu e$ couplings is also ruled out,
as is the simultaneous presence of $htu$ and $htc$
couplings~\cite{Harnik:2012pb}.)

The goal of this paper is to advance collider searches for these potentially
large FCNC Higgs couplings by proposing new search strategies and recasting existing
searches.
In particular, we point out that effective operators like $\mathcal{Q}_u^{ij}$ from
\eqref{eq:Q-uij} can lead to anomalous production channels for di-Higgs final states.
The process we will focus on specifically is $p p \to t h h$,
arising from the operators
\begin{align}
  \mathcal{Q}_u^{31} \equiv \overline{Q_L^3} \tilde{H} u_R^1 (H^\dag H)
    \label{eq:Q-u31}
\intertext{and}
  \mathcal{Q}_u^{13} \equiv \overline{Q_L^1} \tilde{H} u_R^3 (H^\dag H) \,,
    \label{eq:Q-u13}
\end{align}
followed by the decays $t \to b\ell\nu$ and $h \to b\bar{b}$. This process has
an extremely rich signature, outside the scope of present LHC analyses and with
a very low background expectation.  Going beyond the contact operator
approximation, we will discuss the $thh$ final state also in the context of a
type~III two Higgs doublet model (2HDM) with quark flavor violating Yukawa couplings.
Models of this type are emerging as one of the leading UV completions for the
FCNC operators in \cref{eq:Q-dij,eq:Q-uij,eq:Q-eij}~\cite{Bjorken:1977vt,
DiazCruz:1999xe, Han:2000jz, Kanemura:2005hr, Davidson:2010xv, Kopp:2014rva,
Sierra:2014nqa, Crivellin:2015mga, deLima:2015pqa, Omura:2015nja,
Dorsner:2015mja, Altunkaynak:2015twa, Crivellin:2015hha, Altmannshofer:2015esa,
Botella:2015hoa, Arhrib:2015maa, Benbrik:2015evd}.  In a type~III 2HDM, small flavor violating
couplings of the light SM-like Higgs boson $h$ are complemented by large flavor
violating couplings of the heavy Higgs bosons $H^0$, $A^0$, and $H^\pm$, opening up new
production and decay channels for the latter.  The $thh$ signature arises for instance
in the process $pp \to t + (H^0 \to h h)$. We will discuss this process in a
detailed Monte Carlo study, and we will also consider the related processes $p
p \to t + (H^0 \to tu, WW, ZZ)$.

Finally, we will also discuss leptonic Higgs-induced FCNC in the 2HDM,
in particular the process $p p \to H^0 \to \tau\mu$. We will constrain this
process, which could be directly connected to the
tentative hint from CMS, using existing LHC data.

The structure of the paper is as follows: in \cref{sec:effOperator}, we briefly
introduce the effective field theory for FCNC Higgs decays given by operators like
\cref{eq:Q-dij,eq:Q-uij,eq:Q-eij}. We then develop an LHC search for the
$thh$ final state, carefully dissecting the kinematic distributions of the signal
and the various backgrounds, and we estimate the expected sensitivity at the 13~TeV LHC.
In the second part of the paper, \cref{sec:2HDM},
we proceed to a discussion of the quark flavor violating type~III 2HDM. We adapt the
general search strategy for $thh$ production, developed in \cref{sec:effOperator},
to the 2HDM, and we compare its sensitivity to other constraints on the model.
Finally, in \cref{sec:2HDM-Htaumu}, we discuss lepton flavor violation in the
2HDM. We constrain the process $H^0 \to \tau\mu$
using LHC data, and we discuss the implications of these constraints for the
CMS hint in $h \to \tau\mu$. We summarize our results and conclude in
\cref{sec:conclusions}.

\section[$thh$ Production in Effective Field Theory]
        {\texorpdfstring{$thh$}{thh} Production in Effective Field Theory}
\label{sec:effOperator}

\subsection{Effective Field Theory Framework for Flavor Violating Higgs Couplings}
\label{sec:eft-intro}

In the SM, fermions couple to the Higgs doublet via renormalizable dimension
four Yukawa couplings, which, after electroweak symmetry breaking, source both
the fermion mass terms and the Yukawa couplings of the physical Higgs boson.
Therefore, in the SM, no flavor violating couplings of the physical Higgs boson
are possible. In extensions of the SM, however, the mass matrices and Yukawa
couplings can be misaligned in flavor space.  When written in terms of contact
operators, the leading contributions to such misalignment come from the
dimension six operators given in \cref{eq:Q-dij,eq:Q-uij,eq:Q-eij}. In this
section, we focus on up-type quarks only.  FCNC couplings of the Higgs boson to
down-type quarks are tightly constrained~\cite{Blankenburg:2012ex,
Harnik:2012pb}, and the possible LHC signatures of FCNC couplings to leptons
will be addressed in \cref{sec:2HDM-Htaumu}.  The relevant dimension four and
six couplings in the up-type sector are
\begin{align}
  \mathcal{L} \supset
    -\lambda_u^{ij} \overline{Q_L^i} \tilde{H} u_R^j
    -\frac{\lambda_u^{\prime\,ij}}{\Lambda^2} \overline{Q_L^i} \tilde{H} u_R^j (H^\dag H)
    + h.c. \,.
\end{align}
After electroweak symmetry breaking, this Lagrangian becomes
\begin{align}
  \mathcal{L} \supset - m_u^{ij} \, \overline{u_L^i} u_R^j
                      - y_u^{ij} \, \overline{u_L^i} u_R^j h
                      - \frac{f_u^{ij}}{v} \, \overline{u_L^i} u_R^j h^2
                      + \mathcal{O}(h^3 ) + h.c. \,,
  \label{eq:EFT1}
\end{align}
where the mass and coupling matrices are given by
\begin{align}
  m_u^{ij} &= \frac{v}{\sqrt{2}} \bigg( \lambda_u^{ij}
                + \frac{v^2}{\Lambda^2} \frac{\lambda_u^{\prime\,ij}}{2} \bigg) \,, \\
  y_u^{ij} &= \frac{m_u^{ij}}{v}
                + \frac{v^2}{\Lambda^2} \frac{\lambda_u^{\prime\,ij}}{\sqrt{2}} \,, \\
  f_u^{ij} &= \frac{v^2}{\Lambda^2} \frac{3 \lambda_u^{\prime\,ij}}{2\sqrt{2}} \,.
\end{align}
In the above expressions, $i,j = 1,2,3$ are again flavor indices.  To improve
readability, we will also use the notation $i,j=u,c,t, \ldots$ and omit the
subscript $u$ where this is unambiguous, e.g.\ $y^{tu} \equiv y_u^{31}$,
$f^{tu} \equiv f_u^{31}$, etc.
In the up quark mass basis, where $m_u^{ij}$ is diagonal, we see that the
flavor violating couplings satisfy
\begin{align}
  f_u^{ij} &= \frac{3}{2} y_u^{ij} \,.  &  (i \ne j)
\end{align}

Currently, the strongest experimental limits on the off-diagonal elements of $y_u$
come from ATLAS~\cite{Aad:2014dya,Aad:2015pja} and impose the 95\% CL constraints
\begin{align}
  \BR(t \to ch) < 0.0046
  \qquad\qquad\text{and}\qquad\qquad
  \BR(t \to uh) < 0.0045
  \label{eq:BR-tch-limit}
\end{align}
on FCNC top quark decays, which translates into
\begin{align}
  \sqrt{|y^{ct}|^2 + |y^{tc}|^2}  < 0.13 \,
  \qquad\qquad\text{and}\qquad\qquad
  \sqrt{|y^{ut}|^2 + |y^{tu}|^2}  < 0.12 \,.
  \label{eq:yct-limit}
\end{align}
Here, we have used the leading order expression for the branching
ratio~\cite{Greljo:2014dka}, supplemented by a correction factor
$\eta_{QCD} \simeq 1+0.97\alpha_s=1.10$~\cite{Zhang:2013xya,Greljo:2014dka}
accounting for NLO QCD contributions:
\begin{equation}
  \BR(t\to hq)
  = \frac{|y^{tq}|^2 + |y^{qt}|^2} {2\sqrt{2} G_F}
    \frac{(m_t^2 - m_h^2)^2}{(m_t^2 - m_W^2)^2 (m_t^2 + 2 m_W^2)}\eta_{QCD}
    \simeq 0.29 \big( |y^{tq}|^2 + |y^{qt}|^2 \big)\,.
  \label{eq:BR-thq}
\end{equation}
Note that a refined analysis taking into account also the process $p p \to t
h$, which is relevant in the case of $tuh$ couplings (but not for $tch$
couplings) could improve the bounds on $\BR(t \to uh)$ and $\sqrt{|y^{ut}|^2 +
|y^{tu}|^2}$ by about a factor 1.5~\cite{Greljo:2014dka}.  Constraints from CMS
are of the same order as those from ATLAS: CMS obtain
the 95\%~CL bound
$\BR(t \to ch) < 0.0056$ using
searches for
multi-lepton and lepton plus di-photon final states. This bound
translates into $\sqrt{|y^{ct}|^2 + |y^{tc}|^2}
< 0.14$~\cite{CMS:2014qxa,Khachatryan:2014jya,CMS:2015qta}.
The same bound holds also for $tuh$ couplings.  A secondary process
sensitive to anomalous $tuh$ and $tch$ couplings is same-sign top production
through $t$-channel Higgs exchange. The author of
ref.~\cite{Goldouzian:2014nha} has derived limits on $\sqrt{|y^{tq}|^2 +
|y^{qt}|^2}$ from this channel by recasting the CMS same-sign di-lepton plus
$b$ jet measurements~\cite{Chatrchyan:2012paa}.  However, the resulting bounds
are weaker than those from $t \to q h$ decays.  Finally, an effective $tuh$ or
$tch$ coupling may lead to anomalous di-Higgs production, mediated by a top
quark in the $t$-channel. Unfortunately, since the rate of this process is suppressed
by four powers of the small coupling constants $y^{tq}$ and $y^{qt}$ ($q = u, c$),
it is irrelevant in practice. For instance, for $y^{ut} = y^{tu} = 0.08$ at the
upper limit from \cref{eq:yct-limit}, the cross section for $p p \to h h$ is
only $\sim 4$~fb at $\sqrt{s} = 8$~TeV and $\sim 7.4$~fb at $\sqrt{s} = 13$~TeV.

\begin{figure*}
  \includegraphics[width=0.7\textwidth]{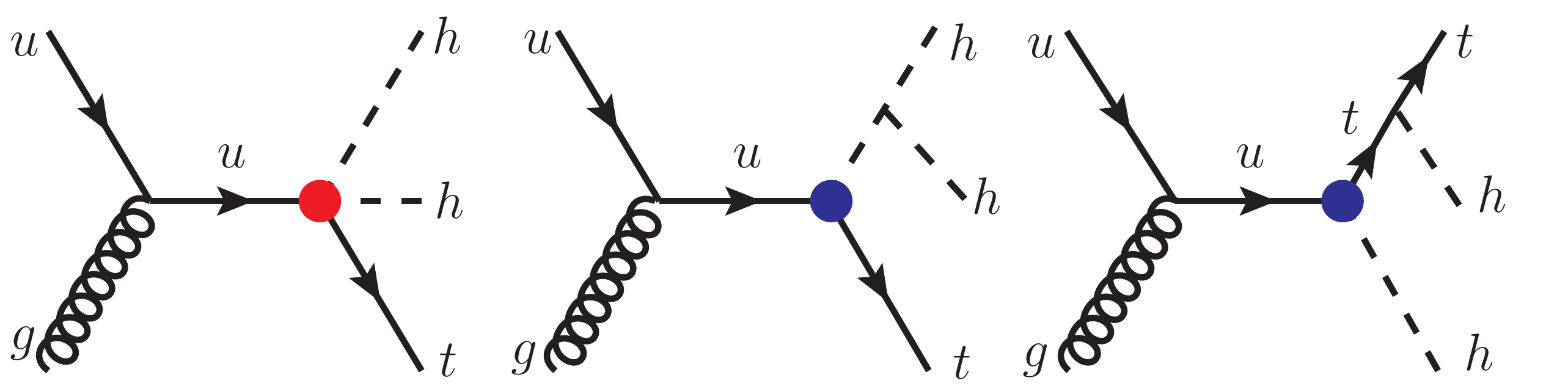} \\[0.2cm]
  \includegraphics[width=0.9\textwidth]{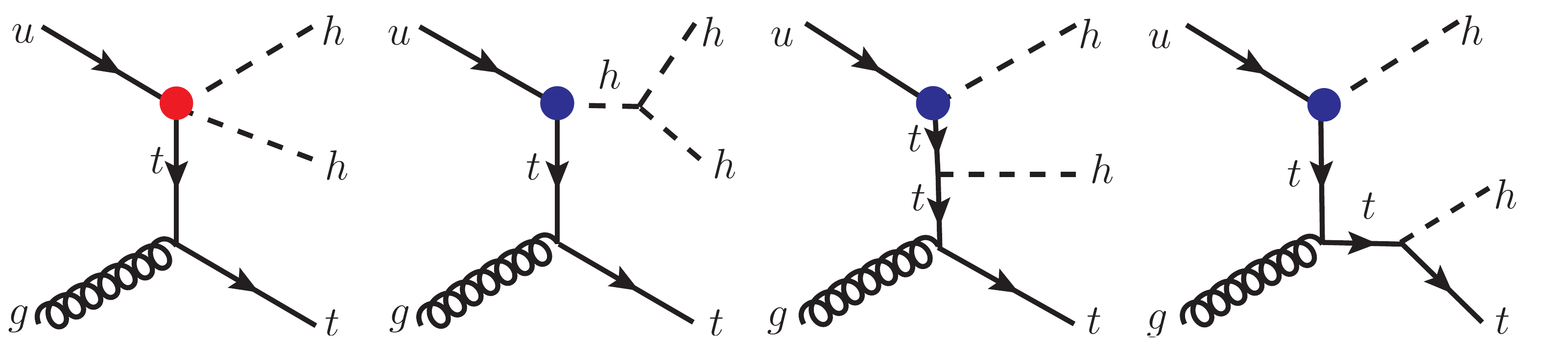}
  \caption{Sample Feynman diagrams for $thh$ production from the
    vertices in the effective Lagrangian in \cref{eq:EFT1}.  Red dots indicate
    the FCNC coupling of two Higgs boson to quarks (third term in
    \cref{eq:EFT1}), which is usually dominant, blue dots stand for the FCNC
    coupling of a single Higgs boson to quarks (second term in \cref{eq:EFT1}),
    the contributions of which are usually subdominant.}
  \label{fig:EFT-feyn}
\end{figure*}

In view of the above constraints, we will in the following use benchmark values
of $y^{tq} = y^{qt} = 0.08$, leading to the expectation of potentially
measurable rates for the process $p p \to t h h$, on which we will focus in the
first part of this paper.  The corresponding Feynman diagrams are given in
Fig.~\ref{fig:EFT-feyn}.

\subsection[LHC Search Strategy for $thh$ Production in EFT]
           {LHC Search Strategy for \texorpdfstring{$thh$}{thh}
            Production in EFT}
\label{sec:simulationEFT}

In the following we investigate the process $p p \to t h h$ in more detail
using Monte Carlo simulations, with the aim of developing a search strategy for
future LHC analyses at 13~TeV and estimating its sensitivity. To maximize the number of
signal events, it is desirable to focus on Higgs decay channels with large
branching ratio, in particular $h \to b\bar{b}$.  Since the final state offers
many kinematic handles (in particular for invariant mass cuts), we
expect to be able to suppress backgrounds efficiently even for hadronic Higgs
decays. We will, however, require the top quark to decay semileptonically to avoid
the QCD multi-jet background. Hence, our final state consists of five
$b$ jets, one hard lepton, and missing
energy.
The dominant background for the $\ell + 5b + \slashed{E}_T$ final
state is inclusive $t \bar{t} + \text{jets}$ production. Other potential
backgrounds such as $W + \text{jets}$ or QCD multi-jet production can be
efficiently suppressed by requiring $b$ tags and by requiring two jet pairs to
have invariant masses close to the Higgs mass $m_h$.  We have also considered
other decay modes of the Higgs bosons and the top quark, but found them to be
less favorable due to small branching ratios and due to the relative softness
of leptons from Higgs decays.

For the simulation of the signal, we use
\textsc{MadGraph~5~v2.3}~\cite{Alwall:2014hca} to compute leading order cross
sections and to generate events, which are then passed to
\textsc{Pythia~6.4}~\cite{Sjostrand:2006za} for parton showering, MLM jet
matching~\cite{Mangano:2002ea} and hadronization. We use a jet matching scale
$q = 30$~GeV. As a detector simulation, we employ
\textsc{Delphes~3.1.2}~\cite{deFavereau:2013fsa}.  Background events from $t
\bar{t} + \text{jets}$ production are generated by
\textsc{Sherpa+OpenLoops}~\cite{Gleisberg:2008ta, Krauss:2001iv,
Cascioli:2011va, Denner:2014gla}.  We simulate events with zero or one jet at
next-to-leading order (NLO) level, and we allow for up to three
jets simulated at leading order (LO) using the MC@NLO multi-jet merging
algorithm.  We treat $c$ and $b$ quarks as heavy in the parton shower.  To
confirm that background other than $t \bar{t} + \text{jets}$ are negligible, we
also simulated $V + \text{jets}$, $V V' + \text{jets}$, $t \bar{t} + V$, $t
\bar{t} + h$ and single-top production, with $V,V'=W, Z$, using
\textsc{Sherpa\,+\,BlackHat}~\cite{Berger:2008sj}.  The first of these
backgrounds is treated at NLO accuracy, while the others are computed at LO. As
expected, all of them are tiny after cuts, mainly due to a lack of
$b$-tagged jets.

Our analysis pipeline starts with a set of preselection cuts: we require
exactly one reconstructed isolated and positively charged lepton with
transverse momentum $p_T \ge 10$~GeV and pseudorapidity $|\eta| \le 2.4$,
and at least five jets with $p_T \ge 20$~GeV
and $|\eta| \le 2.5$. Jets are reconstructed with the anti-$k_T$ algorithm with
cone radius $R=0.5$.
At least four jets need to be $b$-tagged, assuming a tagging-efficiency of
70\% and a misidentification rate of 1\% for jets initiated by a light quark
or a gluon~\cite{Chatrchyan:2012jua,Pardos:2013joa}.
In principle, the desired $t + (h \to b\bar{b})
(h \to b\bar{b})$ final state leads to 5 $b$ jets, but due to the limited
$b$-tagging efficiency, requiring only 4 $b$ jets improves the sensitivity.

The longitudinal momentum component $p_\nu^z$ of the neutrino is reconstructed by
using the on-shell condition for the $W$-boson, $m_{\ell\nu}^2=m_W^2$:
\begin{align}
  p_\nu^z = \frac{1}{2p_{T \ell}^2} \Big[
              (m_W^2 + 2\vec{p}_{T \ell} \cdot \vec{\slashed{p}}_T) p_\ell^z
          \pm E_\ell \sqrt{(m_W^2 + 2 \vec{p}_{T \ell} \cdot \vec{\slashed{p}}_T)^2
            - 4 p_{T \ell}^2 \, \slashed{p}_T^2} \bigg] \,.
  \label{eq:p-nu-z}
\end{align}
Here, $m_{\ell\nu}$ is the invariant mass of the charged lepton and the neutrino,
$E_\ell$ and $\vec{p}_{T \ell}$ are the energy and transverse momentum of the charged
lepton, and
$\vec{\slashed{p}}_T$ is the missing transverse momentum. We assume here that
the neutrino is the only source of missing energy.
Note that \cref{eq:p-nu-z} has two solutions, which may be complex.
To break the ambiguity, and to associate each jet with a particular parent
particle (one of the two Higgs bosons or the top quark), we minimize the
quantity
\begin{align}
  \chi^2 \equiv \frac{(m_{jj}^{(1)} - m_h)^2}{(\Delta m_h)^2}
              + \frac{(m_{jj}^{(2)} - m_h)^2}{(\Delta m_h)^2}
              + \frac{(m_{j\ell\nu} - m_t)^2}{(\Delta m_t)^2} \,
  \label{eq:chi2-combinatorics}
\end{align}
over all possible associations between jets and parent particles and
over the two possible values of $p_\nu^z$.  We use only the five leading jets
in this procedure, and we do not distinguish between $b$ tagged and untagged
jets here.  In the above expression,
$m_{jj}^{(1)}$ and $m_{jj}^{(2)}$ are the invariant masses of jet pairs
and $m_{j\ell\nu}$ is the invariant mass of a jet, the lepton and the neutrino.
For the uncertainties in the denominators we take $\Delta m_h = 12$~GeV
(the mass resolution for $h \to b \bar{b}$ in CMS \cite{Chatrchyan:2012xdj,CMS:yva})
and $\Delta m_t = 1.35$~GeV (the width of top quark~\cite{Agashe:2014kda}).
We have checked that varying $\Delta m_h$ and $\Delta m_t$ by $\mathcal{O}(1)$
factor relative to each other does not alter our results.

To sharpen the signal, and to reduce the background, we impose further cuts.
The thre leading jets $p_T$, denoted as $j_1$, $j_2$ and $j_3$, have to fulfill
$p_{T,j_1} > 140$~GeV, $p_{T,j_2} > 100$~GeV, and $p_{T,j_3} > 60$~GeV.
For the reconstructed
invariant masses $m_{j\ell\nu}$ of the top quark and $m_{jj}^{(1)}$,
$m_{jj}^{(2)}$ of the Higgs bosons, we require $\text{150~GeV} < m_{j\ell\nu} <
\text{200~GeV}$ and $\text{100~GeV} < m_{jj}^{(1,2)} < \text{150~GeV}$.
The Higgs boson with the larger $p_T$, which we will call $h_1$, is required to have
$p_{T,h_1} > 300$~GeV, and the Higgs boson with the smaller $p_T$, called $h_2$,
has to satisfy $p_{T,h_2} > 150$~GeV.  We finally require the angular separation
$\Delta R_{bb}^{h_1,h_2}$
between the two jets associated with the same Higgs boson decay to be not too large:
we impose the cut $\Delta R_{bb}^\text{max} = \max(\Delta R_{bb}^{h_1},
\Delta R_{bb}^{h_2}) < 1.5$.

\begin{figure*}
  \begin{tabular}{ccc}
    \includegraphics[width=0.32\textwidth]{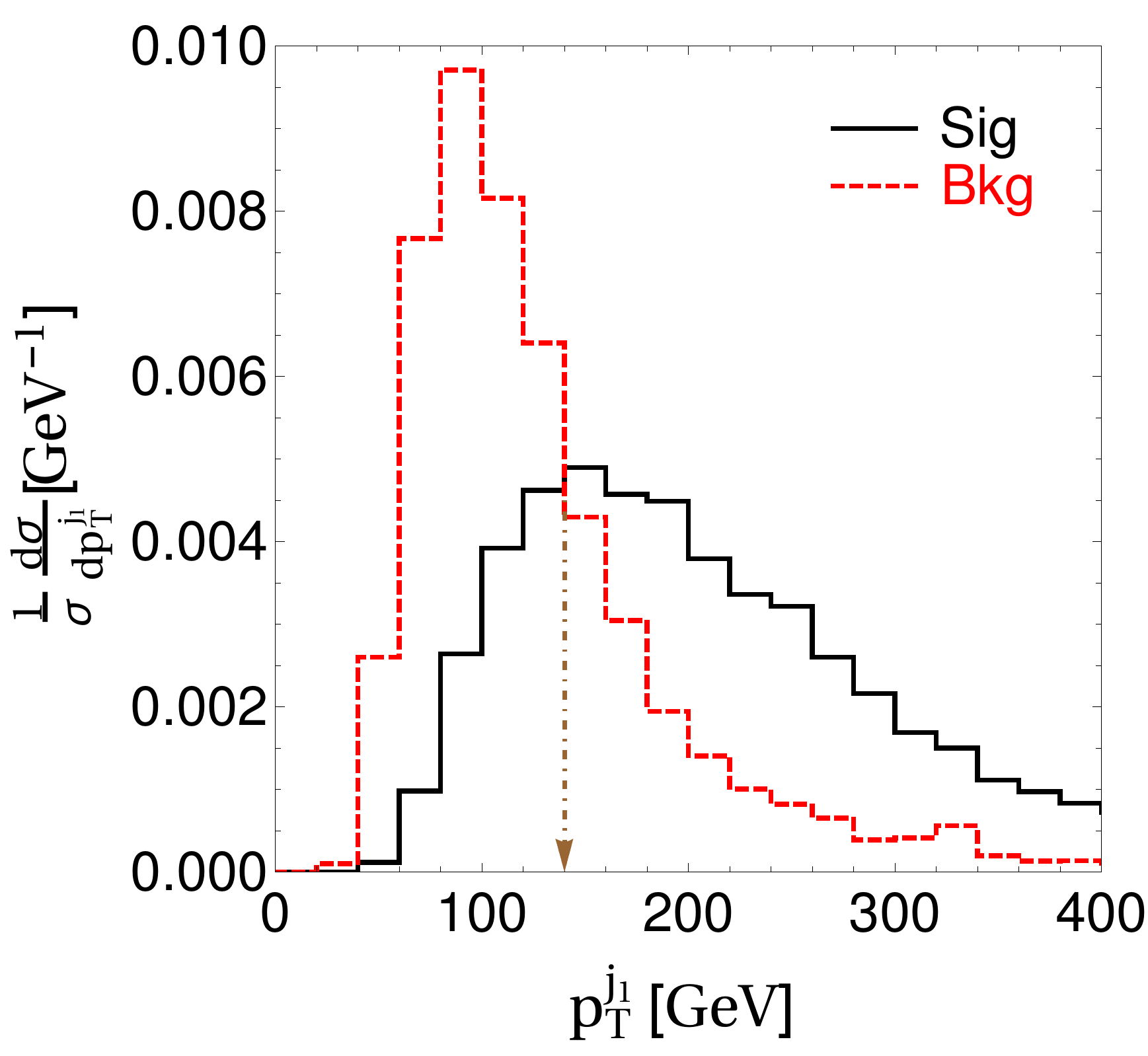} &
    \includegraphics[width=0.32\textwidth]{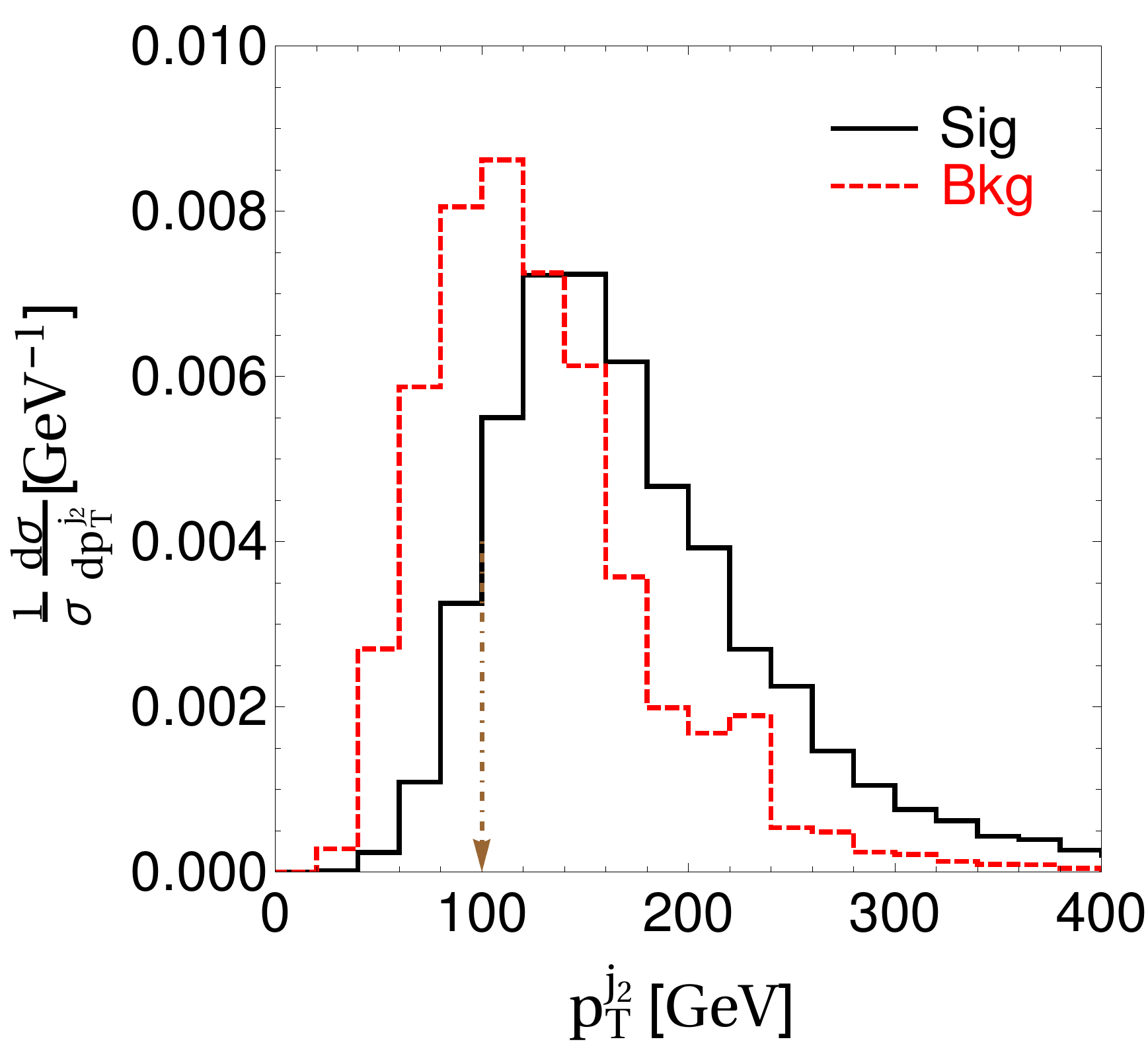} &
     \includegraphics[width=0.32\textwidth]{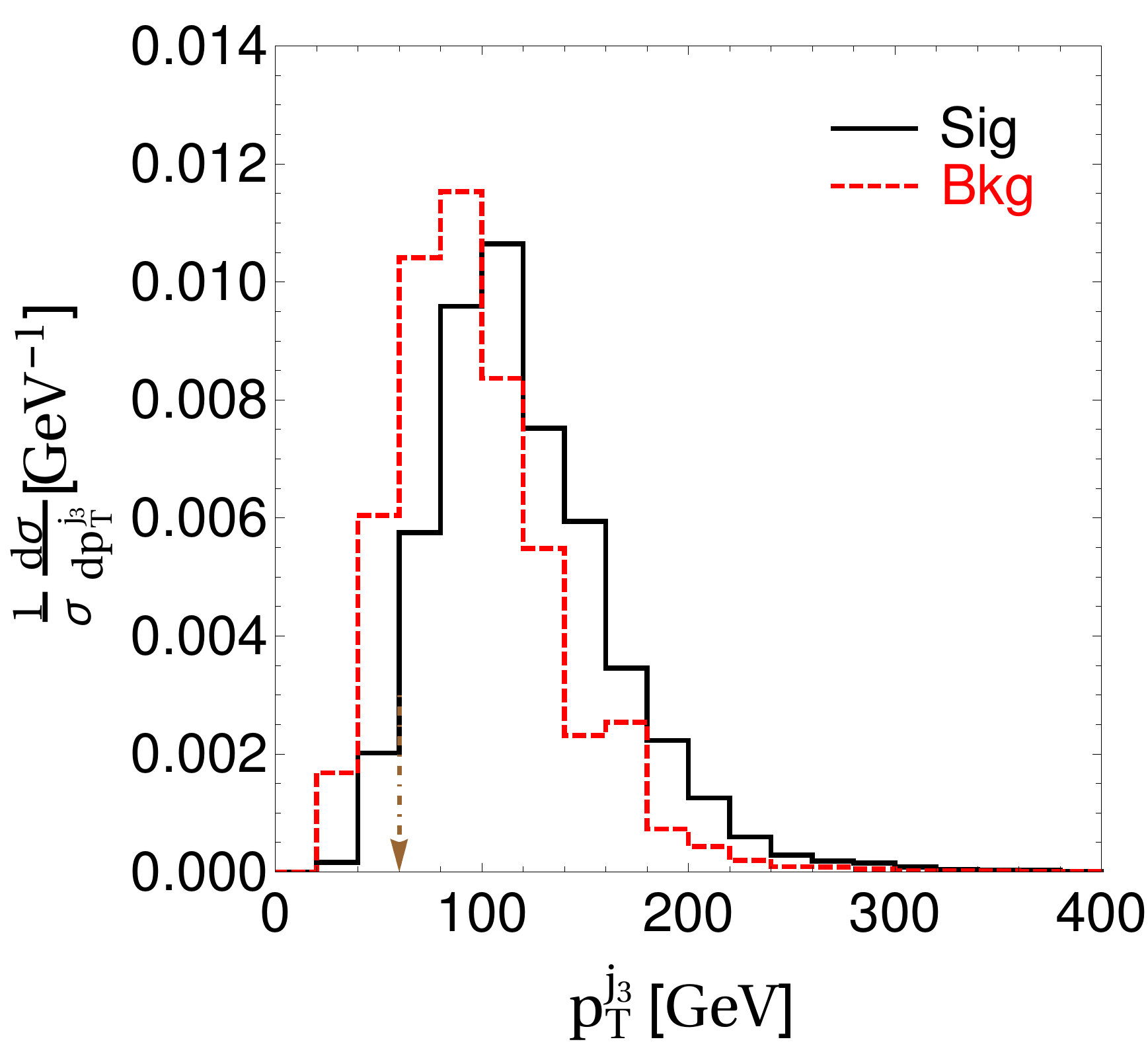}  \\
    (a) & (b) &(c)  \\
    \includegraphics[width=0.32\textwidth]{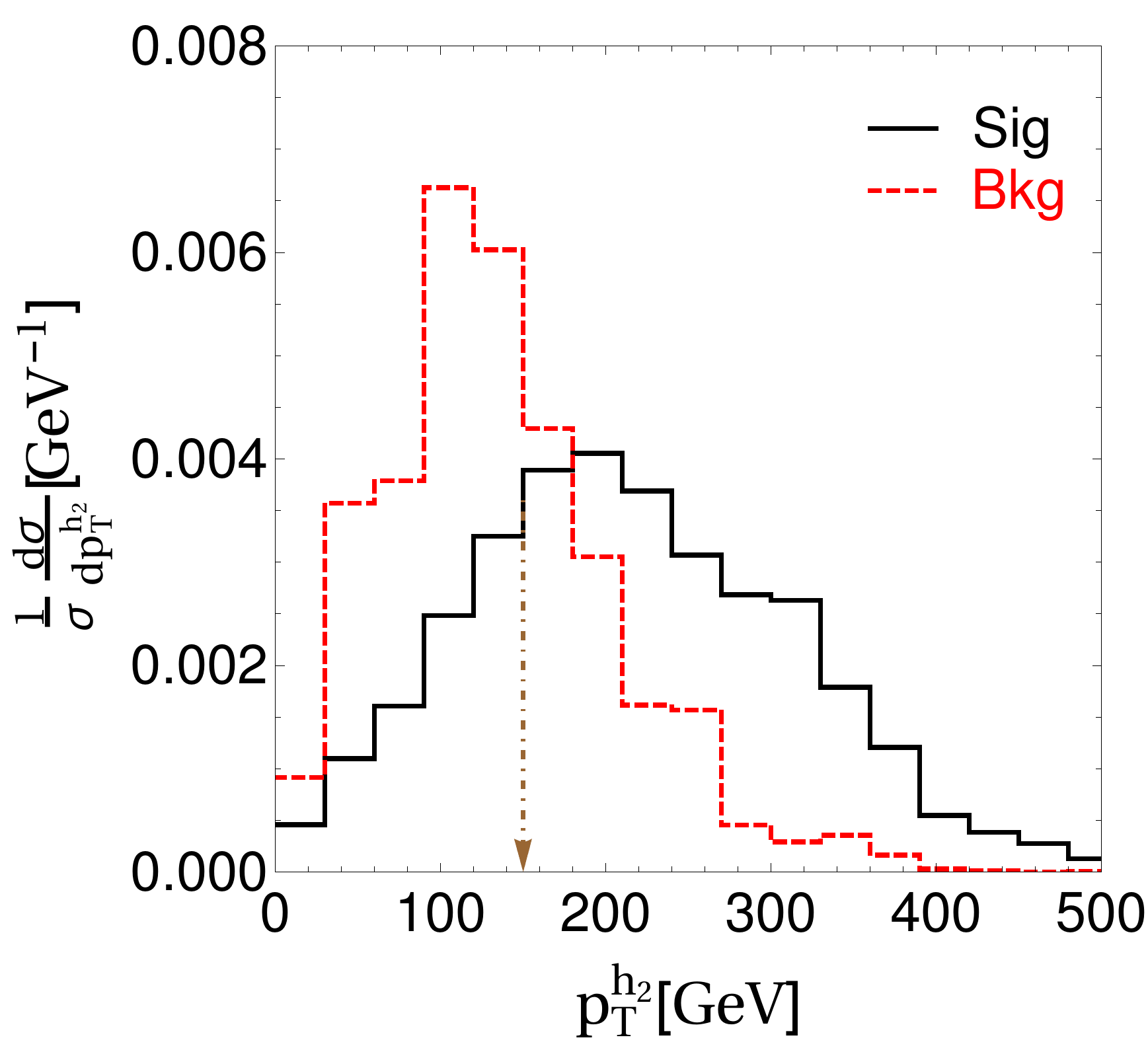} &
    \includegraphics[width=0.32\textwidth]{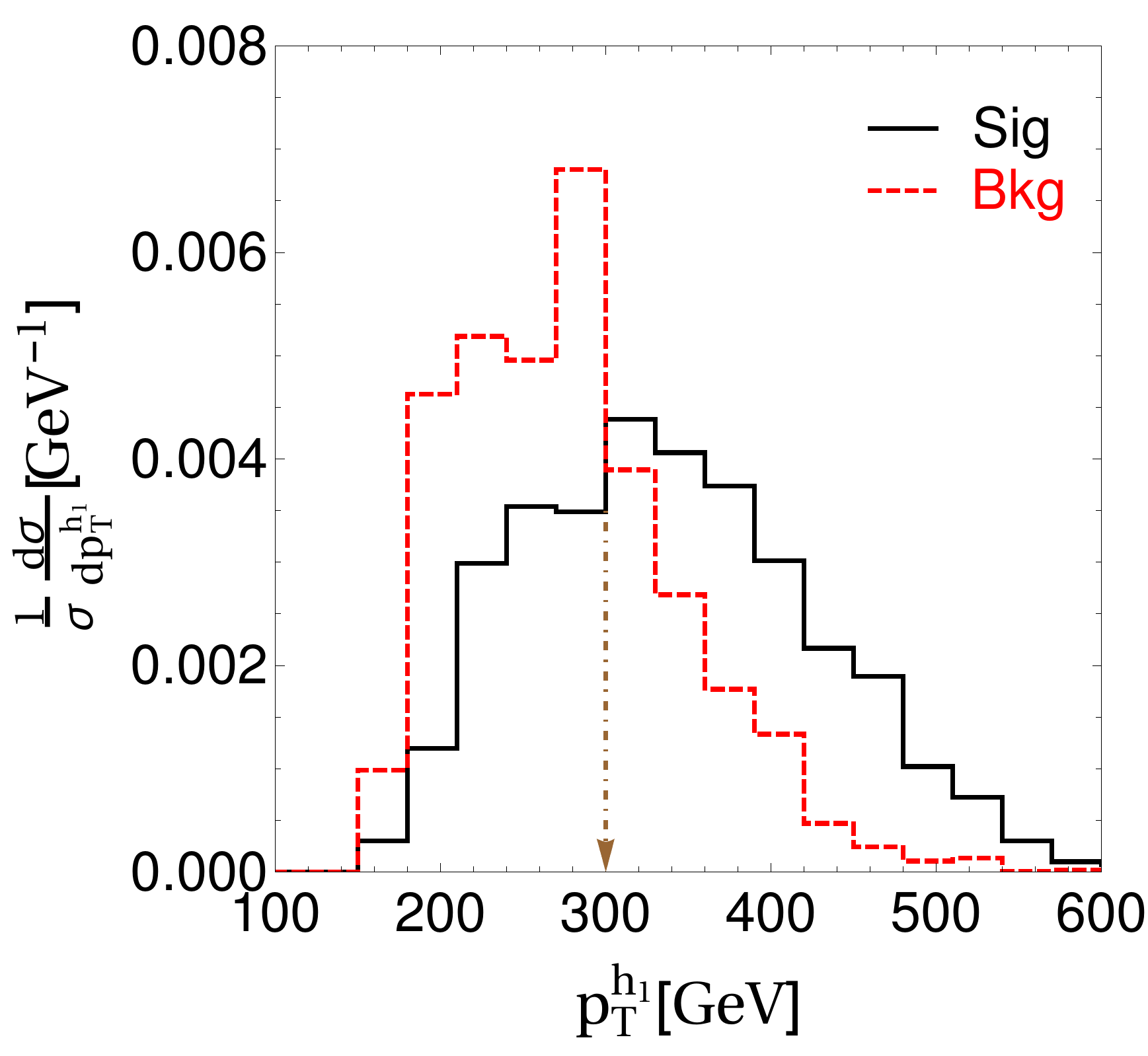} &
        \includegraphics[width=0.32\textwidth]{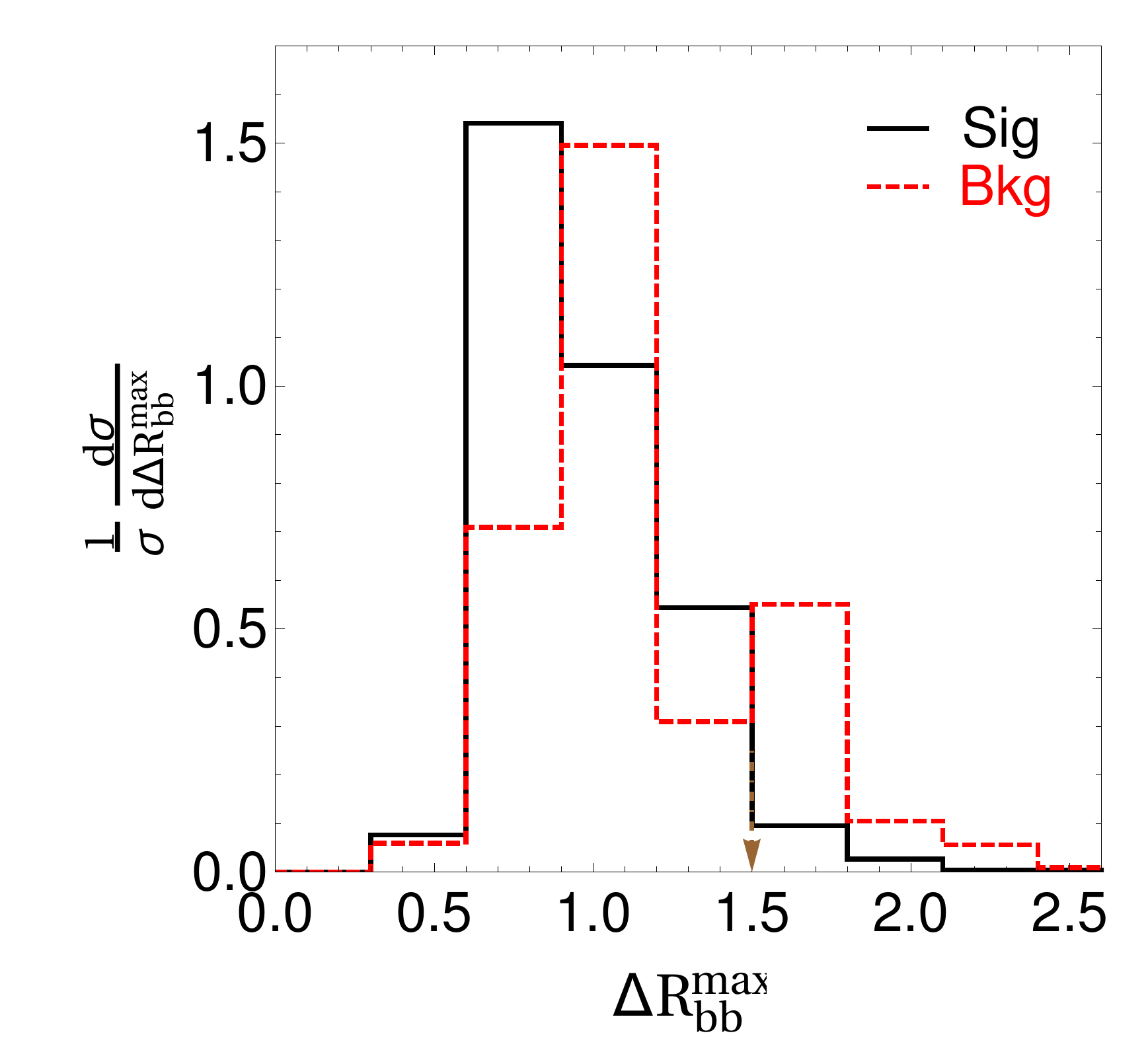} \\
    (d) & (e) &(f)
  \end{tabular}
  \caption{Kinematic distributions of the $t\bar{t}$ background (red dashed) and of the
    $thh$ signal (black solid) in the effective field theory framework defined
    by the Lagrangian \cref{eq:EFT1}. In panels (a), (b)
    and (c), we show the transverse momentum distributions of the three leading
    jets, while panels (d) and (e) display the reconstructed $p_T$
    distributions of the two Higgs bosons.  In panel (f), we define the angular
    separation $\Delta R_{bb}^{h_1 (h_2)}$ between the two $b$ jets from the
    decay of the harder (softer) Higgs boson in an event, and we plot the
    distribution of the larger of the two, $\Delta R_{bb}^\text{max} =
    \max(\Delta R_{bb}^{h_1}, \Delta R_{bb}^{h_2})$.  The vertical arrow in
    each panel indicates the cut we impose on the respective kinematic quantity
    (see \cref{tab:cutflow1}), and in all subsequent panels, this cut is
    applied.  All distributions are normalized to unity.}
  \label{fig:distributionEFT1}
\end{figure*}

\begin{figure*}
  \begin{tabular}{cc}
    \includegraphics[width=0.45\textwidth]{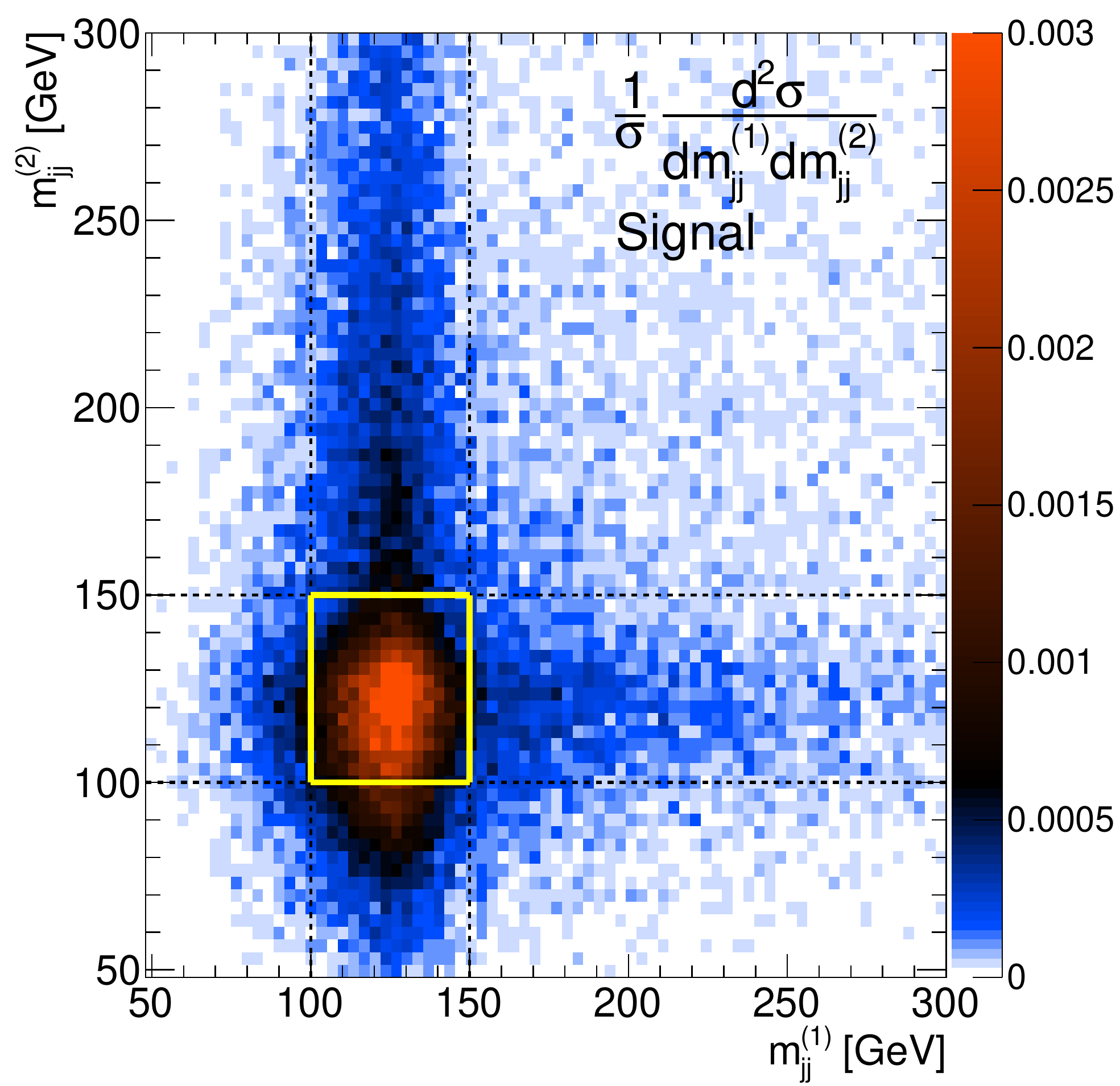} &
    \includegraphics[width=0.45\textwidth]{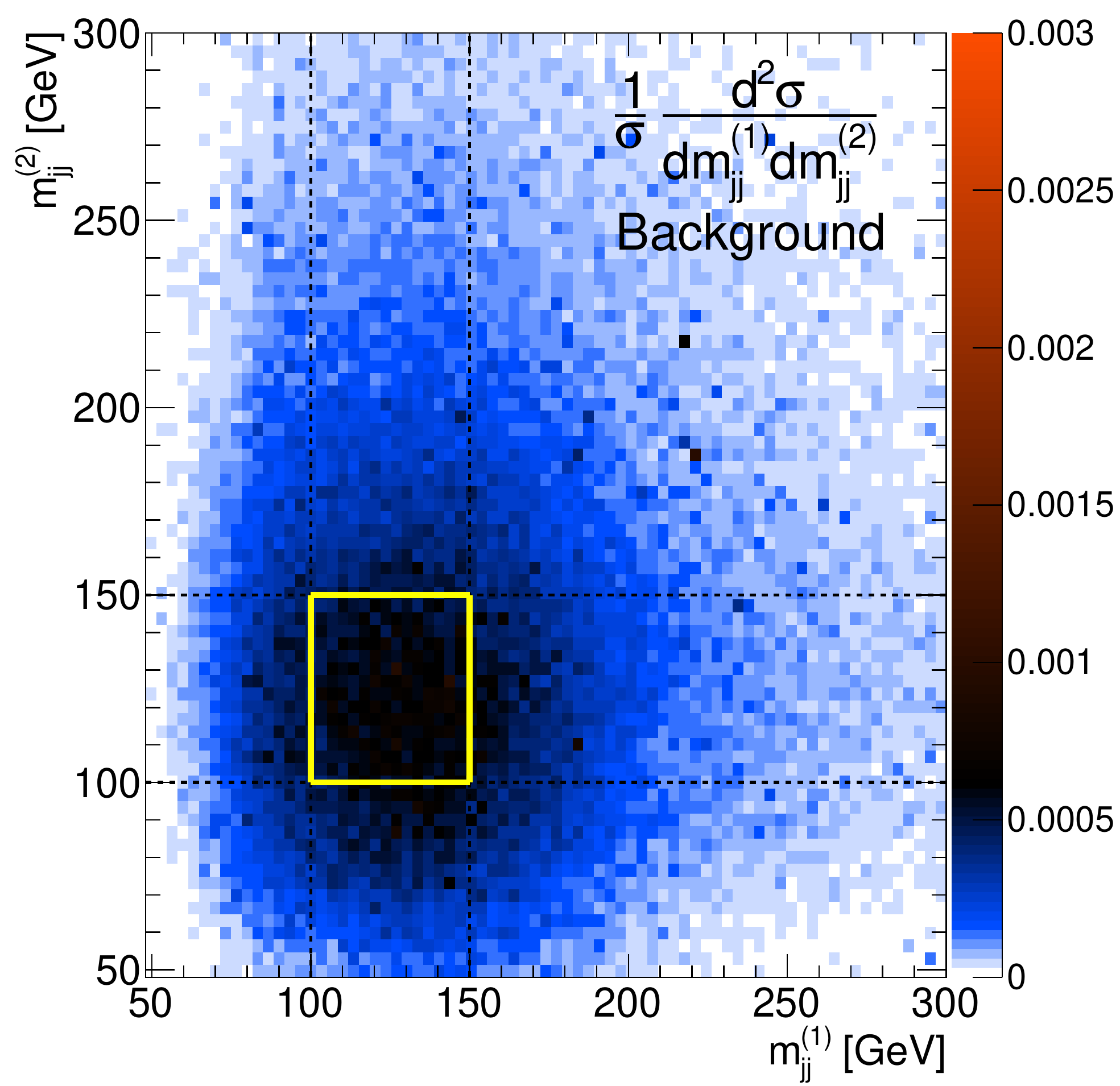} \\
    (a) & (b)
  \end{tabular}
  \caption{Distribution of the reconstructed Higgs masses $m_{jj}^{(1,2)}$
    for (a) $thh$ signal events simulated in the effective field theory defined
    by the Lagrangian \cref{eq:EFT1}, and (b) $t\bar{t}$ background events.
    The yellow square indicates the invariant mass cuts imposed in our analysis.}
  \label{fig:EFT-BloodStain}
\end{figure*}

In \cref{fig:distributionEFT1,fig:EFT-BloodStain}, we show the kinematic
distributions on which we cut, illustrating how our cuts help to separate the $p p
\to t h h$ signal from the $pp \to t\bar{t} + \text{jets}$
background.
From the $p_T$ distributions of the three leading jets (\cref{fig:distributionEFT1}
(a), (b), (c)),  we note that the signal (black solid) is in general harder than the
background (red dashed),
motivating our cuts on the jet transverse momenta. The reason background jets
have on average smaller $p_T$ than signal jets is that pair produced top quarks
are predominantly forward, while the signal is more central.
The same behavior is also reflected in the distribution of the
reconstructed transverse momenta of the two Higgs bosons (\cref{fig:distributionEFT1} (d)
and (e)). Regarding the angular separation $\Delta R_{bb}^\text{max}$, we find
it to be slightly smaller for the signal than for the background, see
\cref{fig:distributionEFT1} (f). The reason is that for the signal, pairs of
$b$ jets originate from the same parent particle, while for the background no such
correlation needs to exist. We have also considered the distribution of the number of jets,
the pseudorapidity distributions of the reconstructed
top quark and Higgs bosons, the angular separations $\Delta R$ between them, and the
invariant mass of the $hh$ system, but have found that these distribution do not offer
additional handles to separate signal from background. Similarly, also attempts
to identify background events based on the presence of two jets without $b$ tags and
with an invariant mass $\sim m_W$ have not lead to an improvement of the sensitivity.

The most critical cuts in suppressing backgrounds in our search are those on
the invariant masses $m_{jj}^{(1,2)}$. (The cut on $m_{j\ell\nu}$ does not reduce
the number of background $t\bar{t}$ events significantly because also background events
contain actual top
quarks.) To illustrate the power of the $m_{jj}^{(1,2)}$ cuts, we show in
\cref{fig:EFT-BloodStain} the two-dimensional distribution of $m_{jj}^{(1)}$
vs.\ $m_{jj}^{(2)}$. We see that, for signal events a pronounced peak is visible
around the true Higgs mass $m_h \sim 125$~GeV. The background distributions
reveal no such peak, but rather a broad
bump around $m_{jj}^{(1,2)} \sim 125$~GeV. This broad bump can be understood if
we consider that
the minimization procedure used to break combinatorial ambiguities (see
\cref{eq:chi2-combinatorics}) favors combinations in which pairs of jets
have invariant masses close to $m_h$ by chance.

\begin{table}
  \centering
  \begin{minipage}{13cm}
  \begin{ruledtabular}
  \begin{tabular}{lcc}
    cut                            & signal ($thh$) & background ($t\bar{t}$) \\
    \hline
    $\sigma_\text{prod}$ [fb]      &  6.1           & $5.9\times 10^5$ \\
    \hline
    preselection                   &  24.0\%        &  \phantom{0}2.20\% \\
    $b$-tagging                    &  19.6\%        &  \phantom{0}0.55\% \\
    $p_T^{j_1} > 140$~GeV          &  76.5\%        &  31.1\% \\
    $p_T^{j_2} > 100$~GeV          &  90.9\%        &  66.3\% \\
    $p_T^{j_3} > 60$~GeV           &  95.7\%        &  84.6\% \\
    Higgs, top mass window         &  24.4\%        &  \phantom{0}8.55\% \\
    $p_T^{h_2} > \text{150~GeV}$   &  73.3\%        &  35.3\% \\
    $p_T^{h_1} > \text{300~GeV}$   &  65.5\%        &  32.3\% \\
    $\Delta R_{b \bar{b}}^{\rm{max}} < 1.5$ & 96.1\% & 77.2\% \\
    \hline
    $\sigma_\text{final}$~[fb]     & 0.022          & 0.093
  \end{tabular}
  \end{ruledtabular}
  \end{minipage}
  \caption{Cut flow table for the $thh$ signal from the effective Lagrangian
    \cref{eq:EFT1}. As in \cref{fig:distributionEFT1}, the quantity $\Delta
    R_{b \bar{b}}^{\rm{max}}$ is defined as $\max(\Delta R_{bb}^{h_1}, \Delta
    R_{bb}^{h_2})$, where $\Delta R_{bb}^{h_1}$ ($\Delta R_{bb}^{h_2}$) is the
    angular separation between the two $b$ jets associated with the harder
    (softer) Higgs boson.}
  \label{tab:cutflow1}
\end{table}

The cut flow of our analysis is summarized in \cref{tab:cutflow1}. We find that
the total leading order cross section for $p p \to thh$ is 56.2~fb at our
benchmark point with $y^{ut} = y^{tu} = 0.08$. Including the branching ratios
for $t \to b \ell\nu$ and $h \to b\bar{b}$, this decreases to 6.1~fb. Our cuts,
especially the $b$ tagging requirements and the mass window cuts, lead to a
signal cross section after cuts of only 0.022~fb.  The reason the $b$-tagging
and mass window cuts reduce not only the background, but also the signal
substantially is that due to combinatorial uncertainties it is likely that one
of the two Higgs bosons is not properly reconstructed, for instance because one
of the jets from its decay is very soft.  Therefore, the final signal cross
section is an order of magnitude smaller than the background cross section.
Nevertheless, with
sufficient luminosity, for instance at the high luminosity LHC, a
discrimination between signal and background may be possible. Using the
$CL_s$ method~\cite{Read:2002hq} and assuming a systematic uncertainty of 30\% on
the signal and the background, we find that a luminosity of about
870~fb$^{-1}$ is needed to exclude our benchmark point at the $95\%$ CL.

This shows that, in the effective field theory framework,  a search for the
$thh$ final state is inferior to the traditional searches for flavor violating
top decays.  In the next section, we will show that this conclusion changes
when considering specific renormalizable models---in particular two Higgs
doublet models---instead of the effective theory.

\section[$pp \to t+H^0$ in the Two Higgs Doublet Model]
        {\texorpdfstring{$pp \to t+H^0$}{pp->t+H^0} in the
         Two Higgs Doublet Model}
\label{sec:2HDM}

\subsection{Brief Introduction to 2HDMs}
\label{sec:2HDM-intro}

Among the most studied extensions of the SM leading to flavor changing
Higgs couplings are Two Higgs Doublet Models (2HDMs).
As their name suggests, 2HDMs augment the
SM with an additional scalar $SU(2)_L$ doublet~\cite{Lee:1973iz} (see
ref.~\cite{Branco:2011iw,Diaz:2002tp} for recent reviews, refs.~\cite{Celis:2013rcs,
Baglio:2014nea} for work in the context of the Higgs discovery, and
refs.~\cite{Bjorken:1977vt, DiazCruz:1999xe, Han:2000jz, Kanemura:2005hr,
Davidson:2010xv, Kopp:2014rva, Sierra:2014nqa, Crivellin:2015mga, deLima:2015pqa,
Omura:2015nja, Dorsner:2015mja, Altunkaynak:2015twa, Crivellin:2015hha,
Altmannshofer:2015esa, Botella:2015hoa, Arhrib:2015maa, Benbrik:2015evd} for studies
related to flavor violating Higgs couplings).
In a 2HDM, we are free to change basis in
the Higgs sector by forming linear combinations of the two doublets, and we will
choose a basis (often called the Georgi basis) where only one of the Higgs doublets
has a vev, $v = 246$~GeV.  In this basis, the two Higgs doublets $\Phi_1$ and $\Phi_2$ can be
decomposed into their component fields as~\cite{Celis:2013rcs},
\begin{align}
  \Phi_1 = \begin{pmatrix}
             G^+ \\
             \frac{1}{\sqrt{2}} (v + h_1 + i G^0)
           \end{pmatrix}
  \qquad
  \Phi_2 = \begin{pmatrix}
             H^+ \\
             \frac{1}{\sqrt{2}} (h_2 + i h_3)
           \end{pmatrix} \,.
  \label{eq:Phi1-Phi2}
\end{align}
Here, $G^0$ and $G^+$ are the Goldstone bosons, $H^+$ is the charged
Higgs boson and $h_1$, $h_2$, $h_3$ are the three neutral Higgs bosons.
Experiments
tell us that $h_1$ behaves approximately like the SM Higgs boson. Note that
$h_1$, $h_2$, and $h_3$ are not physical states yet because there is still
mass mixing among them.

The most general scalar potential for the 2HDM is~\cite{Gunion:2002zf}
\begin{align}
  V &= \mu_1^2 \Phi_1^\dag \Phi_1
     + \mu_2^2 \Phi_2^\dag \Phi_2
     + (\mu_3^2 \Phi_1^\dag \Phi_2  + h.c.) \nonumber \\
   & + \lambda_1 (\Phi_1^\dag \Phi_1 )^2
     + \lambda_2 (\Phi_2^\dag \Phi_2 )^2
     + \lambda_3 (\Phi_1^\dag \Phi_1) (\Phi_2^\dag \Phi_2)
     + \lambda_4 (\Phi_1^\dag \Phi_2) (\Phi_2^\dag \Phi_1 ) \nonumber \\
   & + \big[
      \big(\lambda_5 \Phi_1^\dag \Phi_2
         + \lambda_6 \Phi_1^\dag \Phi_1
         + \lambda_7 \Phi_2^\dag \Phi_2 \big)
      (\Phi_1^\dag \Phi_2) + h.c. ] \,.
      \label{eq:2HDMpotential}
\end{align}
Here, the parameters $\mu_1^2$, $\mu_2^2$, $\lambda_1$, $\lambda_2$, $\lambda_3$ and
$\lambda_4$ must be real, while $\mu_3^2$, $\lambda_5$, $\lambda_6$,
and $\lambda_7$ can be complex. If $V$ contains complex parameters, the quantity
$\mathop{\text{Im}}[\Phi_1^\dag \Phi_2]$ appears in the
potential, violating the CP symmetry.  The three neutral Higgs fields $h_1$, $h_2$
and $h_3$ mix to form three physical states. In the absence of complex
parameters in the scalar potential, i.e.\ in the CP conserving case, these
physical states can be assigned definite CP parity: there is one CP odd Higgs boson
$A^0 = h_3$ and two CP even Higgs bosons, the lighter of which is $h \simeq h_1$
and the heavier of which is $H^0 \simeq h_2$.
The condition that $V$ is at its minimum value when $\Phi_1 = (0, v/\sqrt{2})$,
$\Phi_2 = (0, 0)$,
and that the derivatives of $V$ in any direction in field space must vanish at this
point lead to the relations
\begin{align}
  \mu_1^2 = -\lambda_1 v^2 \, \qquad\text{and}\qquad
  \mu_3^2 = -\lambda_6 \frac{v^2}{2} \,.
  \label{eq:V-min-conditions-2HDM}
\end{align}
From the second derivatives of $V$, the masses of the charged Higgs field $H^\pm$
and the CP odd neutral Higgs $A^0$ follow as
\begin{align}
  m_{H^\pm}^2 = \mu_2^2 + \lambda_3 \frac{v^2}{2} \, \qquad\text{and}\qquad
  m_{A^0}^2   = m_{H^\pm}^2 + v^2 \big( \tfrac{1}{2} \lambda_4 - \lambda_5 \big) \,.
  \label{eq:mHpm-mA}
\end{align}
The two CP even Higgs fields $h_1$ and $h_2$ mix to form the mass eigenstates
$h$ and $H^0$ according to
\begin{align}
  \begin{pmatrix}
    h \\
    H^0
  \end{pmatrix} = \begin{pmatrix}
                    \cos\alpha  &  \sin\alpha \\
                   -\sin\alpha  &  \cos\alpha
                  \end{pmatrix}
                  \begin{pmatrix}
                    h_1 \\
                    h_2
                  \end{pmatrix} \,,
  \label{eq:higgs-mixing}
\end{align}
with the mixing angle $\alpha$ given by
\begin{align}
  \tan 2\alpha = \frac{-2 \lambda_6 v^2}{m_A^2 + 2 v^2 (\lambda_5 - \lambda_1)} \,.
  \label{eq:tan-alpha}
\end{align}
The masses of the physical CP even Higgs bosons are then
\begin{align}
  m_{h,H^0}^2 = \frac{1}{2} m_{H^\pm}^2
            + \frac{1}{2} v^2 \big (2 \lambda_1 + \tfrac{1}{2} \lambda_4 + \lambda_5 \big)
          \pm \frac{1}{2} \sqrt{[m_A^2 + 2 v^2 (\lambda_5 - \lambda_1)]^2
                              + 4 v^4 \lambda_6^2} \,.
  \label{eq:mh-mH}
\end{align}
From the requirement that the potential must be bounded from below, one can derive
several further constraints on its parameters~\cite{Gunion:2002zf}:
\begin{align}
  \lambda_1 > 0 \,,\qquad\quad
  \lambda_2 > 0 \,,\qquad\quad
  \lambda_3 > -2 \sqrt{\lambda_1 \lambda_2} \,,\qquad\quad
  \lambda_3 + \lambda_4 - 2 \lambda_5 > -2 \sqrt{\lambda_1 \lambda_2} \,.
\end{align}
If $\lambda_6 = \lambda_7 = 0$, $\lambda_5$ should be replaced by $|\lambda_5|$
in the last inequality.
Perturbativity moreover requires that $|\lambda_j| \ll 4\pi$ for $j=1\ldots7$,
and finally tree level unitarity imposes constraints, which we handle using
2HDMC~\cite{Eriksson:2009ws}.

\begin{figure*}
  \begin{tabular}{cc}
    \includegraphics[width=0.45\textwidth]{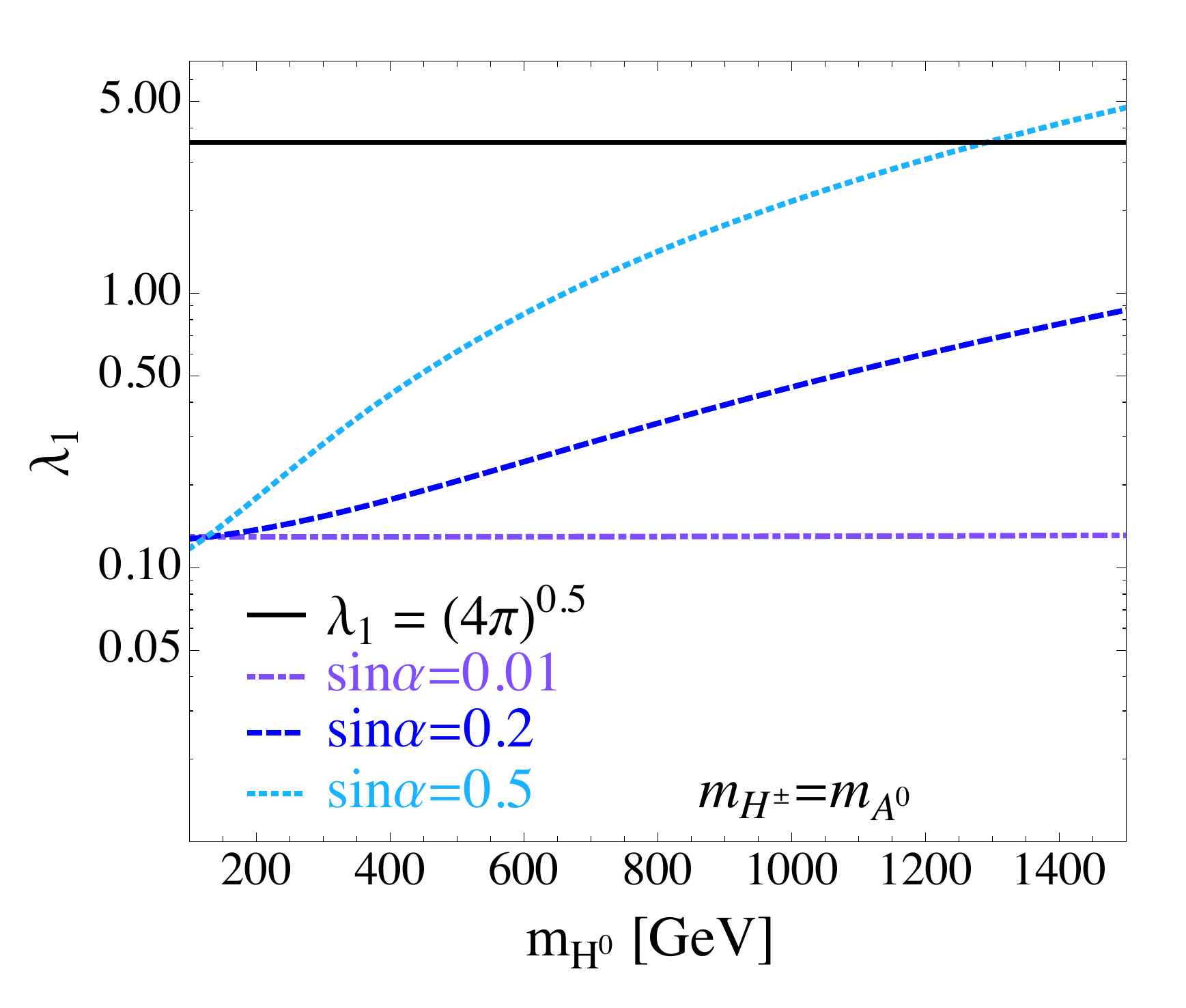} &
    \includegraphics[width=0.45\textwidth]{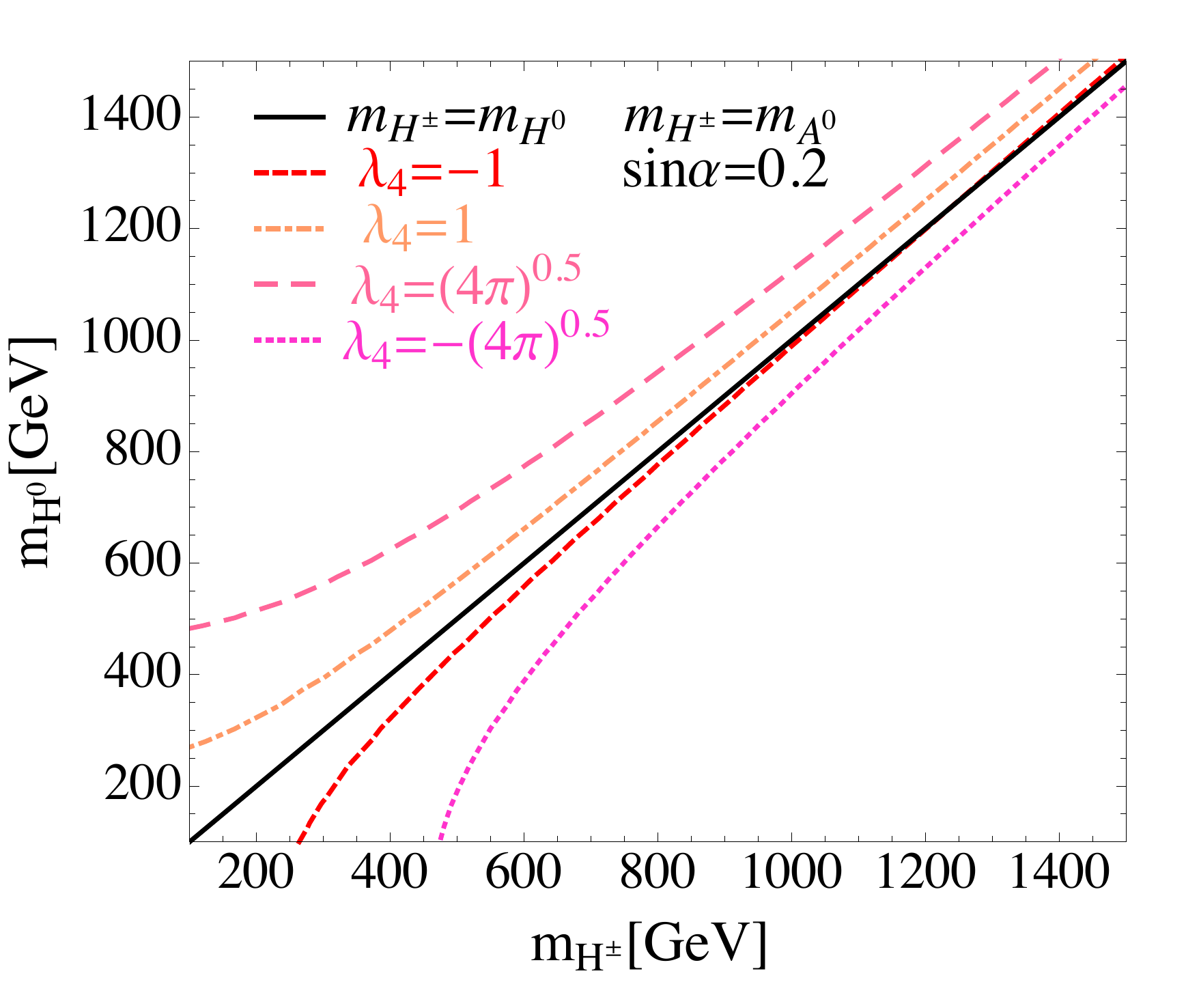}  \\
    (a) & (b)
  \end{tabular}
  \caption{\textit{Left panel:} The $\Phi_1$ quartic coupling $\lambda_1$
    as a function the heavy Higgs boson masses, assuming $m_{H^0} = m_{H^\pm} = m_{A^0}$,
    for different values of the Higgs mixing $\sin\alpha$.
    \textit{Right panel:} Contours of the mixed quartic coupling $\lambda_4$ as
    a function of the charged Higgs boson mass $m_{H^\pm}$ and the mass of the
    heavy CP-even neutral Higgs boson $m_{H^0}$.}
 \label{fig:lambdacontour}
\end{figure*}

It is often convenient to express the free parameters of the 2HDM in terms of physical
observables to the extent possible. First, $\mu_2^2$ can be expressed
in terms of $m_{H^\pm}$ and $\lambda_3$ by virtue of \cref{eq:mHpm-mA}.  The
quartic coupling $\lambda_2$ is irrelevant to us because it only affects
four-scalar interactions, which we are not interested in in this paper.
$\lambda_6$ can be expressed in terms of $\lambda_1$, $\lambda_5$ and the
mixing angle $\alpha$ using \cref{eq:tan-alpha}.
$\lambda_5$ can be eliminated if we assume $m_{A^0} = m_{H^\pm}$, which is
preferred by custodial symmetry. In this case, we have $\lambda_5 = \lambda_4 / 2$,
see \cref{eq:mHpm-mA}.  $\lambda_1$ and $\lambda_4$
can then be expressed in terms of $m_h$ and $m_{H^0}$ according to
\cref{eq:mh-mH}. This leads to the relations
\begin{align}
  \lambda_1^\pm &= \frac{m_{H^0}^2 + m_h^2 \pm \cos 2\alpha \,
                         (m_{H^0}^2 - m_h^2)}{4 v^2} \,,
                                               \label{eq:lambda1} \\
  \lambda_4^\pm &= \frac{m_{H^0}^2 + m_h^2 - 2 m_{H^\pm}^2 \mp \cos 2\alpha \,
                         (m_{H^0}^2 - m_h^2)}{2 v^2} \,,
                                               \label{eq:lambda4} \\[0.2cm]
  \lambda_5     &= \lambda_4 / 2 \,,           \label{eq:lambda5} \\
  \lambda_6     &= -\sin 2\alpha \, \frac{m_{H^0}^2 - m_h^2}{2 v^2} \,.
                                               \label{eq:lambda6}
\end{align}
Note that there are two solutions for $\lambda_1$ and $\lambda_4$. The $+$
($-$) solution is valid for $\cos 2\alpha < 0$ ($>0$). Since the SM-like nature
of the Higgs boson observed at the LHC requires $\alpha$ to be small, we use
the second solution, with the minus sign on the right hand side of the
expression for $\lambda_1$ and the plus sign on the right hand side of the
expression for $\lambda_4$.  We illustrate the dependence of $\lambda_1$ and
$\lambda_4$ on the physical mass variables in \cref{fig:lambdacontour}.  From
the distributions in panel (a), we read off that $\lambda_1$, which is only a
function of $m_h$, $m_{H^0}$ and $\sin\alpha$, is always well within the
perturbative regime if $\sin\alpha \ll 0.5$ (as required by LHC constraints) and
$m_{H^0} \leq 1.5$~TeV.  From the contour plot of $\lambda_4$ as a function of
$m_{H^0}$ and $m_{H^\pm}$, \cref{fig:lambdacontour} (b), we can see that when
$m_{H^\pm} \sim m_{H^0}$, then $\lambda_4$ is small as well. In our analysis,
we choose $m_{H^\pm} = m_{H^0}$ to ensure $\lambda_4^2 \ll 4\pi$ and to
eliminate one additional parameter.

We briefly summarize how we have reduced the parameter space of the scalar
potential.  We started with 10 parameters, namely $\lambda_1, \dots \lambda_7$ and
$\mu_1, \dots \mu_3$.  Of these, $\mu_1$ and $\mu_3$ are eliminated by the
conditions minimizing the potential, \cref{eq:V-min-conditions-2HDM}.
$\lambda_2$ is important only for Higgs boson self-interactions, which are
irrelevant for our purposes.  We have seen that the seven remaining parameters
can be reexpressed in terms of $\sin\alpha$, $m_h$, $m_{H^0}$, $m_{A^0}$,
$m_{H^{\pm}}$, $\lambda_3$ and $\lambda_7$. Knowing that the SM Higgs mass is
about 125~GeV, and assuming $m_{H^0} = m_{H^\pm} = m_{A^0}$ we have reduced the
number of free parameters in the Higgs potential to four.

Most important to us in this paper is the Yukawa sector of the 2HDM.
In the most general case (type~III 2HDM), both Higgs doublets can couple to all
fermions via arbitrary complex Yukawa matrices. While in the Georgi basis only
the couplings of $\Phi_1$ contribute to fermion masses, both Higgs doublets
contribute to the Yukawa couplings of the physical Higgs bosons.
Thus, the mass matrices and the Yukawa coupling matrices can be easily
misaligned in flavor space, inducing flavor violation.
Specifically, for the up-type quarks, the Yukawa couplings are
\begin{align}
  \mathcal{L}_\text{up}
    &= - \eta_{u,1}^{ij} \overline{Q_L^i} \tilde\Phi_1 u_R^j
       - \eta_{u,2}^{ij} \overline{Q_L^i} \tilde\Phi_2 u_R^j
       + h.c. \,,
\end{align}
with $\tilde\Phi_k \equiv i \sigma^2 \Phi_k^\dag$.
As before, we will use the notations $i,j = 1,2,3$ and $i,j = u,c,t,\dots$
interchangeably, and we will omit the subscript $u$ on $\eta_{u,1}^{ij}$
and $\eta_{u,2}^{ij}$ where doing so is
unambiguous, e.g.\ $\eta_2^{ut} \equiv \eta_{u,2}^{13}$.
We will assume for simplicity that all flavor violating Yukawa couplings
are real.
After electroweak symmetry breaking, and working in the mass basis where
$\eta_{u,1}^{ij} \propto \delta^{ij}$, this becomes
\begin{align}
  \mathcal{L}_\text{up}
     &= - m_i \overline{u_L^i} u_R^i
        - y^{ij}_{u,h} \overline{u_L^i} u_R^j h
        - y^{ij}_{u,H} \overline{u_L^i} u_R^j H^0
        + h.c. \,,
\intertext{with $m_i  = \eta_{u,1}^{ii} v / \sqrt{2}$ and}
  y^{ij}_{u,h} &=   \frac{m_i}{v} \delta^{ij} \cos\alpha
                  + \frac{1}{\sqrt{2}} \eta_{u,2}^{ij} \sin\alpha
  \label{eq:yuh-2hdm} \\
  y^{ij}_{u,H^0} &= - \frac{m_i}{v} \delta^{ij} \sin\alpha
                  + \frac{1}{\sqrt{2}} \eta_{u,2}^{ij} \cos\alpha \,.
  \label{eq:yuH-2hdm}
\end{align}
In the alignment limit $\sin\alpha \simeq 0$, $h$ has SM-like couplings while
large flavor violation can arise for $H^0$.  The decay rates of $h$
and $H^0$ into the flavor violating final states $\bar{t} u$ and $\bar{u} t$
are given by
\begin{align}
  \Gamma_{h \to \bar{t}u} =
  \Gamma_{h \to \bar{u}t} &=
    \frac{3}{32\pi} m_h \Big(1 - \frac{m_{H^0}^2}{m_h^2} \frac{x_t}{4} \Big)^2 \sin^2 \alpha
    \Big[ \big|\eta_2^{ut}\big|^2 + \big|\eta_2^{tu}\big|^2 \Big] \,,
  \label{eq:htu-2HDM} \\[0.2cm]
  \Gamma_{H^0 \to \bar{t}u} =
  \Gamma_{H^0 \to \bar{u}t} &=
    \frac{3}{32\pi} m_{H^0} \Big(1 - \frac{x_t}{4} \Big)^2 \cos^2 \alpha
    \Big[ \big|\eta_2^{ut}\big|^2 + \big|\eta_2^{tu}\big|^2 \Big] \,,
  \label{eq:Htu-2HDM}
\end{align}
with $x_t \equiv 4 m_t^2 / m_{H^0}^2$. Other important decay rates of the heavy Higgs
boson $H^0$ are
\begin{align}
  \Gamma_{H^0 \to t \bar{t}} &=
    \frac{3}{8\pi} m_{H^0} \bigg( - \sin\alpha \frac{m_t}{v}
                              + \cos\alpha \frac{\eta_2^{tt}}{\sqrt{2}} \bigg)^2
                       (1 - x_t)^{3/2} \,,
  \label{eq:Htt-2HDM} \\
  \Gamma_{H^0 \to WW} &=
    \frac{1}{64\pi} \frac{m_{H^0}^3}{v^2} \sin^2\alpha \sqrt{1 - x_W}
    (4 - 4x_W  + 3x_W^2) \,,
  \label{eq:HWW-2HDM} \\
  \Gamma_{H^0 \to ZZ} &=
    \frac{1}{128\pi} \frac{m_{H^0}^3}{v^2} \sin^2\alpha \sqrt{1 - x_Z}
    (4 - 4 x_Z + 3 x_Z^2) \,,
  \label{eq:HZZ-2HDM} \\
  \Gamma_{H^0 \to hh} &=
    \frac{1}{8\pi} \frac{g_{H^0hh}^2 v^2}{m_{H^0}} \sqrt{1 - x_h} \,.
  \label{eq:Gamma-Hhh}
\end{align}
where again $x_a \equiv 4 m_a^2 / m_{H^0}^2$ with $a = t, W, Z, h$. In the last
expression, $g_{H^0hh}$ is the coupling constant of the term
\begin{align}
  \mathcal{L}_{H^0hh} =  g_{H^0hh} \,v\, H^0 h h \,.
\end{align}
It is given by
\begin{align}
  g_{H^0 h h}
  &= 3 \sin\alpha \cos\alpha \bigg( \frac{\lambda_7}{2} \sin\alpha
                                  - \lambda_1 \cos\alpha \bigg)
   + \tfrac{1}{2} \big( \lambda_3 + \lambda_4 + 2\lambda_5 \big) \sin\alpha \,
     \big( 3 \cos^2\alpha - 1 \big) \nonumber\\
  &\hspace{9cm}
   + \tfrac{3}{2} \lambda_6 \cos\alpha \, \big( 1 - 3 \sin^2 \alpha \big)
                                                                  \nonumber \\
  &\simeq \sin\alpha \bigg( \lambda_3  - \frac{3m_{H^0}^2}{2v^2} \bigg)
   + \tfrac{3}{2} \lambda_7 \sin^2 \alpha
   + \mathcal{O}(\sin^3\alpha) \,.
 \label{eq:gHhh}
\end{align}
In the second equality, we have used the relations in
\cref{eq:lambda6,eq:lambda5,eq:lambda1,eq:lambda4} and the assumption
$m_{H^0} = m_{H^\pm} = m_{A^0}$, to express $g_{H^0hh}$ in terms of $\sin\alpha$,
$m_{H^0}$, $\lambda_3$, and $\lambda_7$.  We have also expanded
in $\sin\alpha$, and we see that in this case, $g_{H^0hh}$ is dominantly
determined by $m_{H^0}$ and $\lambda_3$.

Note that, since $\Phi_2$ does not acquire a vev, the decay rates for $H^0 \to
WW\!$, $ZZ$ are just the corresponding decay rates in the SM, with an additional
suppression factor $\sin^2\alpha$ arising from $H^0$--$h$ mixing. In the following,
we will for simplicity assume that the diagonal entries of $\eta_{u,2}$, i.e.\ the
diagonal couplings of $\Phi_2$ to up-type quarks, vanish.  In this case, also the rate
for $H^0 \to t\bar{t}$ is given by the SM rate, suppressed by $\sin^2\alpha$.

\subsection{Constraints on the Quark Flavor Violating 2HDM}
\label{sec:2HDM-constraints}

\begin{figure*}
  \begin{tabular}{ccc}
    \includegraphics[width=0.33\textwidth]{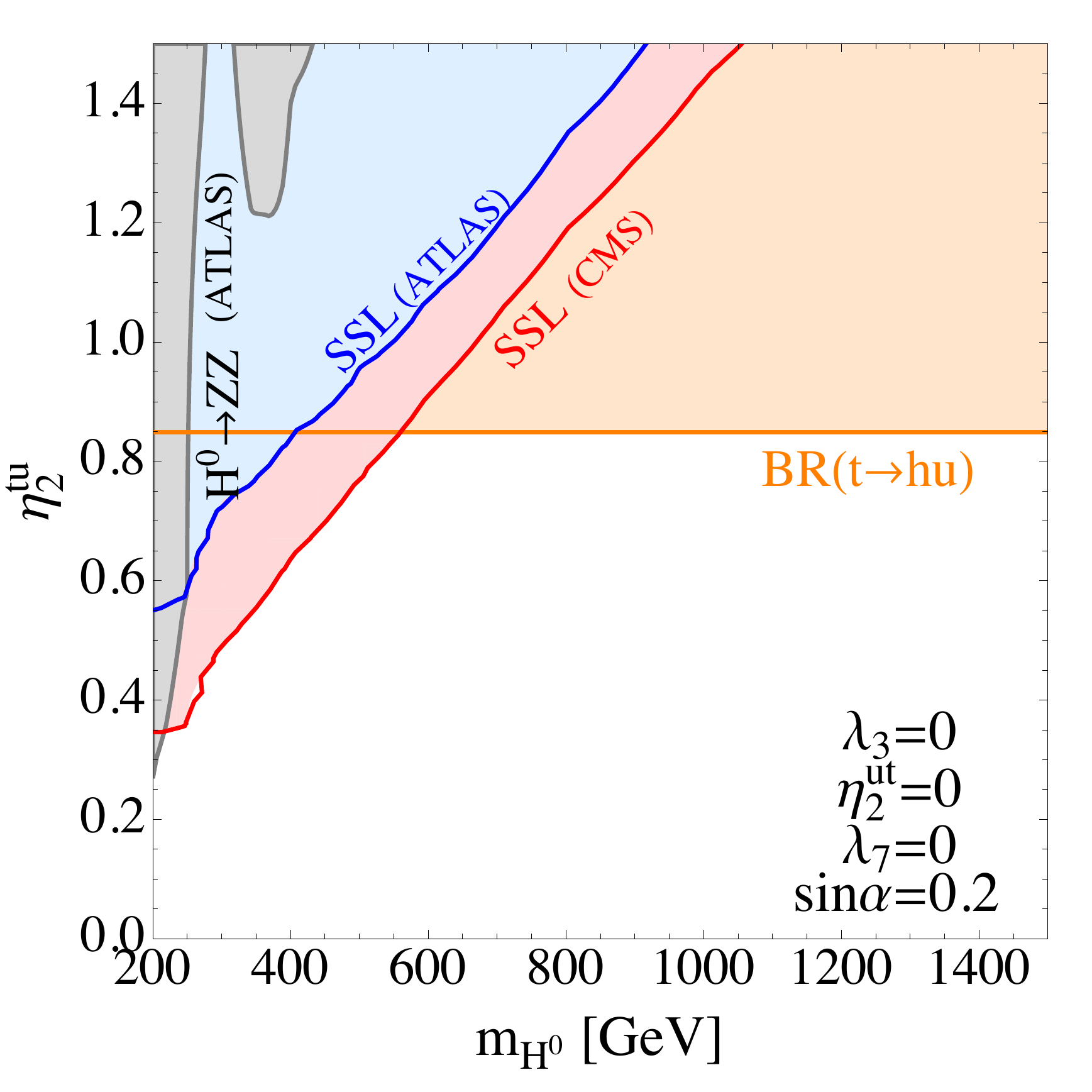} &
    \includegraphics[width=0.33\textwidth]{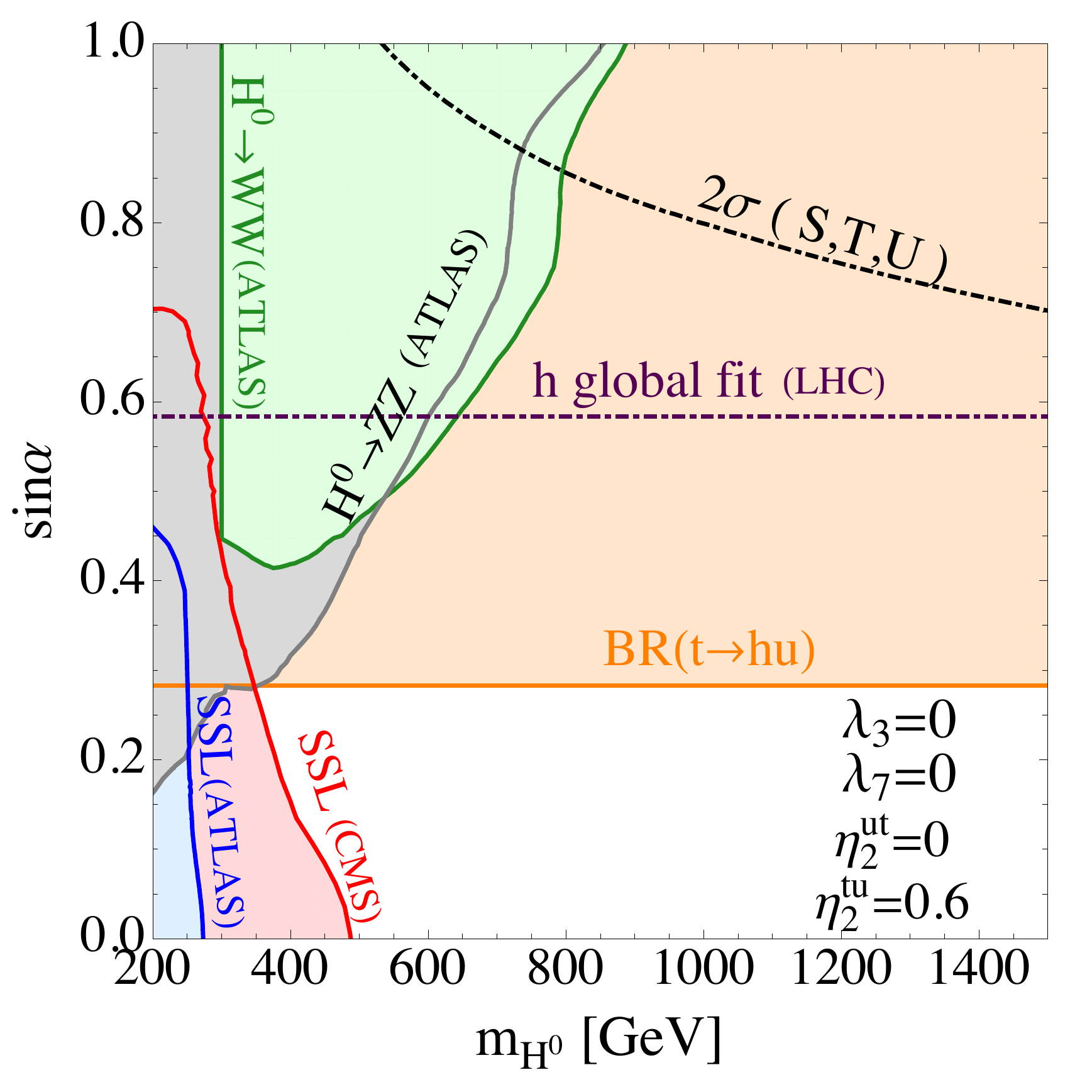} &
    \includegraphics[width=0.33\textwidth]{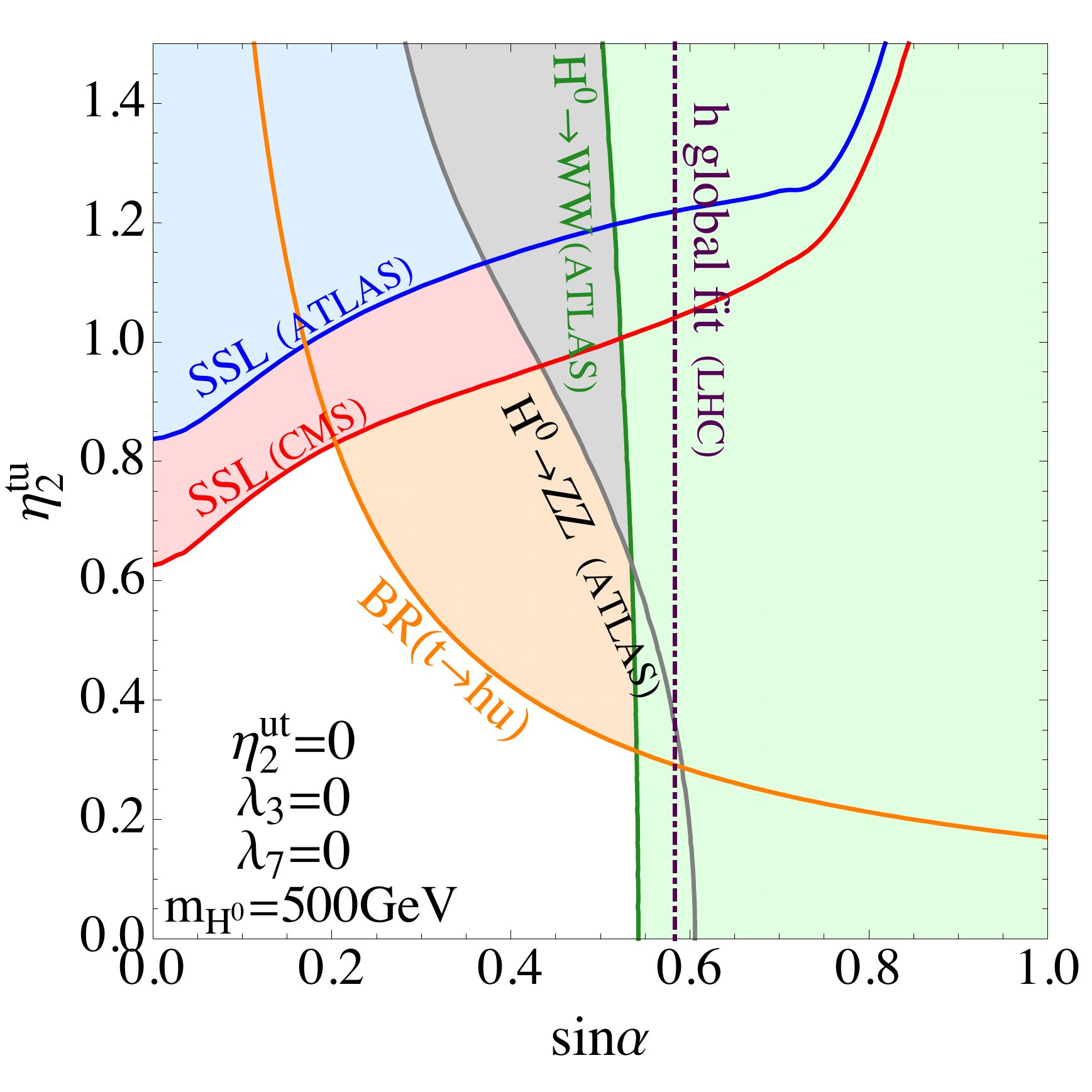} \\
    (a) & (b) & (c)
  \end{tabular}
  \caption{95\% CL constraints on the quark flavor violating Two Higgs Doublet
    Model of type~III as a function of the heavy CP-even Higgs mass $m_{H^0}$, the
    flavor violating Yukawa coupling $\eta_2^{tu} $ of the
    second Higgs doublet, and the
    neutral CP-even Higgs mixing angle $\sin\alpha$. We show results from a
    recasting of the same-sign di-lepton (SSL) + $b$ jet searches in
    ATLAS~\cite{TheATLAScollaboration:2013jha} (blue) and
    CMS~\cite{Chatrchyan:2012paa} (red), from ATLAS
    searches for heavy Higgs bosons in the $H^0 \to WW, ZZ$ final
    states~\cite{TheATLAScollaboration:2013zha,Aad:2015kna} (green, gray), from the ATLAS
    search for $t \to h q$~\cite{Aad:2015pja} (orange), from a global fit to the data on
    the SM-like Higgs boson (purple dot-dashed line in panels (b) and
    (c))~\cite{No:2013wsa}, and from electroweak precision data (black
    dot-dashed curve in panel (b))~\cite{Eriksson:2009ws}.
  }
  \label{fig:2hdm-constraints}
\end{figure*}

Low energy flavor experiments impose the strongest limits on flavor violating
couplings \emph{not} involving the top quark (or the $\tau$ lepton).
The only flavor constraints relevant to us are those from neutral meson mixing
and radiative $b$ quark decays.
In particular $B_d$--$\bar{B}_d$ receives contributions from a $t$--$H^\pm$--$W$
loop, which leads to the constraints $|\eta_2^{tu}| \lesssim \mathcal{O}(1)$
and $|\eta_2^{ut}| \lesssim \mathcal{O}(0.01)$ for $m_{H^+}=500$~GeV.
The radiative decay $b \to d \gamma$ further constrains $|\eta_2^{ut}| \lesssim \text{few}
\times 10^{-3}$ for $m_{H^+}=500$ GeV~\cite{Crivellin:2011ba, Crivellin:2013wna}.
In the following, we choose $\eta_2^{ut} = 0$ to avoid this flavor constraints.
With this choice, the most important bounds on the flavor violating Yukawa coupling
$\eta_2^{tu}$ come from LHC searches.  As we have seen in
\cref{sec:effOperator}, direct searches for $t \to q h$ decays require
$|y^{tu}_h| < 0.12$.  This can be translated into a
limit on $\eta_2^{tu}$ and the neutral CP-even Higgs mixing
$\sin\alpha$ using \cref{eq:yuh-2hdm}.  This limit is shown as an orange
exclusion region in \cref{fig:2hdm-constraints}. 
Panel (a) in this figure shows
constraints as a function of the heavy Higgs mass $m_{H^0}$ and of $\eta_2^{tu}
$, panel (b) as a function of $m_{H^0}$ and $\sin\alpha$, and
panel (c) as a function of $\sin\alpha$ and $\eta_2^{tu} $.  In
all three panels, we have assumed $\eta_2^{tc}=\eta_2^{ct}=0$.

Current CMS and ATLAS analyses do not directly search for flavor violating
couplings of heavy Higgs bosons. However, CMS~\cite{Chatrchyan:2012paa} and
ATLAS~\cite{TheATLAScollaboration:2013jha} have searched for final states with
same-sign di-leptons and $b$ jets at 8~TeV. In the type~III 2HDM, this final
state could arise due to the top--up--$H^0$ interaction, for instance in the
process $g + u \to (t \to b\ell\nu) + (H^0 \to WW, ZZ, \bar t u, \bar u t)$.
We have recast the CMS and ATLAS searches from refs.~\cite{Chatrchyan:2012paa}
and \cite{TheATLAScollaboration:2013jha}, generating $p p \to t t$ and $p p \to
t H^0$ events with the same simulation tools as in \cref{sec:simulationEFT}.
In the presence of top flavor violating Higgs couplings, the first of these
processes (same-sign top production) receives contributions from $t$-channel
exchange of $h$, $H^0$, while in the second process, the dominant parton level
interaction is $g u \to t H^0$, with an up quark in the $s$-channel. Note also
that a $p p \to t t$ signal would dominate by far over the corresponding $p p
\to \bar{t} \bar{t}$ signal because the former is initiated by valence quarks.
For the same reason, also $p p \to t H^0$ is much stronger than $p p \to
\bar{t} H^0$.  Finally, note that the process $p p \to t h$ is irrelevant
because its cross section is suppressed by $\sin^4\alpha$ compared to the cross
section for $p p \to t H^0$, and because $h$ can only decay to $WW^*$, $ZZ^*$,
but not to on-shell gauge bosons.  The results of our recasting are shown as red
and blue exclusion regions in \cref{fig:2hdm-constraints}.  From the figure, we
see that the CMS limit using only $10.5$~fb$^{-1}$ of integrated luminosity is
somewhat stronger than the ATLAS limit using $14.3$~fb$^{-1}$ in the same
channel.  The reason is that CMS separates the data into events with two
positive leptons (the dominant final state for our signal) and events with two
negative leptons, while ATLAS only shows them combined, thus diluting the
sensitivity.

Additional constraints on 2HDMs arise from direct searches for heavy Higgs
bosons decaying to $WW$ or
$ZZ$~\cite{TheATLAScollaboration:2013zha,ATLAS:2013nma,
Chatrchyan:2013yoa,Aad:2015kna}. We have recast the ATLAS searches
\cite{TheATLAScollaboration:2013zha,Aad:2015kna} (green and gray regions
in \cref{fig:2hdm-constraints}), considering $H^0$ production both through gluon
fusion and through the flavor violating processes $pp \to tH^0, \bar t H^0$.
Since gluon fusion is possible even without flavor violating couplings,
non-trivial constraints can be expected even for $\eta_2^{tu} =
0$.  In their $H^0 \to ZZ$ search~\cite{Aad:2015kna}, ATLAS have separated
their event sample according to the $H^0$ production mode (gluon fusion vs.\
vector boson fusion (VBF)) and according to the decay channel ($4 \ell$, $2
\ell 2\nu$ and $2q2\ell + 2q 2\nu$). We find that in our model, where
production is dominated by $p p \to t H^0$ (except at very small $\eta_2^{tu} $),
 the strongest constraint is achieved for VBF events in the
$4\ell$ category. The reason is that our signal is often vetoed in the gluon
fusion channels which require low jet multiplicity, while cuts are
rather loose in the $4\ell$ category.  The ATLAS $H^0 \to WW$ search
\cite{TheATLAScollaboration:2013zha} also separates the data into gluon fusion
and VBF categories, with the former required to have a jet multiplicity $N_j =
0$ or $1$, and the latter to have $N_j \geq 2$.  To purify the VBF sample, a
strong cut  $m_{jj} > 500$~GeV is imposed on the invariant mass of the two
leading jets in the $N_j \geq 2$ sample, and their rapidity difference is
required to be be $|\Delta \eta_{jj}| > 2.8$.  These strong cuts make the VBF
event sample rather insensitive to our signal, and we therefore include
only the $N_j = 0$ and $N_j = 1$ event categories in our recasting.  The CMS search
for heavy Higgs bosons in ref.~\cite{Chatrchyan:2013yoa} includes both the $H^0
\to WW$ and $H^0 \to ZZ$ channels, but is based on less data (up to
5.1~fb$^{1}$ at $\sqrt{s} = 7$~TeV and up to 5.3~fb$^{-1}$ at $\sqrt{s} =
8$~TeV) than the ATLAS searches
\cite{Aad:2015kna,TheATLAScollaboration:2013zha} which employ about
20~fb$^{-1}$ of 8~TeV data each.  Therefore, we do not include CMS results
here.

One may wonder whether relevant constraints may be obtained from non-resonant
di-Higgs production in the process
$u \bar{u} \to h h$, mediated by a $t$-channel top quark. We have computed the
cross section for this process, but find it to be negligibly small.
In fact, ref.~\cite{Aad:2015xja} gives a $95\%$~CL upper
limit of $0.69$~pb on the cross section of non-resonant di-Higgs production. This
translates into the limit $\eta_2^{tu} \sin\alpha < 1.09$, which is much weaker
than  the constraint from the exotic top decay $t \to h q$.

The neutral Higgs boson mixing angle $\alpha$ is also constrained because of
the $\cos^2\alpha$ suppression of the $h W^+ W^-$ and $h Z Z$ couplings.
A global analysis of Higgs couplings at the LHC suggests
$\sin^2 \alpha < 0.34$ at $95\%$~CL~\cite{No:2013wsa}.

Finally, we have checked the electroweak precision constraints using the
oblique parameters $S$, $T$ and $U$~\cite{Peskin:1991sw, LEP:2005ema,
Agashe:2014kda} (black dot-dashed curves in \cref{fig:2hdm-constraints}). We
employed the program 2HDMC~\cite{Eriksson:2009ws} for this comparison.  Since
the second Higgs doublet $\Phi_2$ does not violate custodial symmetry and since we have
assumed $m_{H^0} = m_{A^0} = m_{H^\pm}$, the resulting constraints are rather weak.

In summary, \cref{fig:2hdm-constraints} shows that LHC
searches for $t \to h u$ impose the strongest constraints on the quark flavor
violating 2HDM at $m_{H^0} \gtrsim \text{few~100~GeV}$. At smaller masses
same-sign di-lepton searches and direct searches for heavy Higgs bosons
become more important.  Overall, values of $\eta_2^{tu}
\lesssim \text{few} \times 10^{-1}$ are still allowed.

\subsection[LHC Sensitivity to $thh$ Production in the 2HDM]
           {LHC Sensitivity to \texorpdfstring{$thh$}{thh}
            Production in the 2HDM}
\label{sec:2HDM-thh}

\subsubsection{Production Cross Sections and Decay Rates}
\label{sec:2HDM-thh-xsecs}

Let us consider again the $thh$ final state arising from quark flavor violating
couplings in the Higgs sector.  We will demonstrate in the following that, in
the 2HDM, this final state may offer superior sensitivity to quark flavor
violating Higgs couplings already in Run II of the LHC.

\begin{figure*}
  \includegraphics[width=0.45\textwidth]{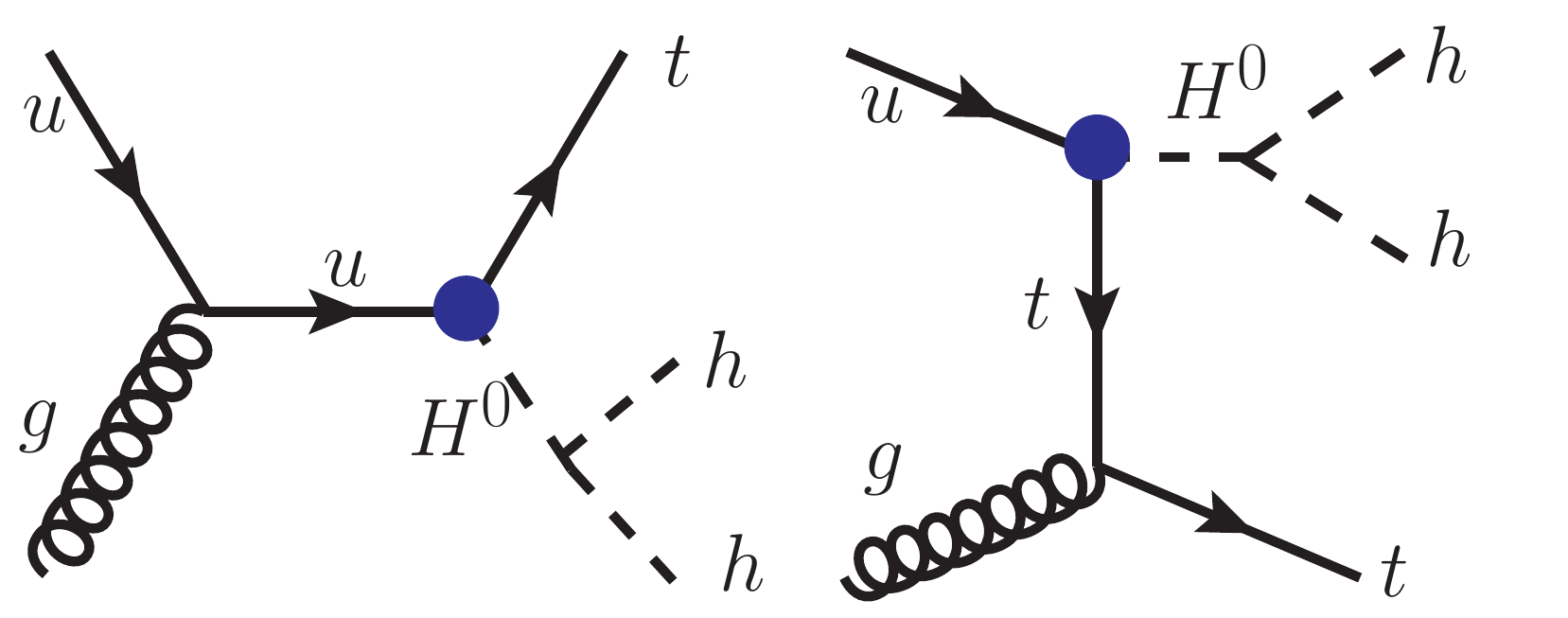}
  \caption{The Feynman diagrams for the process $p p \to t + (H^0 \to hh)$
    in the 2HDM. The blue dot indicates the flavor violating Yukawa coupling
    proportional to $\eta_2^{tu}$ or $\eta_2^{ut*}$.}
  \label{fig:2HDM-feyn}
\end{figure*}

If the heavy neutral CP even Higgs boson $H^0$ is significantly heavier than
the light one, $h$, it can be integrated out, reducing the 2HDM to the
effective Lagrangian \cref{eq:EFT1}.  However, it follows from relations
\eqref{eq:lambda1}--\eqref{eq:lambda4} and from the requirement $\lambda_j <
4\pi$ that going to the limit $m_{H^0}, m_{H^\pm} \gg h$ requires fine-tuning.
If $H^0$ is not too heavy, it can be produced on-shell and decay to two $h$
particles, so that the process $p p \to t + (H^0 \to h h)$, which does not
exist in the effective theory from \cref{sec:effOperator}, contributes to the
$thh$ final state as well.  The corresponding Feynman diagrams are given in
\cref{fig:2HDM-feyn}.

In order to study $thh$ production in the 2HDM in detail, we will consider two
benchmark points within the parameter space of the model.  (We will also
discuss how our results are affected when departing from these benchmark
points.)  To find suitable benchmark points, we first use the fact that the
SM-like nature of the light Higgs boson $h$ indicates small mixing
$\sin\alpha$. Based on \cref{fig:2hdm-constraints}, we use the benchmark value
$\sin\alpha = 0.2$.  To further simplify the high dimensional parameter space,
note that $\lambda_7$ affects the coupling $g_{H^0hh}$ in \cref{eq:gHhh} only at
order $\sin^2\alpha$.  Therefore, we always take $\lambda_7 = 0$ in the
following for simplicity.  For $\lambda_3$, which enters $g_{H^0hh}$ at
$\mathcal{O}(\sin\alpha)$, we use the benchmark values $0$ and $-3$.  For
positive values of $\lambda_3$, partial cancellation would occur in the
coefficient of $\sin\alpha$ in \cref{eq:gHhh}, reducing the $H^0 \to h h$
branching ratio compared to the $\lambda_3 \leq 0$ case.  For negative
$\lambda_3$, this branching ratio is enhanced.  Note also that both $\lambda_7$
and $\lambda_3$ affect only the $H^0 \to hh$ rate, but not their decay rates
into other final states or the $H^0$ production cross section.  Finally, to
best illustrate the impact of the flavor violating decay channel of $H^0$, we
always choose $\eta_2^{tu}$ close to the current upper limit
from \cref{fig:2hdm-constraints}.  For $\sin\alpha=0.2$, we take $\eta_2^{tu} = 0.6$ as our benchmark value.  As we can see from
\cref{fig:2hdm-constraints}, these values are somewhat above the upper limit
for $m_{H^0} \lesssim 500$~GeV, and below the upper limit for $m_{H^0} \gtrsim
500$~GeV. We summarize our benchmark assumptions in table~\ref{tab:benchmark}.

\begin{table}
  \centering
  \begin{minipage}{13cm}
  \begin{ruledtabular}
  \begin{tabular}{lccl}
                                & Benchmark~1 & Benchmark~2 & Comments \\ \hline
    $\sin\alpha$                &     0.2     &     0.2     &          \\
    $\eta_2^{ut}$               &     0       &     0       & $b \to d \gamma$ constraint \\
    $\eta_2^{tu} $              &     0.6     &     0.6     &
                 see \cref{fig:2hdm-constraints} \\
    $\lambda_7$                 &     0       &     0       &
                 enters $g_{H^0hh}$ only at $\mathcal{O}(\sin^2\alpha)$ \\
    $\lambda_3$                 &     0       &   $-3$      &
                 influences $g_{H^0hh}$ \\
    $m_{A^0}$                   & $m_{H^\pm}$ & $m_{H^\pm}$ &
                 preferred by custodial symmetry \\
    $m_{H^\pm}$                 &  $m_{H^0}$  &   $m_{H^0}$ &
                 preferred by perturbativity (see \cref{fig:lambdacontour} (b)) \\
  \end{tabular}
  \end{ruledtabular}
  \end{minipage}
  \caption{Benchmark points for the quark flavor violating 2HDM.}
  \label{tab:benchmark}
\end{table}

\begin{figure*}
  \centering
  \includegraphics[width=0.45\textwidth]{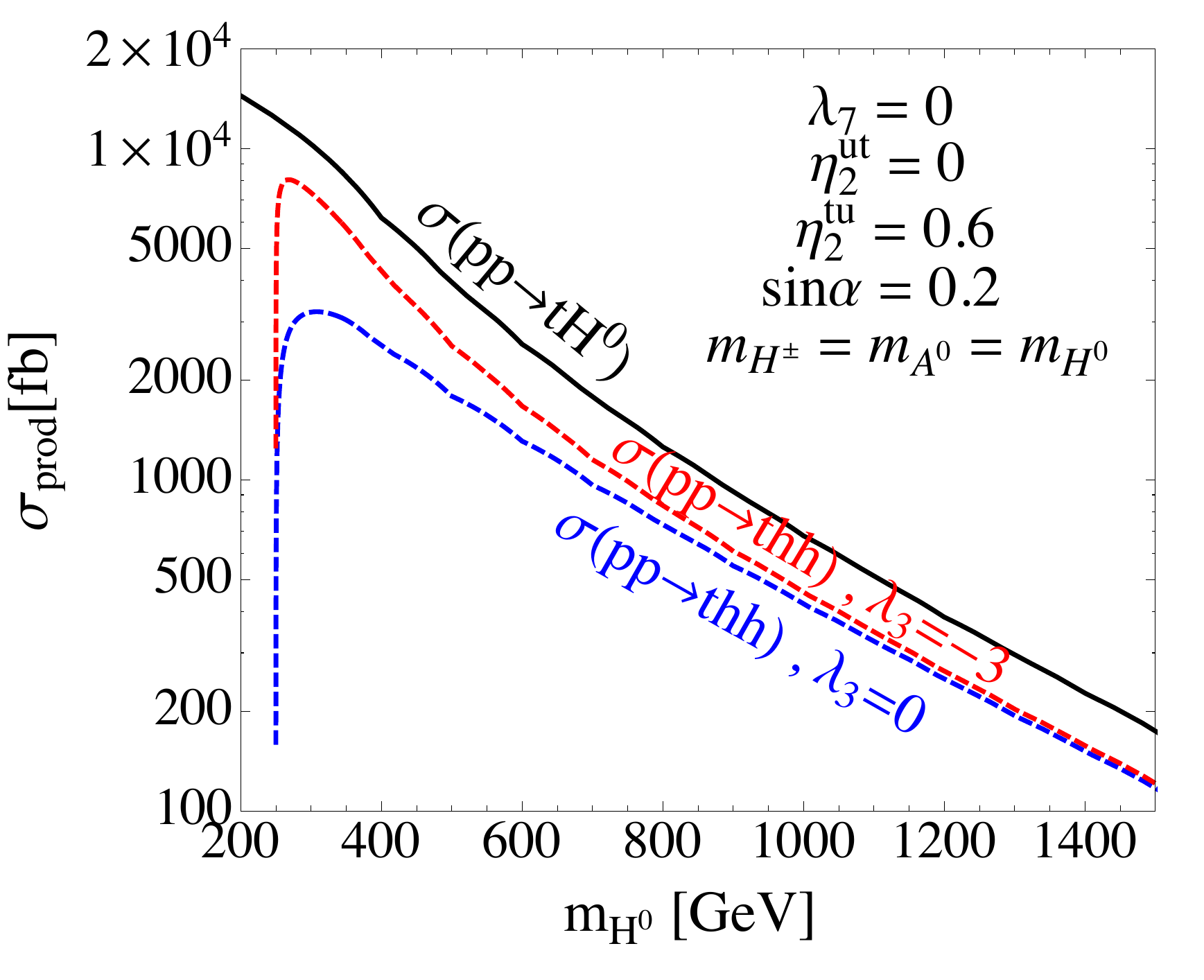}
  \caption{The production cross section of the heavy CP even neutral Higgs boson
    $H^0$ in the 2HDM (black solid curve), and the cross sections for the
    production + decay process $p p \to t + (H^0 \to h h)$ (red dashed and blue dashed
    curves). Production of $H^0$ is dominantly mediated by the
    flavor violating Yukawa coupling $\eta_2^{tu}$ here.}
  \label{fig:thhXSection}
\end{figure*}

\begin{figure*}
  \begin{tabular}{cc}
    \includegraphics[width=0.45\textwidth]{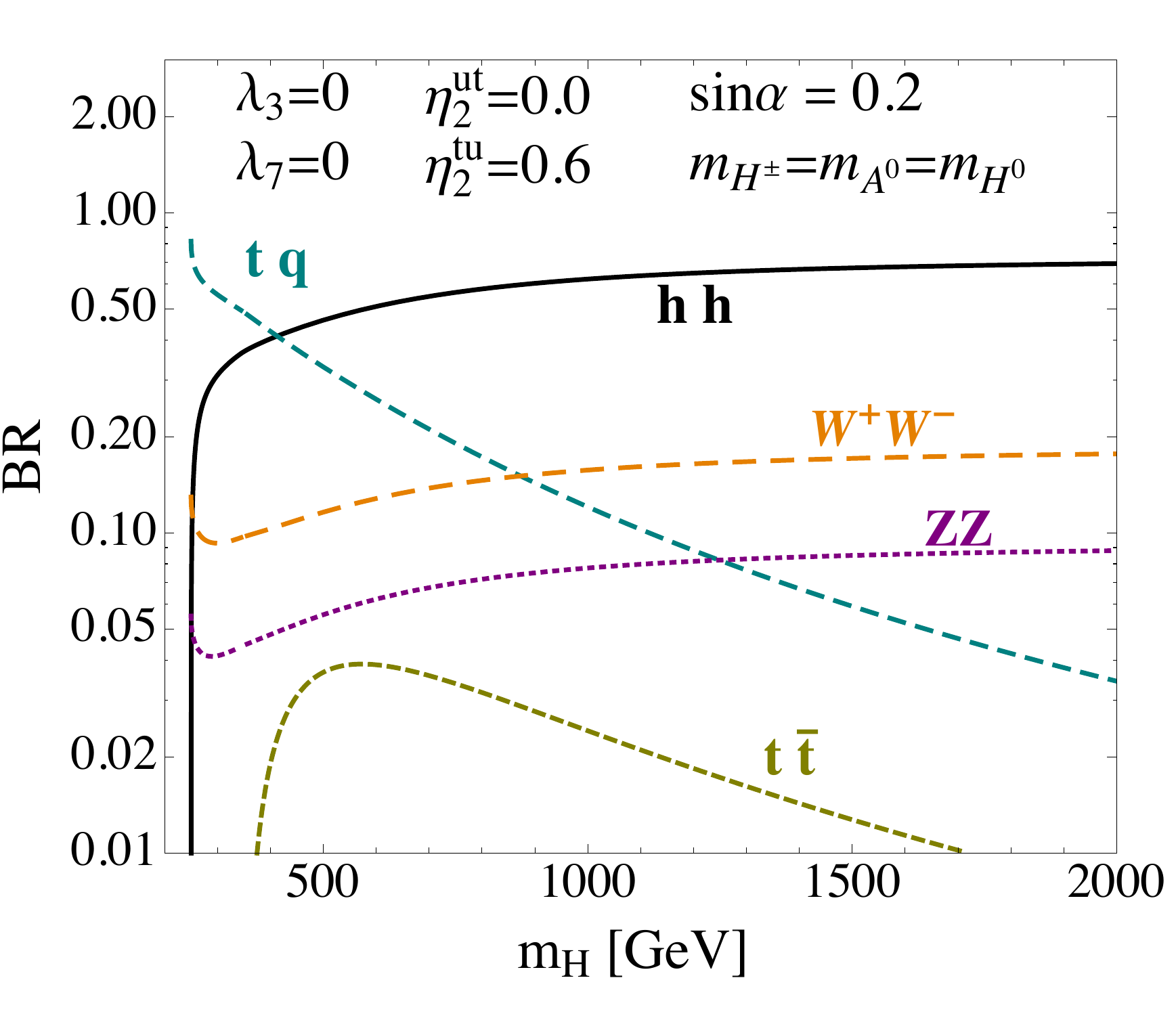} &
    \includegraphics[width=0.45\textwidth]{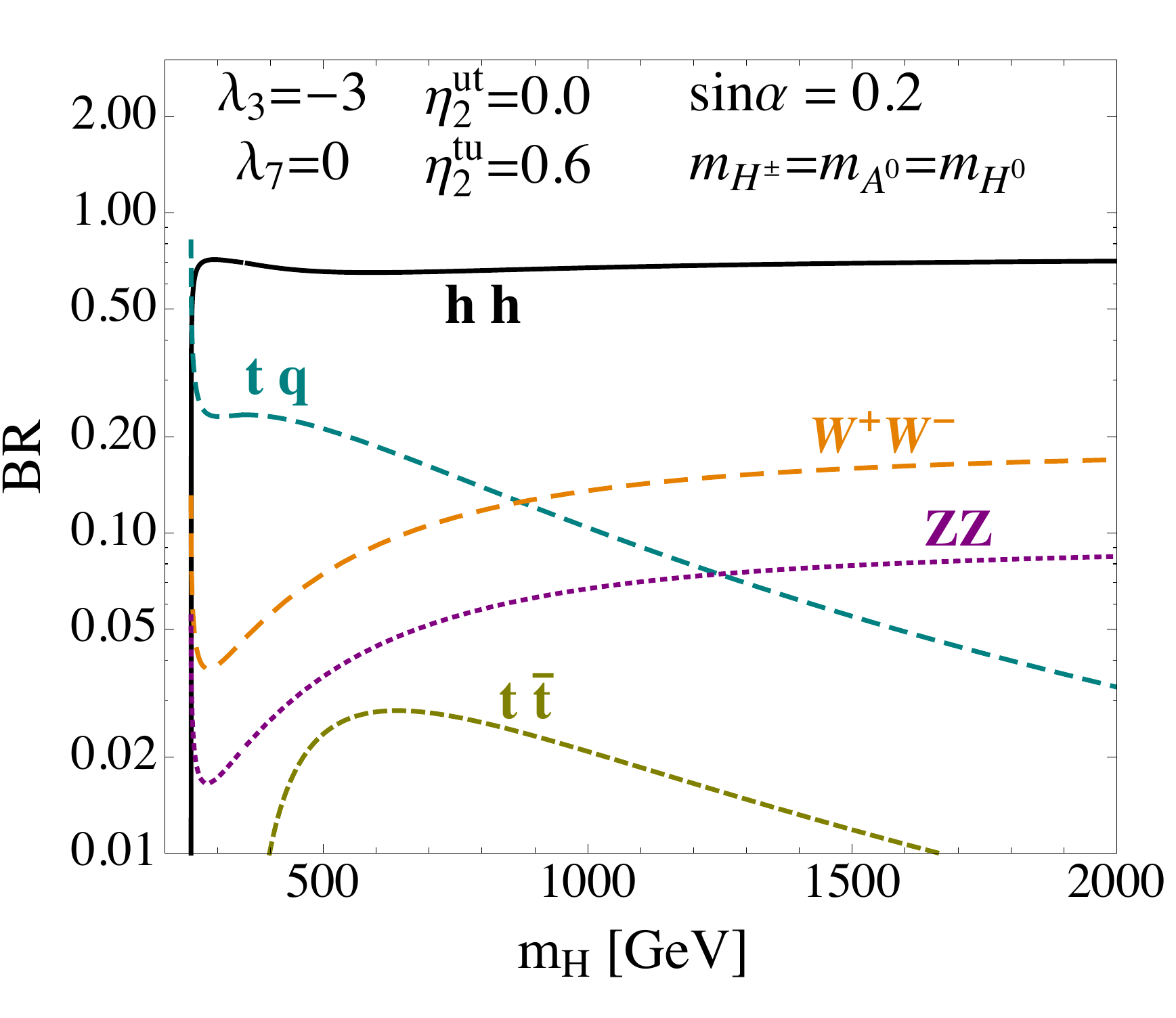} \\
    (a) & (b)
  \end{tabular}
  \caption{The branching ratios of the different $H^0$ decay modes for two
    different parameter points of the 2HDM. In both cases, we assume a large
    flavor violating top--up--$H^0$ coupling $\eta_2^{tu}
    = 0.6$. This value is slightly above the current upper limit for
    $m_{H^0} \lesssim 500$~GeV and below the upper limits for $m_{H^0} \gtrsim 500$~GeV
    (see \cref{fig:2hdm-constraints} (b)). Choosing different values
    for $\eta_2^{tu}$ would change the branching ratio
    of $H^0 \to t u$ relative to the other channels.}
  \label{fig:heavyHiggsDecay}
\end{figure*}

The $t+H^0$ production cross section and the cross section for the process $p p
\to t + (H^0 \to h h)$ are plotted in \cref{fig:thhXSection} as a function of
$m_{H^0}$. The dependence of the branching ratios of $H^0$ on its mass is
illustrated in \cref{fig:heavyHiggsDecay}.  We see that the $H^0 \to t\bar u,
\bar t u$ channels dominate for not too large $m_{H^0}$ since the branching
ratios for all other channels are proportional to $\sin^2 \alpha$.  At very
large $m_{H^0}$, the decay channels to bosonic final states, $H^0 \to  WW, ZZ,
hh$ become dominant since their rates are proportional to $m_{H^0}^3 / v^2$,
i.e.\ they grow with the third power of $m_{H^0}$.

\subsubsection{Search Strategy and Sensitivity Estimates for the 13~TeV LHC}
\label{sec:2HDM-thh-simulation}

We now adapt our single top plus di-Higgs search from \cref{sec:simulationEFT}
to the 2HDM, including in particular on-shell $t+H^0$ production through flavor
violating couplings, followed by the decays $t \to b \ell \nu$, $H^0 \to h h$,
and $h \to b \bar{b}$.
In principle, we should also
consider other $H^0$ decay channels. In fact, as we saw in \cref{fig:heavyHiggsDecay}, the
dominant decay channel is the flavor violating one, $H^0 \to t q$, in vast
regions of parameter space. However, if at least one of the top quarks in the final state
decays hadronically, the background from $t\bar{t}$ production is huge. If both
top quarks decay semileptonically, $pp \to t + (H^0 \to t q)$ contributes to the
same-sign top sample. Unfortunately, the small branching ratio into the semileptonic
top decay channel renders the total event number tiny.
Similarly, $t + (H^0 \to WW, ZZ)$ events are difficult to extract because of
small event numbers and/or large backgrounds (see our recasting of the corresponding
ATLAS and CMS searches~\cite{TheATLAScollaboration:2013zha, ATLAS:2013nma,
Chatrchyan:2013yoa, Aad:2015kna} in \cref{fig:2hdm-constraints}).

For the $thh$ final state, we apply similar cuts as in
\cref{sec:simulationEFT}.  However, since the signal is dominated
by events with an on-shell $H^0$, we add a cut on the
invariant mass $m_{hh}$ of the two reconstructed Higgs bosons. More precisely,
we require $|m_{hh} - m_{H^0}^\text{test}| < 0.1 m_{H^0}^\text{test}$,
where $m_{H^0}^\text{test}$ is the heavy Higgs mass being tested.
We also modify the $\Delta R_{b\bar{b}}^\text{max}$ cut slightly, and we soften
the cut on $p_T^{h_1}$. This softening helps to keep more signal events, while
background events are still suppressed efficiently by the $m_{H^0}$ cut which could
not be used in the EFT case.
The cut flow for this adapted analysis is shown in \cref{tab:cut-flow-2}.

\begin{table}
  \begin{minipage}{13cm}
  \begin{ruledtabular}
  \begin{tabular}{lcc}
    cut                          & signal ($thh$) & background ($t\bar{t}$) \\
    \hline
    $\sigma_\text{prod}$ [fb]    & 273.6          & $5.9\times 10^5$ \\
    \hline
    preselection                 & 28.5\%         &  \phantom{0}2.20\% \\
    $b$-tagging                  & 18.4\%         &  \phantom{0}0.55\% \\
    $p_T^{j_1} > 140$~GeV        & 90.6\%         & 31.1\% \\
    $p_T^{j_2} > 100$~GeV        & 93.9\%         & 66.3\% \\
    $p_T^{j_3} > 60$~GeV         & 97.3\%         & 84.6\% \\
     Higgs, top mass window      & 14.3\%         &  8.6\% \\
    $p_T^{h_2} > \text{150~GeV}$ & 71.9\%         & 35.3\% \\
    $p_T^{h_1} > \text{200~GeV}$ & 94.4\%         & 90.3\% \\
    $0.9\le \Delta R_{b\bar{b}}^\text{max} < 2.1$
                                 & 89.8\%         & 67.8\% \\
    $m_{H^0}$ mass window        & 69.9\%         & 31.1\% \\
    \hline
    $\sigma_\text{final}$ [fb]   & 0.72         & 0.071
  \end{tabular}
  \end{ruledtabular}
  \end{minipage}
  \caption{Cut flow table for the $thh$ signal in the 2HDM for
    $m_{H^0} = m_{H^\pm} = m_A = 500$~GeV, $\sin\alpha = 0.2$,
    $\lambda_3 = -3$, $\lambda_7 = 0$, $\eta_2^{ut} = 0$ and $ \eta_2^{tu} = 0.6$.
    If we use $\lambda_3 = 0$ instead, we find a signal cross section before cuts
    of $\sigma_\text{prod} = 192.93$~fb, and a signal cross section after cuts
    of $\sigma_\text{final} = 0.508$~fb. The cut efficiencies remain unchanged.}
  \label{tab:cut-flow-2}
\end{table}

\begin{figure*}
  \begin{tabular}{cc}
    \includegraphics[width=0.48\textwidth]{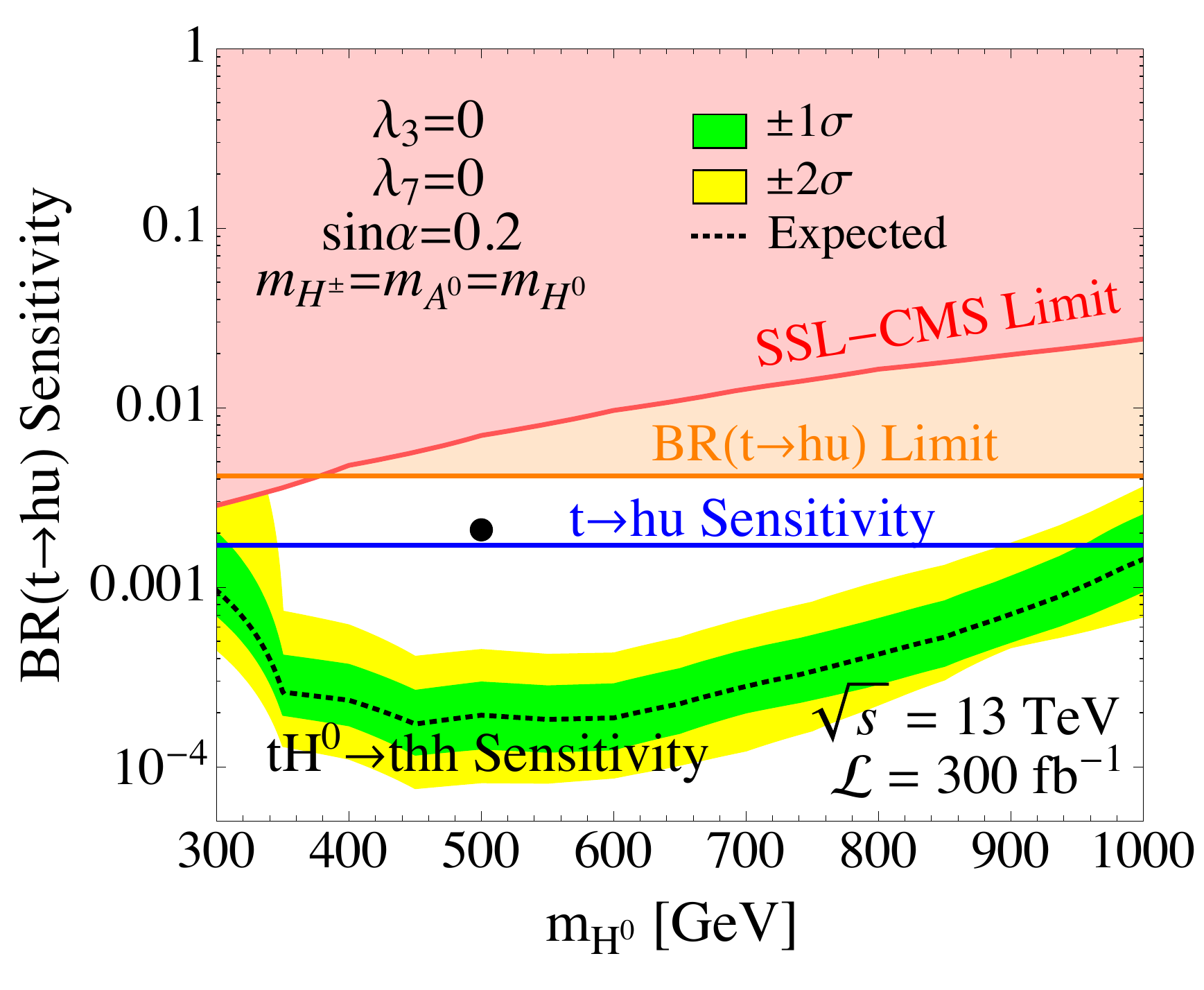} &
    \includegraphics[width=0.48\textwidth]{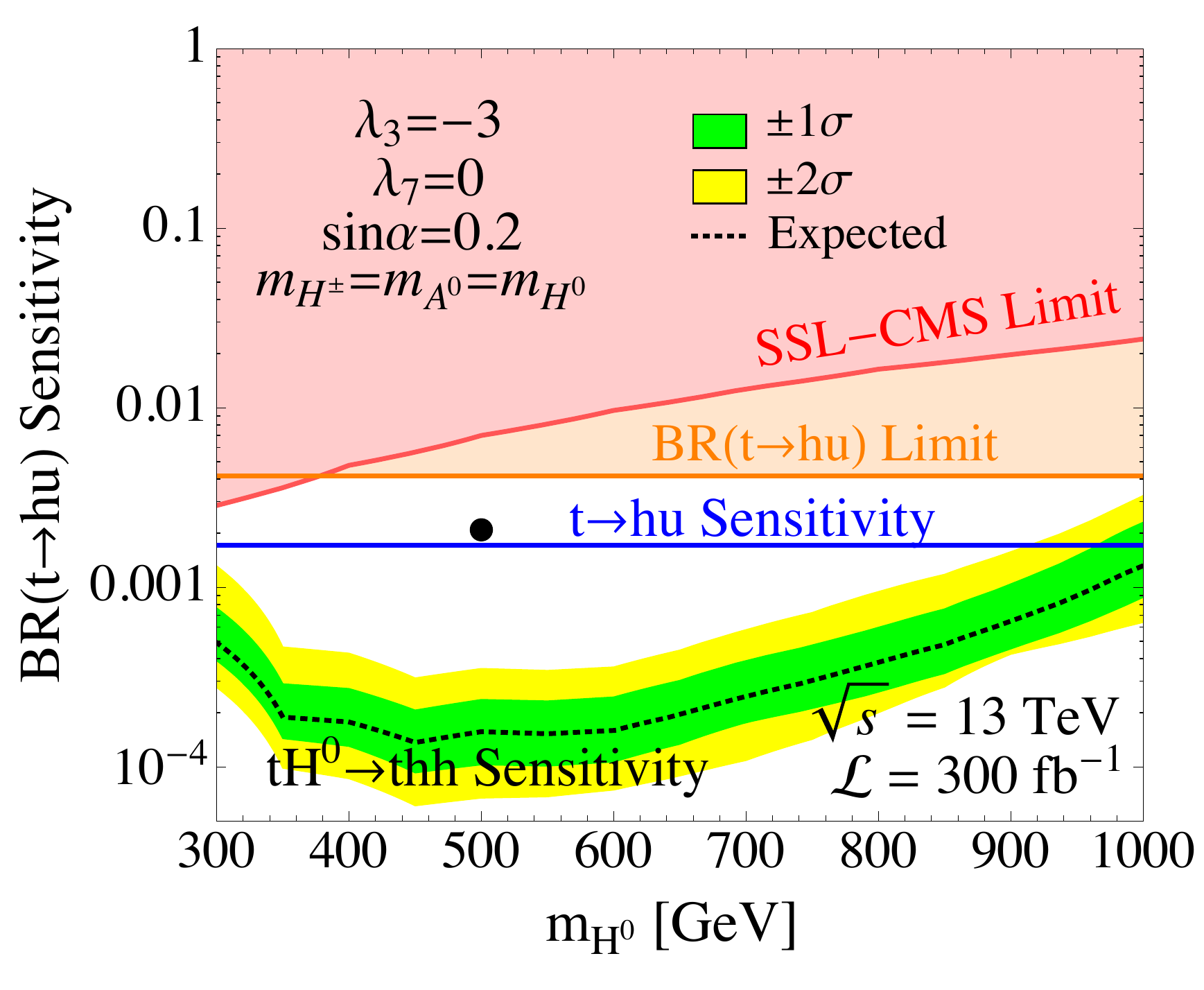} \\
    (a) & (b)
  \end{tabular}
  \caption{
    In the context of the 2HDM, we show the 95\% CL sensitivity of the proposed search
    for the $\text{top} + \text{di-Higgs}$ ($thh$) final state to quark flavor violating
    Higgs couplings, expressed here in terms of the branching ratio of the
    decay $t \to h u$ (Brazilian bands).  Comparing to the projected
    sensitivity of a direct search for $t \to h u$ from \cite{Greljo:2014dka}
    (horizontal blue lines), we find that the $thh$ search is more sensitive in a wide
    range of heavy Higgs masses $m_{H^0}$. For comparison, we also show the
    current limit on $\BR(t \to h u)$ from ref.~\cite{Aad:2015pja} (horizontal
    orange lines) and the current limits on $p p \to t H^0$ from a recasting of
    the CMS search for same-sign di-leptons (SSL) + $b$
    jets~\cite{Chatrchyan:2012paa}, see \cref{sec:2HDM-constraints} (red shaded
    regions). The black dots are the two benchmark points in \cref{tab:benchmark}.}
  \label{fig:BR-htu-limit}
\end{figure*}

We display the expected 95\%~CL sensitivity of our search
in \cref{fig:BR-htu-limit} as a function of
$m_{H^0}$ for the two parameter points from \cref{tab:benchmark}.
We assume signal and background uncertainties of 30\%.
We see that discovery prospects are best at $m_{H^0}
\sim 400$--$800$~GeV.
In particular, a search for the $t + (H^0 \to h h)$ final state
would be superior to the traditional searches for $t \to h u$
in this mass range.
At smaller $m_{H^0}$, the branching ratio for $H^0 \to h h$
limits the sensitivity (see \cref{fig:heavyHiggsDecay}), while at larger $m_{H^0}$,
the production cross section peters out (see \cref{fig:thhXSection}).

\section[$H^0 \to \tau\mu$ Decay in the 2HDM]
        {\texorpdfstring{$H^0 \to \tau\mu$}{H^0 -> tau mu} Decay in the 2HDM}
\label{sec:2HDM-Htaumu}

We now move from Higgs couplings violating quark flavor to couplings violating
lepton flavor. This is motivated for instance by the recent CMS excess in the
$h \to \tau\mu$ channel~\cite{Khachatryan:2015kon}.  Since in the 2HDM, any
such flavor violation would be related to a misalignment in flavor space
between the Yukawa couplings of $\Phi_1$ and $\Phi_2$, we expect flavor
violating effects also for the heavy neutral Higgs boson $H^0$.  Moreover, in
view of the required smallness of $\sin\alpha$, an observable $\tau \mu h$ coupling
requires a much larger $\tau \mu H^0$ coupling and thus a large branching ratio for
$H^0 \to \tau\mu$. Our goal in the following is to constrain this decay channel
of the heavy Higgs boson using LHC data.

\subsection{Production Cross Sections, Decay Rates and Indirect Constraints}
\label{sec:2HDM-tau-mu-xsecs}

We consider the lepton Yukawa couplings
\begin{align}
  \mathcal{L}_\ell \supset - \eta_{\ell,1}^{ij} \overline{L_L^i} \Phi_1 e_R^j
                           - \eta_{\ell,2}^{ij} \overline{L_L^i} \Phi_2 e_R^j + h.c.\,.
  \label{eq:L-ell-1}
\end{align}
Here, $L_L^i$ are the left-handed lepton doublets and $e_R^j$ are the
right handed charged lepton fields.
Since we will see that the coupling of $H^0$ to top quarks is crucial for the
phenomenological of lepton flavor violating processes, we will also allow
$\eta_{u,2}^{tt}$ to be nonzero (see \cref{sec:2HDM-intro}).

After electroweak symmetry breaking, the lepton Yukawa couplings turn into
\begin{align}
\mathcal{L}_\ell
  &=  - \bar{e}_L^i e_R^j \bigg[
         h \bigg(  \frac{m_i}{v} \delta^{ij} \cos\alpha
                 + \frac{\eta_{\ell,2}^{ij}}{\sqrt{2}} \sin\alpha \bigg)
       + H^0 \bigg( -\frac{m_i}{v} \delta^{ij} \sin\alpha
                 + \frac{\eta_{\ell,2}^{ij}}{\sqrt{2}} \cos\alpha \bigg) \bigg] + h.c.
    \label{eq:L-ell-2} \\
  &\equiv -\bar{e}_L^i e_R^j \Big[ {y^{ij}_{\ell,h} h + y^{ij}_{\ell,H} H^0} \Big] + h.c. \,,
    \label{eq:L-ell-3}
\end{align}
where $i,j = 1,2,3$  or $e, \mu, \tau$. We will again omit the subscript $\ell$
to unclutter the notation where this is unambiguous.
We will also assume for simplicity that the only
non-zero elements of $\eta_{\ell,2}$ are $\eta_{\ell,2}^{\mu\tau}$ and
$\eta_{\ell,2}^{\tau\mu}$, and that both are identical and real.  The decay
rates for $H^0 \to \tau\mu$ and $h\to\tau\mu$ are
\begin{align}
  \Gamma(H^0 \to \tau^+\mu^-) = \Gamma(H^0 \to \tau^-\mu^+)
    &= \frac{1}{32\pi} m_{H^0} \cos^2\alpha \Big(
         |\eta_2^{\mu\tau}|^2 + |\eta_2^{\tau\mu}|^2 \Big) \,, \\
  \Gamma(h \to \tau^+\mu^-) = \Gamma(h \to \tau^-\mu^+)
    &= \frac{1}{32\pi} m_h \sin^2\alpha \Big(
         |\eta_2^{\mu\tau}|^2 + |\eta_2^{\tau\mu}|^2 \Big) \,.
\end{align}
The CMS search for $h \to \tau^{\pm} \mu^{\mp}$~\cite{Khachatryan:2015kon}
constrains the combined branching ratio for these channel to be $< 1.51\%$ at 95\%~CL.
This translates into a bound
$\sin\alpha \, (|\eta_2^{\mu\tau}|^2 + |\eta_2^{\tau\mu}|^2)^{1/2} \leq 0.0050$.
The best fit value for the branching ratio is
$\BR(h \to \tau\mu) = (0.84^{+0.39}_{-0.37})\%$.

The strongest indirect constraint on $\eta_2^{\mu\tau}$ and
$\eta_2^{\tau\mu}$ comes from searches for the rare decay $\tau \to
\mu\gamma$~\cite{Blankenburg:2012ex, Harnik:2012pb,Aubert:2009ag}.  To quantify
this constraint, we follow refs.~\cite{Harnik:2012pb, Kopp:2014rva} and work with
the effective operators
\begin{align}
  \mathcal{L}_{\text{eff,\,} \tau\to\mu\gamma} = c_L Q_{L\gamma} + c_R Q_{R\gamma} \,,
  \label{eq:L-tau-mu-gamma}
\end{align}
where $c_L$, $c_R$ are Wilson coefficients, and the dimension-5 operators
$Q_{L\gamma}$, $Q_{R\gamma}$ are given by
\begin{align}
  Q_{L\gamma,R\gamma} = \frac{e}{8\pi^2} m_\tau \big(
                          \bar{\mu} \sigma^{\alpha\beta} P_{L,R} \tau
                        \big) F_{\alpha\beta} \,.
  \label{eq:QLR-tau-mu-gamma}
\end{align}
$c_L$ and $c_R$ receive contributions from 1- and 2-loop diagrams involving
$h$, $H^0$, $A^0$, and $H^\pm$. The 2-loop contributions are comparable to the
1-loop terms because the latter are suppressed by the small $\tau$ Yukawa
coupling~\cite{Blankenburg:2012ex,Harnik:2012pb}.  The loop diagrams involving
only $h$ or only $H^0$ are given by the expressions in the appendix of
ref.~\cite{Harnik:2012pb} (adopted from \cite{Chang:1993kw}), with the Yukawa
matrices in these formulas identified with $y_{\ell,h}$ and $y_{\ell,H}$ from
\cref{eq:L-ell-3} for leptons and with $y_{u,h}^{tt}$ and $y_{u,H}^{tt}$ from
\cref{eq:yuh-2hdm,eq:yuH-2hdm} for top quarks.
Note that diagrams containing $h$ and $H^0$ tend to cancel
each other.  The reason is that each diagram contains one flavor violating
Yukawa coupling and one flavor conserving one. According to \cref{eq:L-ell-2},
this implies that most of the diagrams involving $h$ differ from their
counterparts involving $H^0$ by a minus sign. (The only exception can be
diagrams with $h$ or $H^0$ coupled to a top quark loop if $(m_t / v) \sin\alpha
< \eta_2^{tt} / \sqrt{2}$.) The diagrams involving only $A^0$ are obtained in a
similar way~\cite{Omura:2015nja},
by replacing the up quark Yukawa couplings in the expressions from
ref.~\cite{Harnik:2012pb} by $-i \eta_{u,2}^{ij} / \sqrt{2}$ and the lepton
Yukawa couplings by $i \eta_{\ell,2}^{ij} / \sqrt{2}$. Since we assume that the
only nonzero Yukawa couplings of $\Phi_2$ are $\eta_2^{\mu\tau}$,
$\eta_2^{\tau\mu}$ and $\eta_2^{tt}$, and that $A^0$ does not mix with the
other neutral Higgs bosons, the only relevant diagrams containing $A^0$ are the
2-loop Barr--Zee diagrams with a top quark loop.  Regarding the neglected
diagrams involving $H^\pm$, we estimate that their contribution is of the same
order as that of diagrams involving only $H^0$ or only $A^0$, and we will
include this uncertainty in our plots.

\subsection[LHC Constraints on $H^0 \to \tau\mu$ in the 2HDM]
           {LHC Constraints on \texorpdfstring{$H^0 \to \tau\mu$}
            {H0 -> tau mu} in the 2HDM}
\label{sec:2HDM-tau-mu-constraints}

\begin{figure}
  \centering
  \begin{tabular}{c@{\qquad}c}
    \includegraphics[width=0.42\textwidth]{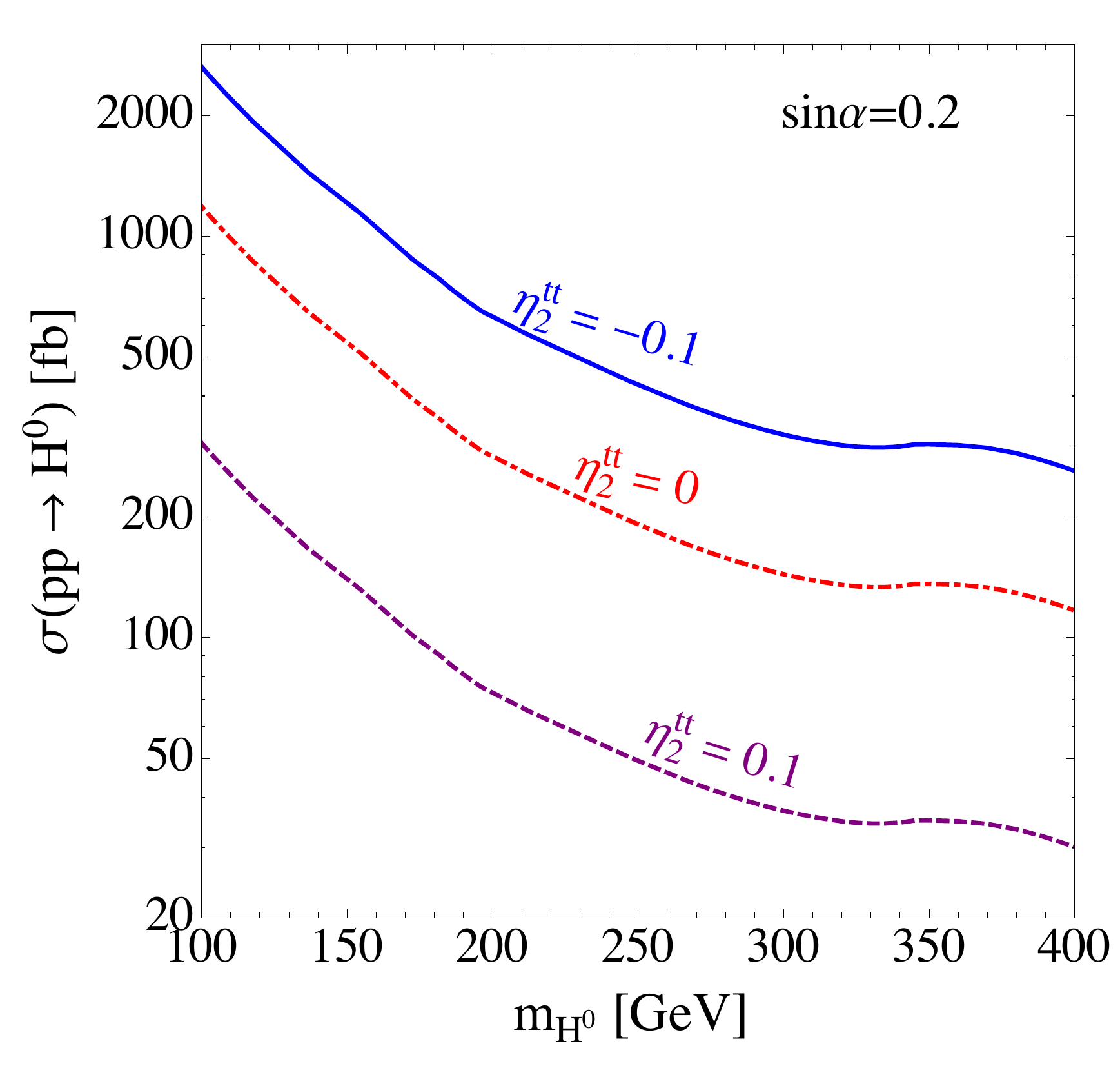} &
    \includegraphics[width=0.47\textwidth]{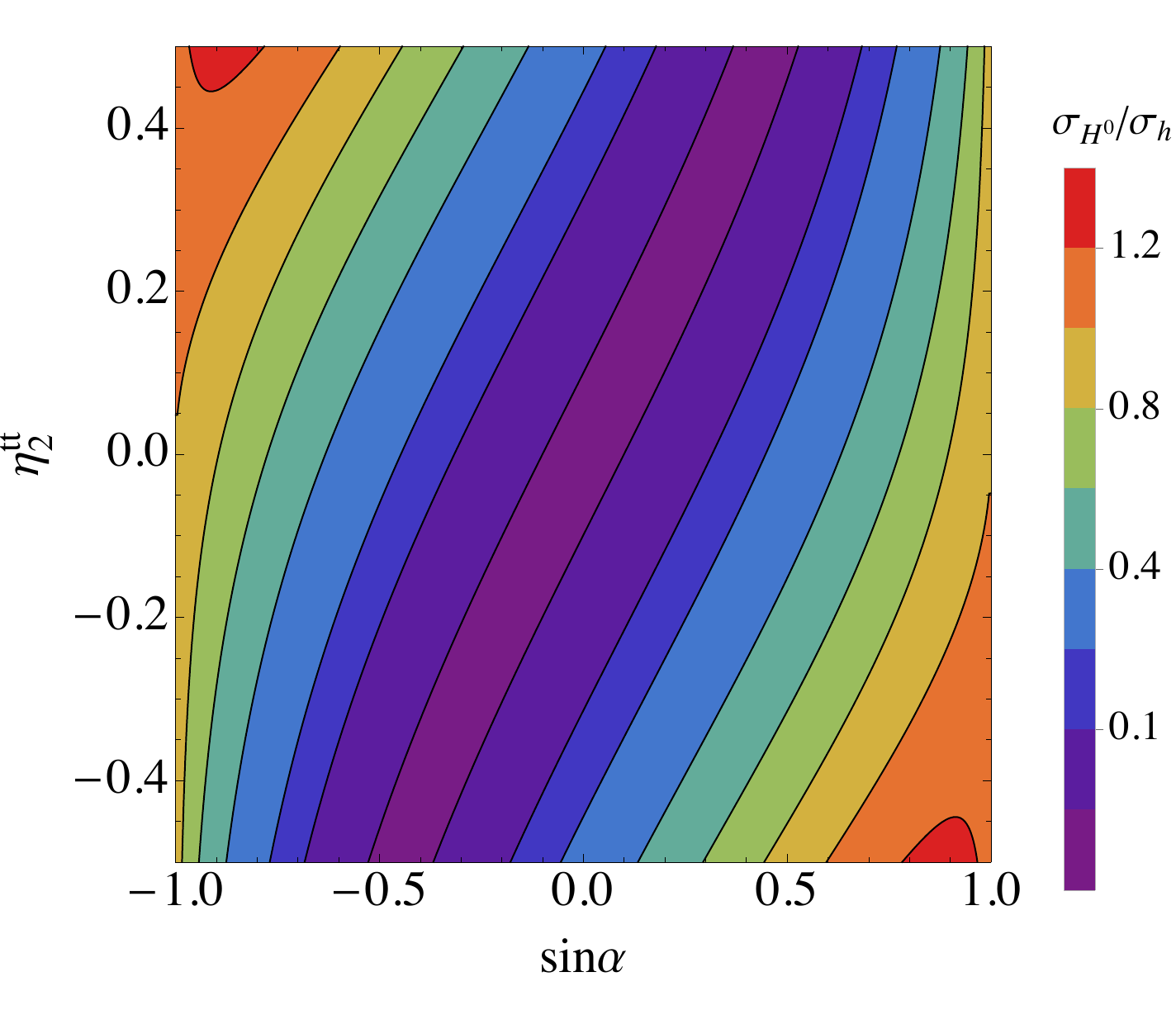} \\
    (a) & (b)
  \end{tabular}
  \caption{(a) The production cross section of the heavy Higgs boson $H^0$ via gluon
    fusion (see \cref{eq:xsec-ggH}) as a function of $m_{H^0}$ in the lepton
    flavor violating 2HDM.  The shape of the curves follows that of the SM Higgs
    production cross section, taken from~\cite{Dittmaier:2011ti}.
    (b) Ratio of the $H^0$ production cross section to the
    production cross section of a SM-like Higgs at the same mass as a function
    of the neutral Higgs boson mixing angle $\sin\alpha$ and the Yukawa coupling
    of the second Higgs doublet to top quarks, $\eta_2^{tt}$. Note that this
    ratio is independent of $m_{H^0}$.}
  \label{fig:xsec-ggH}
\end{figure}

\begin{figure*}
  \begin{tabular}{cc}
    \includegraphics[width=0.48\textwidth]{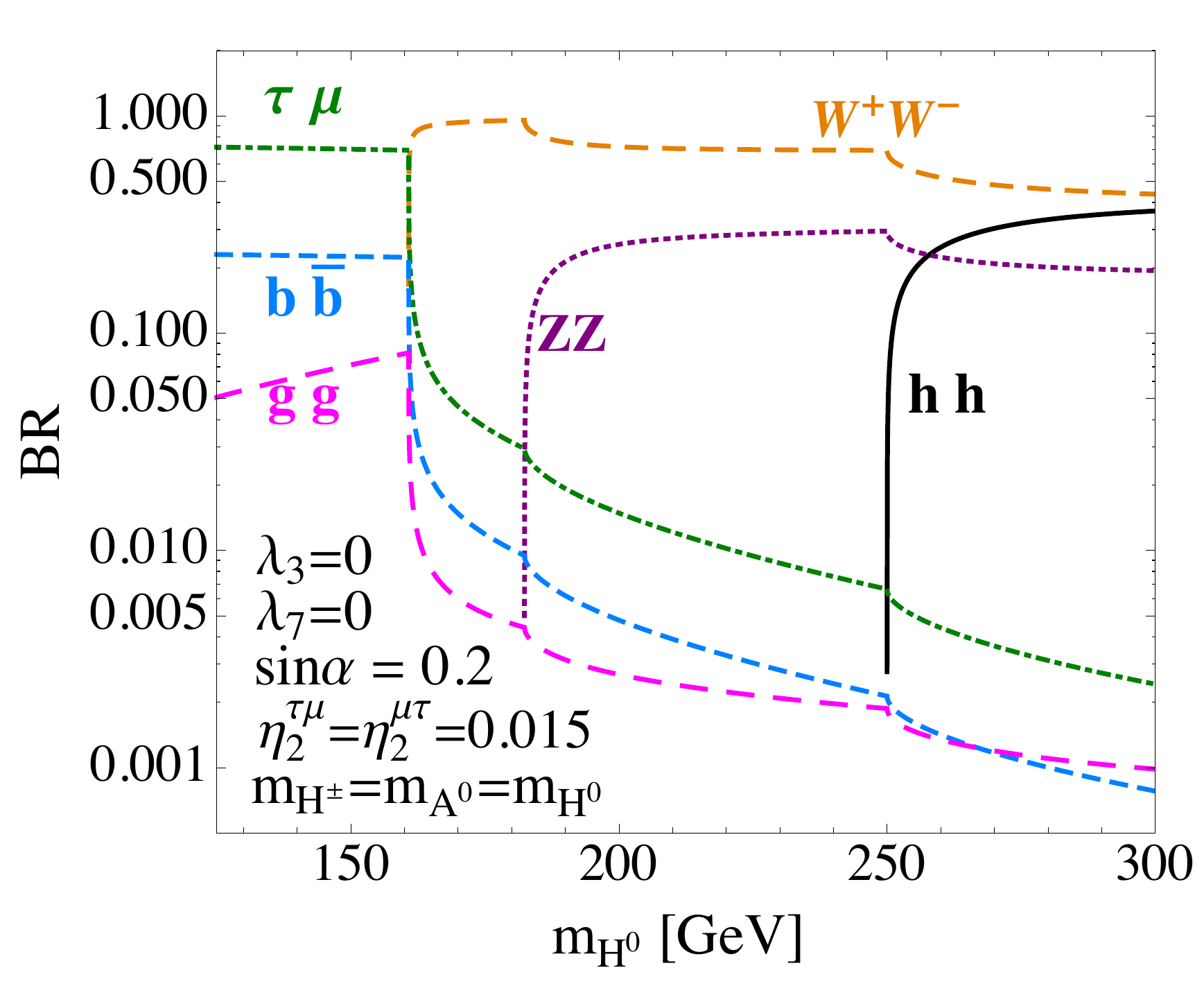} &
    \includegraphics[width=0.48\textwidth]{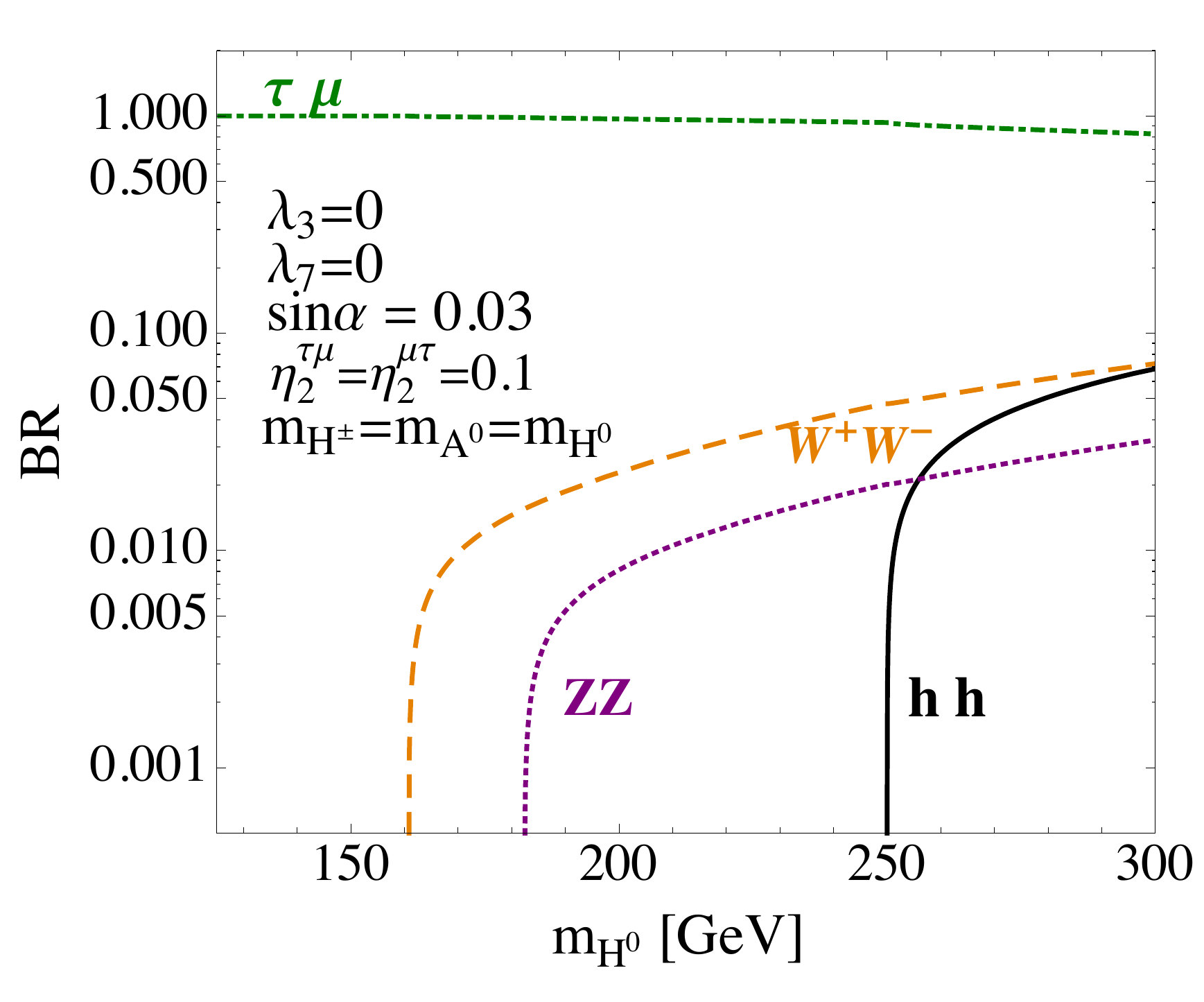} \\
    (a) & (b)
  \end{tabular}
  \caption{Branching ratios of the heavy neutral CP-even Higgs boson $H^0$ as a function
    of its mass $m_{H^0}$ for two different parameter points of the lepton flavor violating
    2HDM. We assume here a scenario with large lepton flavor violation in the
    $\mu$--$\tau$ sector, as expressed by the Yukawa couplings $\eta_2^{\mu\tau}$,
    $\eta_2^{\tau\mu}$ of the second Higgs doublet.}
  \label{fig:Hbranching-taumu}
\end{figure*}

If we assume the direct Yukawa couplings of the second Higgs doublet $\Phi_2$ to
light quarks to be small like for the SM-like Higgs, the dominant
$H^0$ production channel in the lepton flavor violating 2HDM is gluon fusion,
with the cross section
\begin{align}
  \sigma(pp \to H^0) \simeq
    \bigg(\!\sin\alpha - \eta_2^{tt} \cos\alpha \frac{v}{\sqrt{2} m_t} \bigg)^2
    \times \sigma(gg \to h)\big|^\text{SM}_{m_h=m_{H^0}} \,.
  \label{eq:xsec-ggH}
\end{align}
Here, $\sigma(gg \to h)\big|^\text{SM}_{m_h=m_{H^0}}$ is the gluon fusion cross
section in the SM if the Higgs mass is set to $m_{H^0}$~\cite{Dittmaier:2011ti}.
The first term in
parentheses arises from mixing between the two CP even neutral Higgs bosons
$h_1$ and $h_2$ (see \cref{eq:Phi1-Phi2}), while the second one is related to
the direct coupling of a top quark loop to $h_2$.  If other diagonal Yukawa
couplings of $\Phi_2$ to quarks besides $\eta_2^{tt}$ are $\gtrsim \sin\alpha$,
they would contribute as well.
For instance, if $\Phi_2$ has a large coupling $\eta_2^{bb}$ to bottom quarks,
the factor in parentheses in \cref{eq:xsec-ggH} would receive an extra contribution
\begin{align*}
  -\eta_2^{bb} \cos\alpha \frac{v}{\sqrt{2} m_t} \times
    \frac{f(4 m_b^2 / m_{H^0}^2)}{f(4 m_t^2 / m_{H^0}^2)} \,,
\end{align*}
where $f(x) = x [1 + (1 - x) F(x)]$ is a loop function~\cite{Ellis:1975ap,
Shifman:1979eb,Kniehl:1995tn}. The factor
$F(x)$ is given by
\begin{align*}
  F(x) = \begin{cases}
           \arcsin^2(1/\sqrt{x})                  & \text{for $x \ge 1$} \\
           \frac{1}{2} \left[ \log\left( \frac{1 + \sqrt{1-x }}{1 - \sqrt{1-x}} \right)
                            - i\pi  \right]^2     & \text{for $x < 1$}
         \end{cases} \,.
\end{align*}
We will, however, not consider
this possibility here and assume $\eta_2^{tt}$ to be the largest diagonal element of
$\eta_2$ in the following.  With the additional simplifying assumptions $\eta_2^{\mu\tau} =
\eta_2^{\tau\mu}$ (see \cref{sec:2HDM-tau-mu-xsecs}), the overall process $p p
\to H^0 \to \mu \tau$ depends on four parameters: $m_{H^0}$, $\sin\alpha$,
$\eta_2^{tt}$, and $\eta_2^{\mu\tau} = \eta_2^{\tau\mu}$ We show the production
cross section $\sigma(pp \to H^0)$ in \cref{fig:xsec-ggH}, and the branching
ratios into the most relevant decay channels for two benchmark points in
\cref{fig:Hbranching-taumu}.

\begin{figure*}
  \begin{tabular}{cc}
    \includegraphics[width=0.45\textwidth]{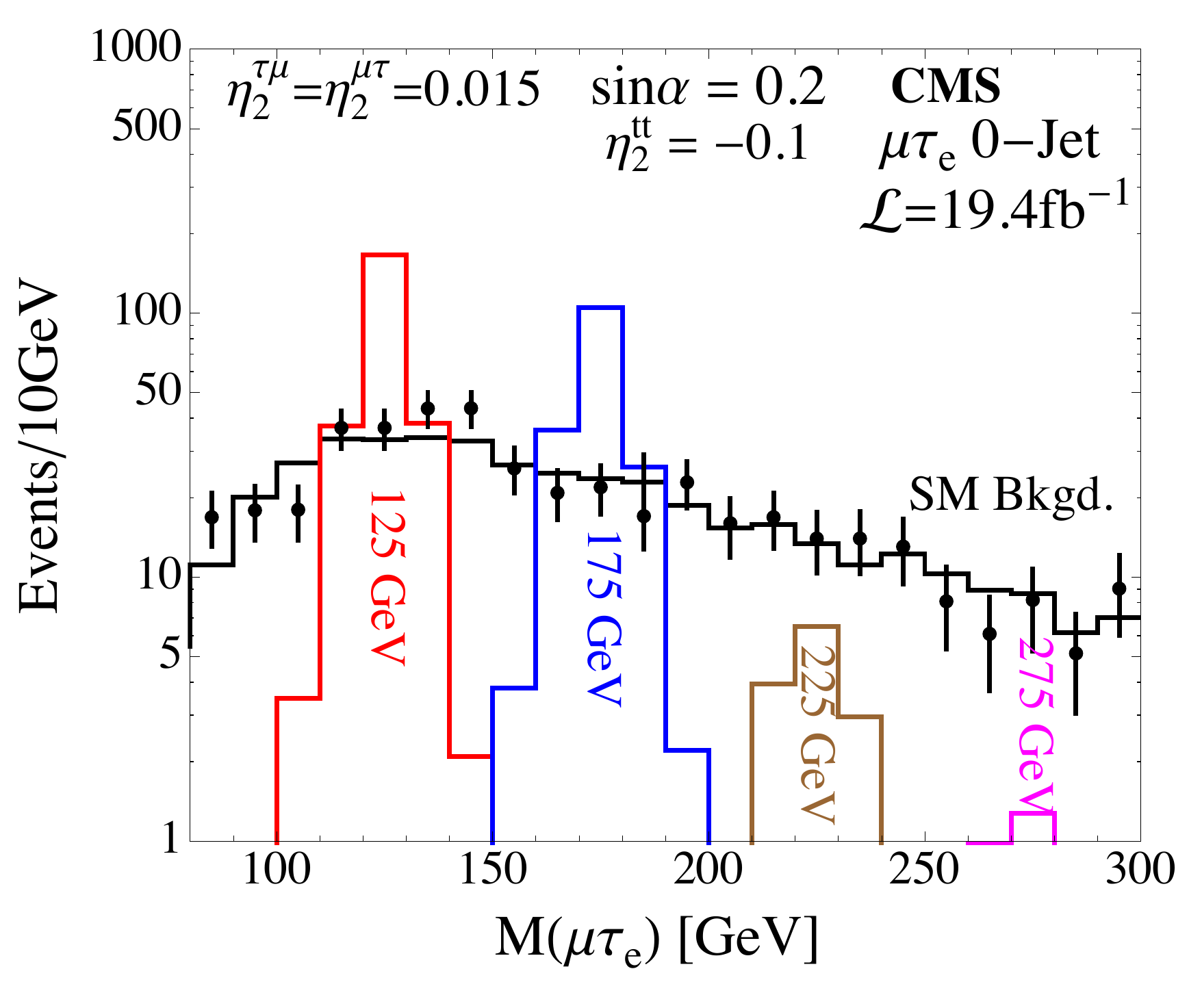} &
    \includegraphics[width=0.45\textwidth]{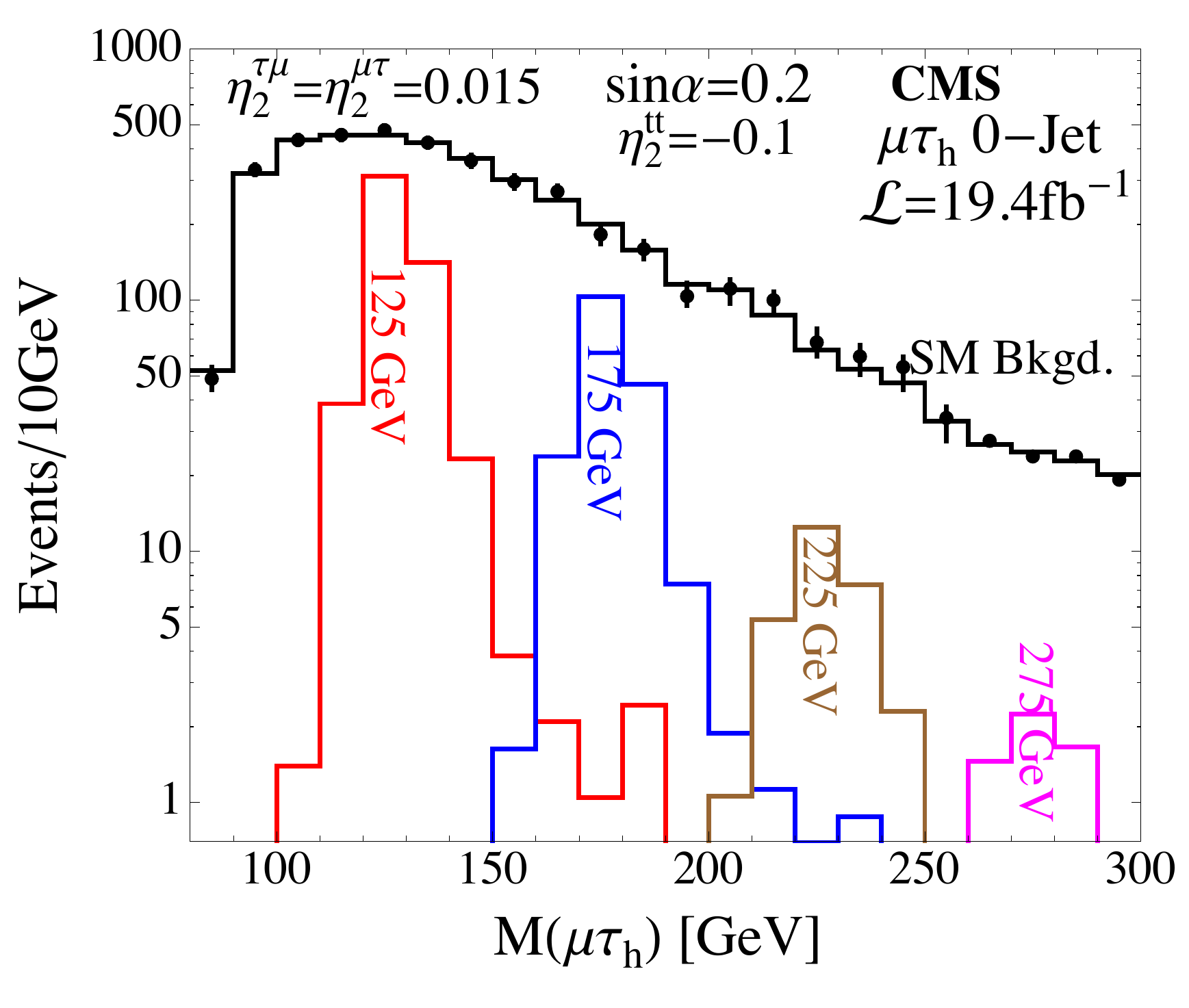} \\
    (a) & (b) \\[0.2cm]
     \includegraphics[width=0.45\textwidth]{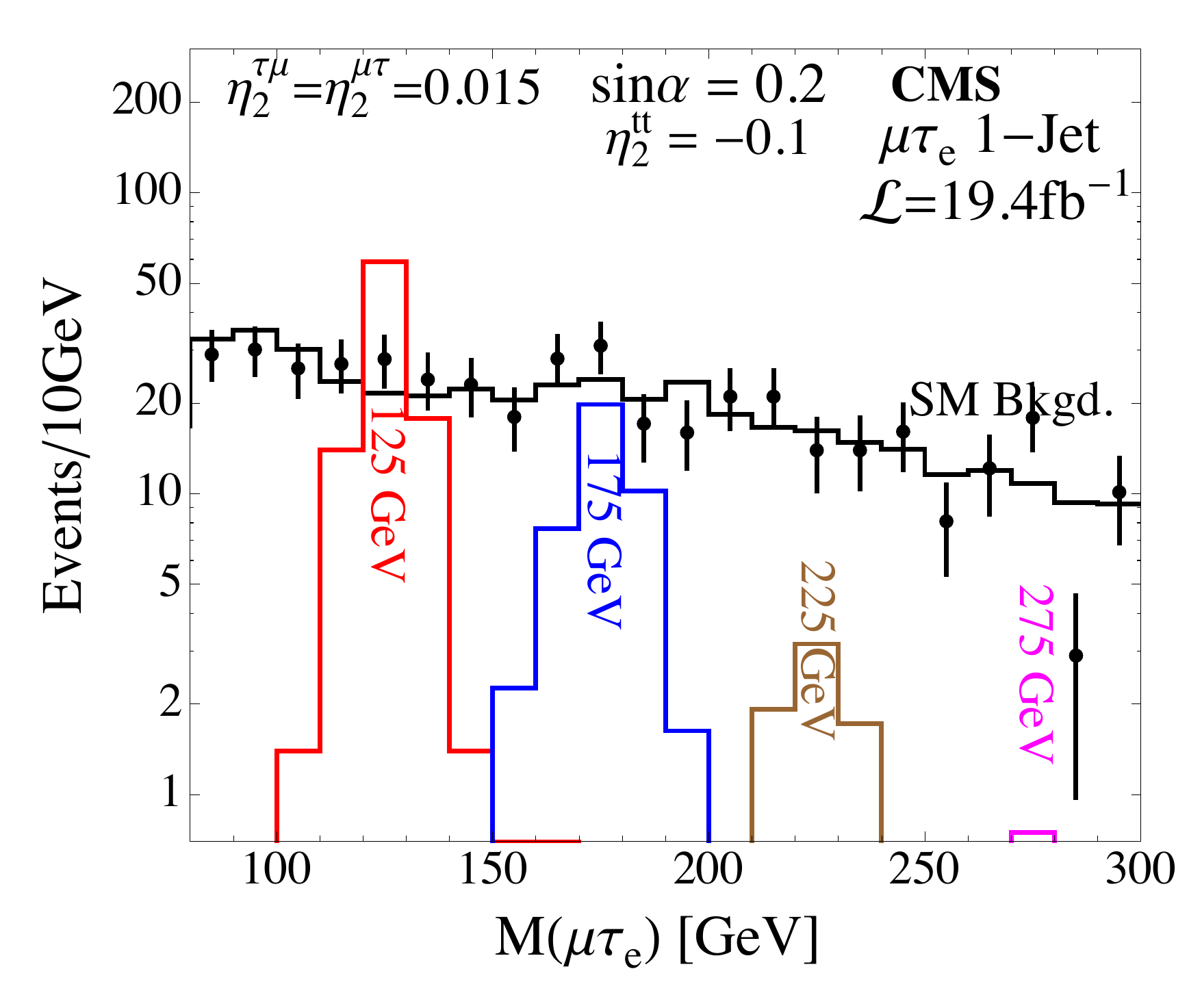} &
    \includegraphics[width=0.45\textwidth]{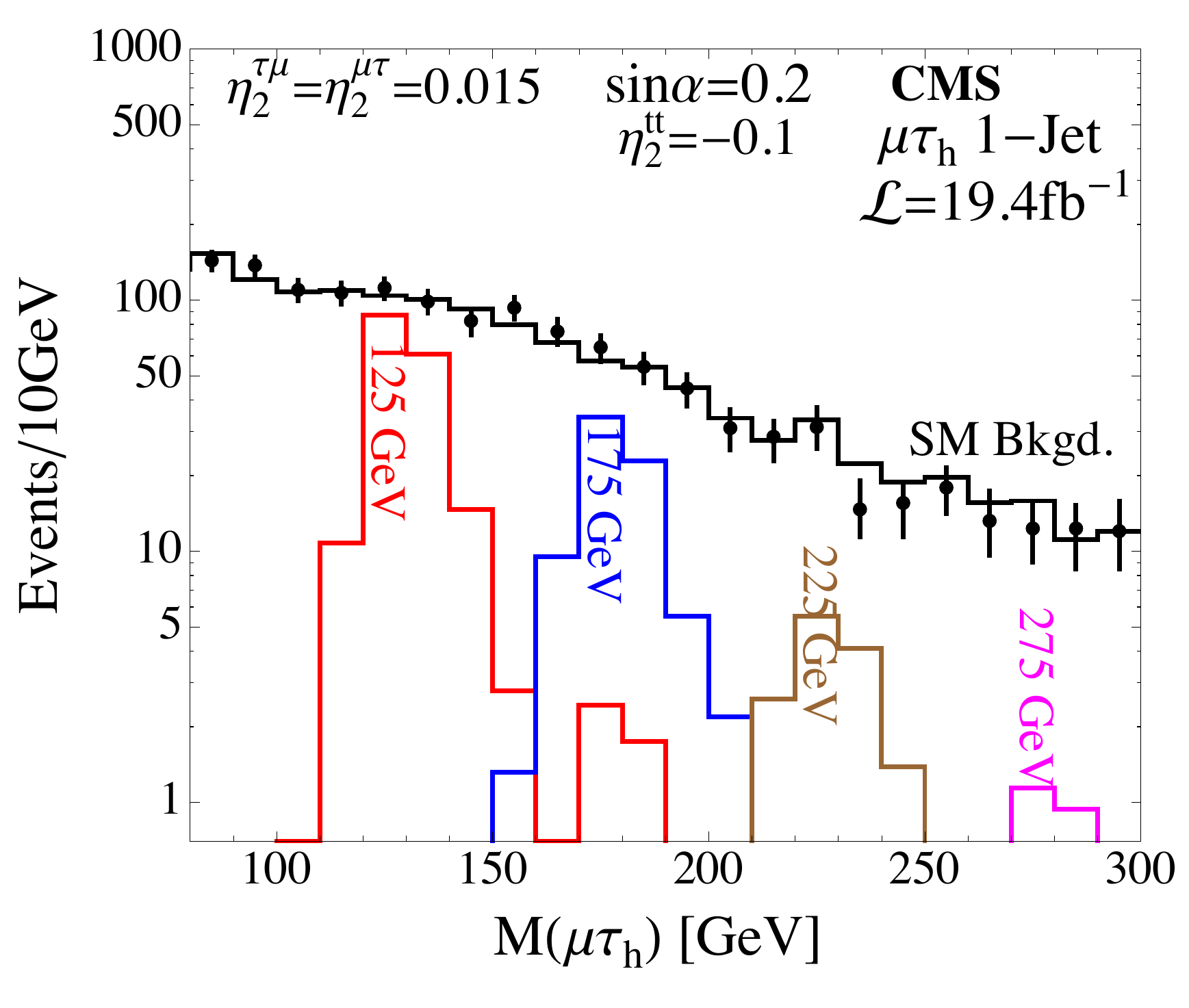} \\
    (c) & (d)
  \end{tabular}
  \caption{Distribution of the collinear mass $m^\text{coll}_{\mu\tau}$
    after cuts for the SM background from \cite{Khachatryan:2015kon}
    and for a hypothetical $H^0 \to \tau\mu$ signal at various values of $m_{H^0}$.
    We work in the context of the lepton flavor violating 2HDM,
    with the parameters indicated in the plots.
    The panels on the left are for events with leptonic $\tau$ decays
    $\tau \to e \nu\nu$, denoted here as $\tau_e$, while the panels on the
    right include only events with hadronic $\tau$ decays, denoted as $\tau_h$.
    In the upper row we show events with no extra jets with $p_T > 30$~GeV,
    $|\eta| < 4.7$, while in the bottom row we require exactly one such jet.
    We do not include event categories with $\geq 2$ jets in our analysis.}
 \label{fig:h-tau-mu-mcoll}
\end{figure*}

In the following, we use the data from the CMS search for $h \to
\tau\mu$~\cite{Khachatryan:2015kon} to constrain also the decay $H^0 \to
\tau\mu$. In particular, we compare CMS' measured distributions of the reconstructed
collinear mass $m^\text{coll}_{\mu\tau}$ of the $\mu$--$\tau$ system to the
predicted signals at various values of $m_{H^0}$. This comparison is shown graphically
in \cref{fig:h-tau-mu-mcoll}, which leads us to expect that constraints will
be competitive for $m_{H^0} \lesssim 250$~GeV, while at larger masses the signal
will be too small compared to the background. To produce \cref{fig:h-tau-mu-mcoll}
and for the subsequent statistical analysis, we have
simulated inclusive samples of $p p \to H^0 \to \tau\mu$ events and applied the
cuts from \cite{Khachatryan:2015kon}.
We have used the same simulation tools as in \cref{sec:simulationEFT,sec:2HDM-constraints}.
The predicted SM background distributions are taken from
\cite{Khachatryan:2015kon}. For search channels with hadronic $\tau$ decays
(denoted here as $\tau_h$),
the background is dominated by events with fake leptons, while for events with leptonic
$\tau$ decays (denotes as $\tau_e$), the dominant backgrounds are $Z \to \tau\tau$,
di-boson production, fake leptons, and, in event categories that allow for extra
jets, $t\bar{t}$.  We have checked that,
when setting $m_{H^0} = 125$~GeV and $\sin\alpha = 1$, we obtain a limit that is about
15\% weaker than the CMS limit $\BR(h \to \mu\tau) < 0.015$,
and we also confirm that the null hypothesis of
the branching ratio being zero is disfavored at $\sim 2\sigma$.
We surmise that the reason why our limit
is slightly weaker than the CMS limit is the omission of the event
categories with $\geq 2$ jets in our analysis.

\begin{figure*}
  \includegraphics[width=0.45\textwidth]{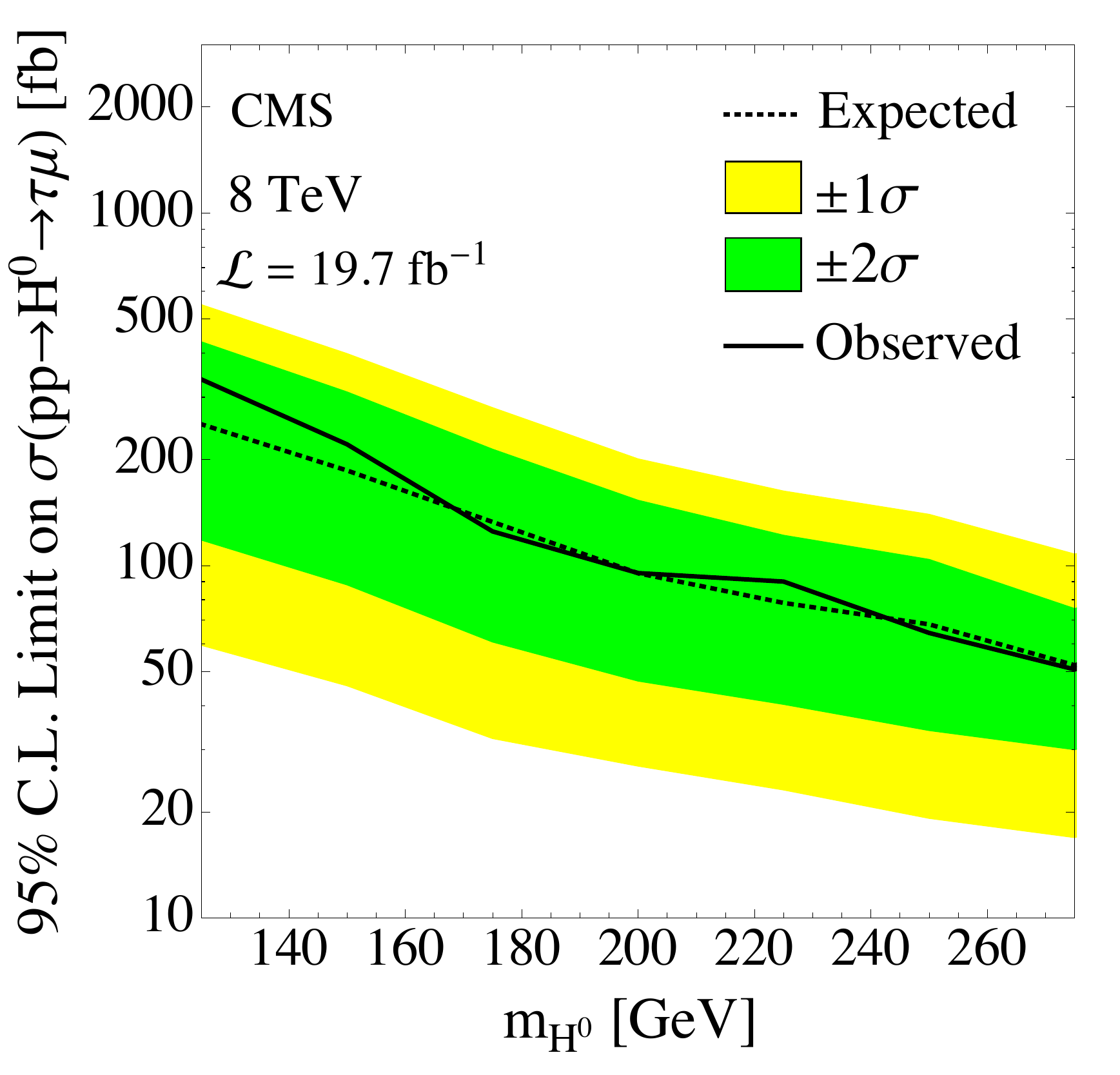}
  \caption{95\%~CL limit on the cross section for a $p p \to H^0 \to \tau\mu$
    signal as a function of $m_{H^0}$, obtained by recasting the results from
    \cite{Khachatryan:2015kon}.}
  \label{fig:xsec-limit-H-tau-mu}
\end{figure*}

\begin{figure*}
  \begin{tabular}{ccc}
    \includegraphics[width=0.33\textwidth]{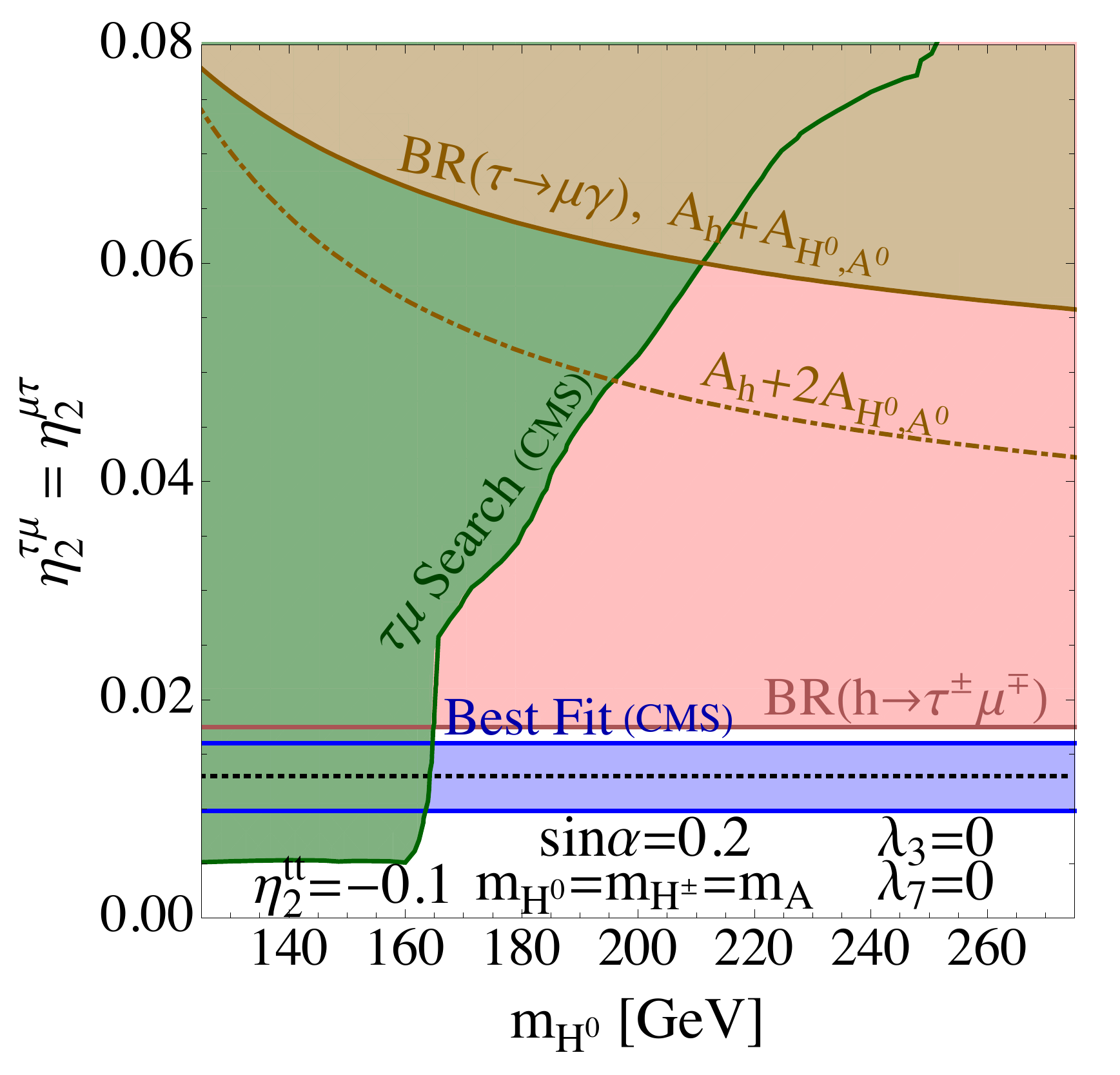} &
    \includegraphics[width=0.33\textwidth]{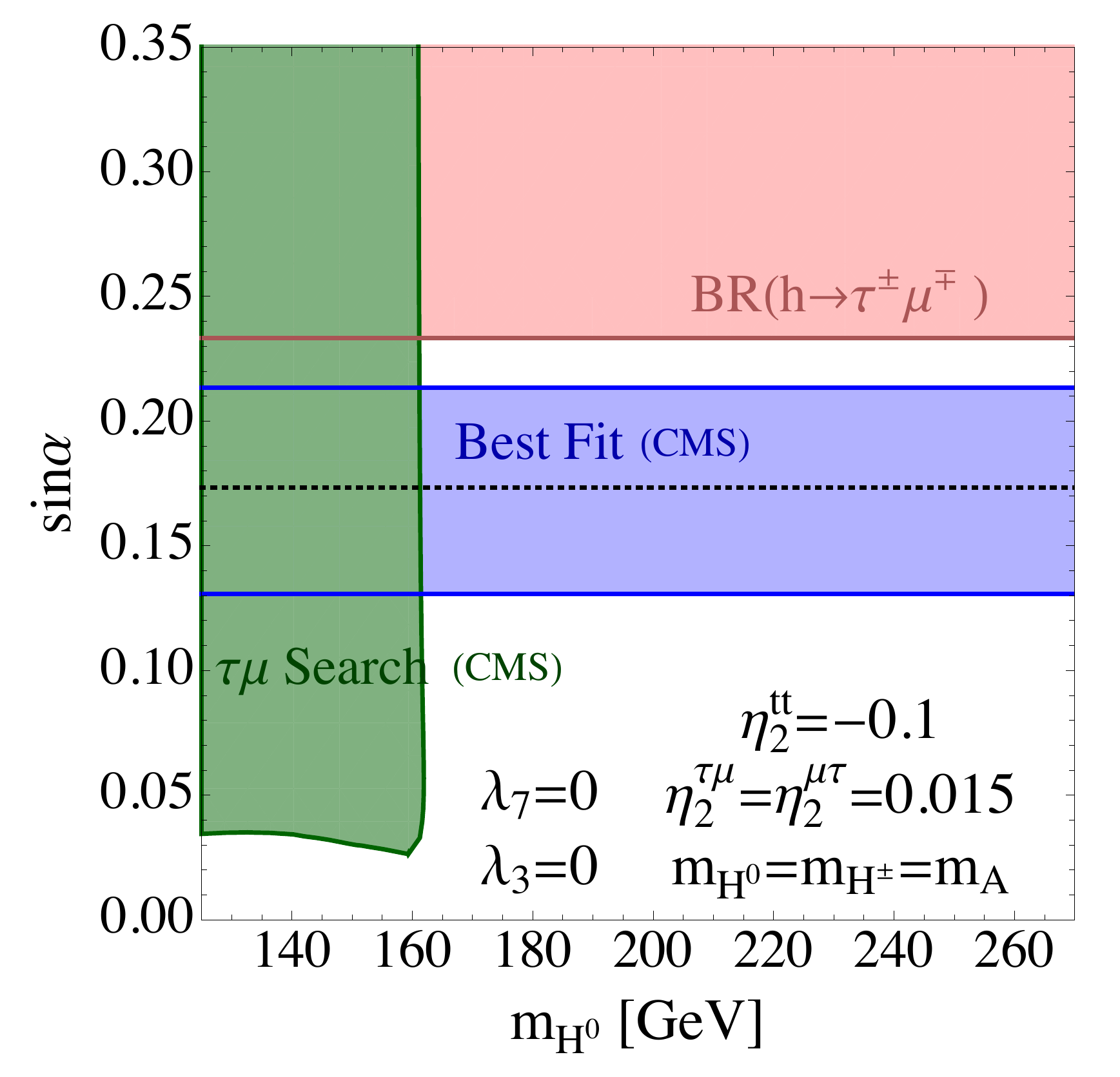} &
    \includegraphics[width=0.33\textwidth]{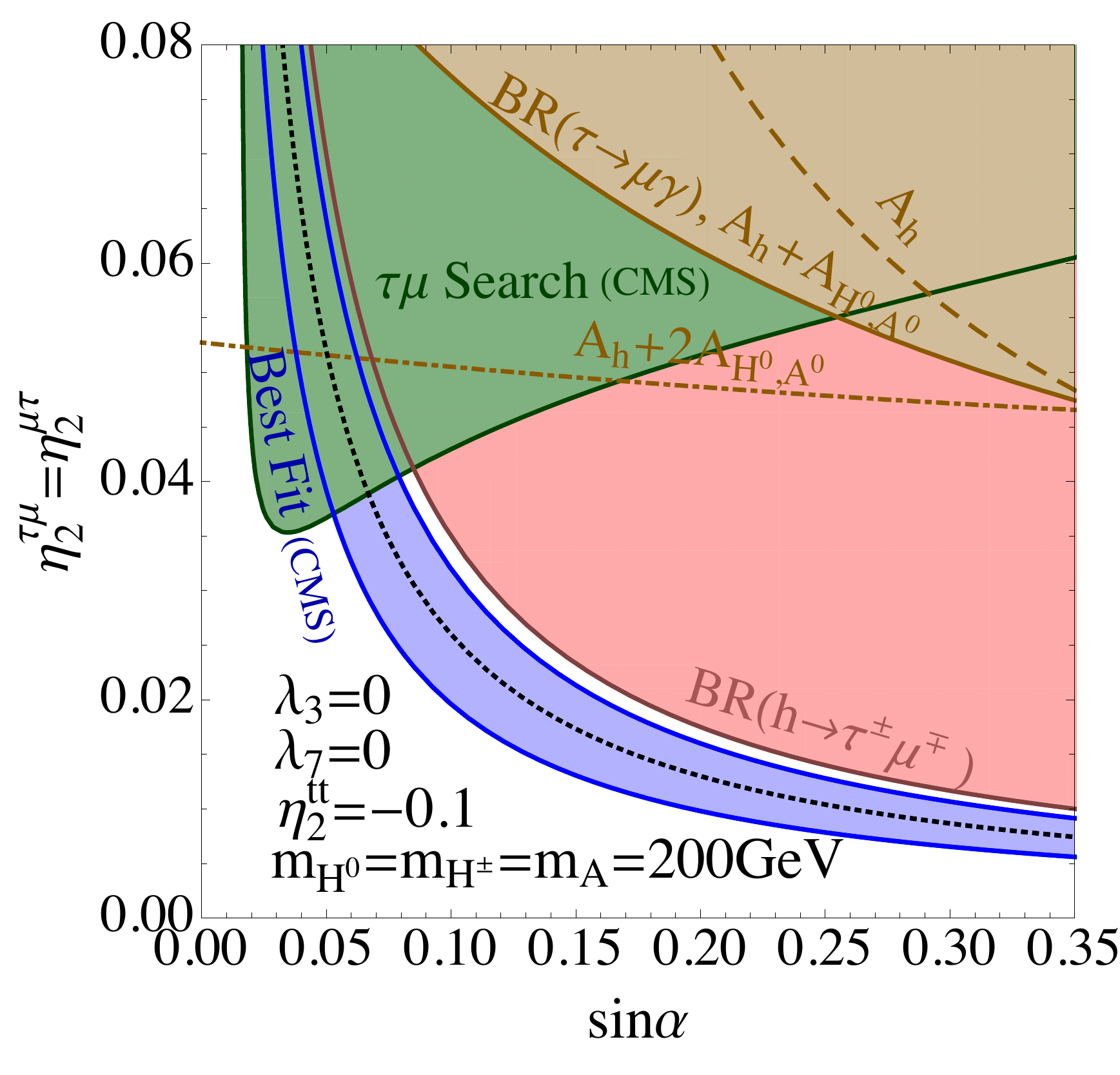} \\[-0.1cm]
    (a) & (b) & (c) \\[0.2cm]

    \includegraphics[width=0.33\textwidth]{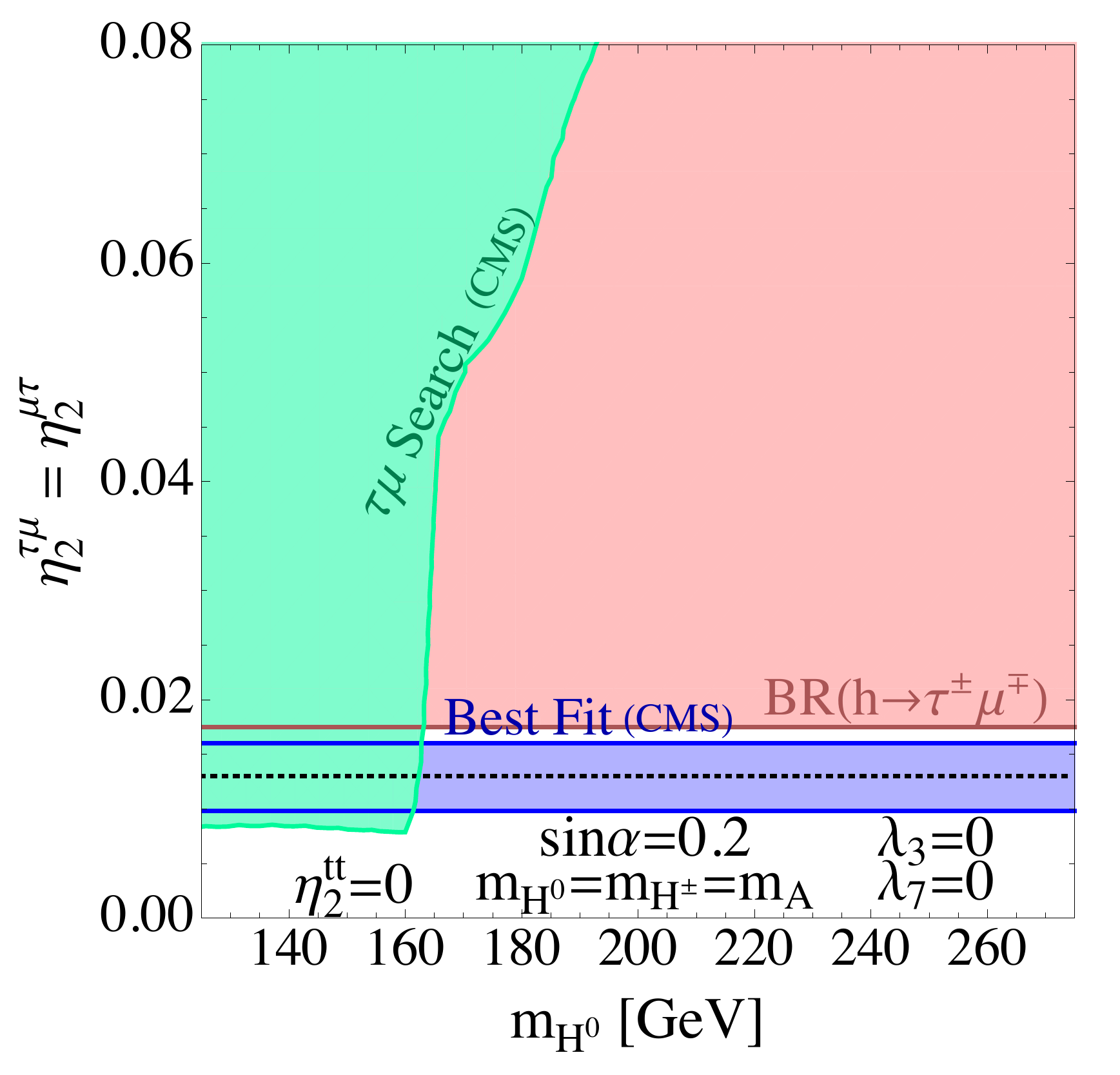} &
    \includegraphics[width=0.33\textwidth]{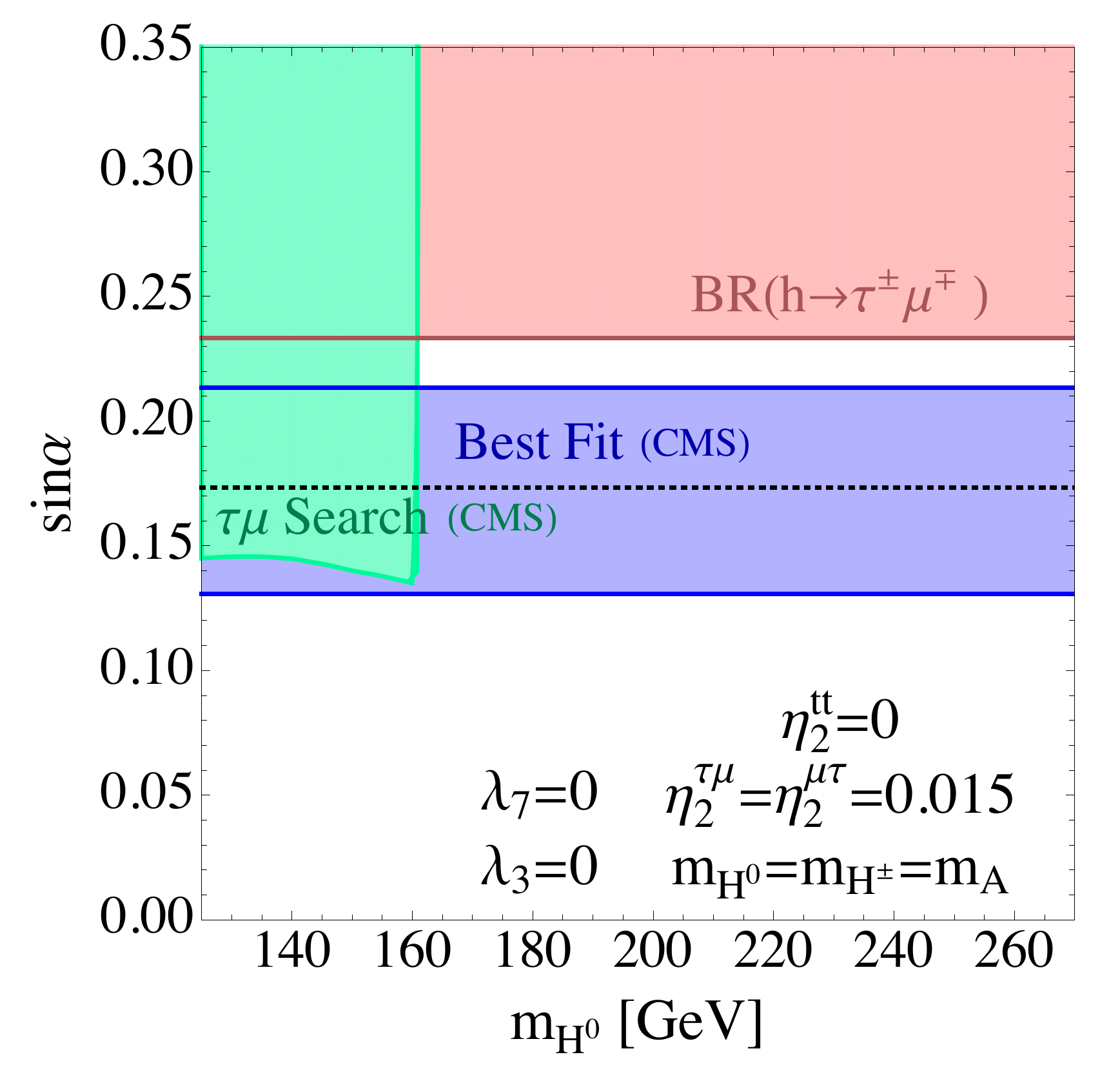} &
    \includegraphics[width=0.33\textwidth]{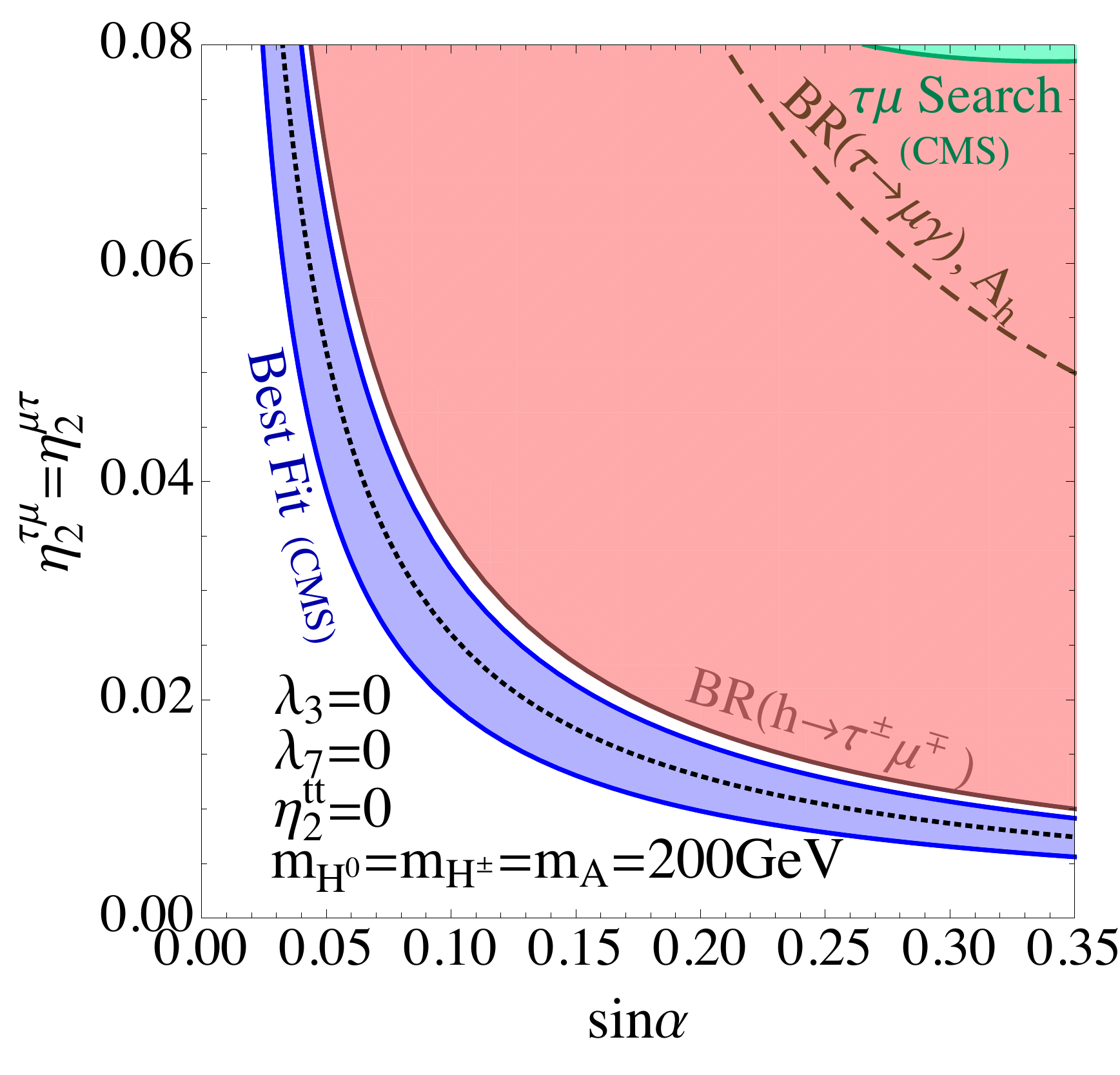} \\[-0.1cm]
    (d) & (e) & (f) \\[0.2cm]

    \includegraphics[width=0.33\textwidth]{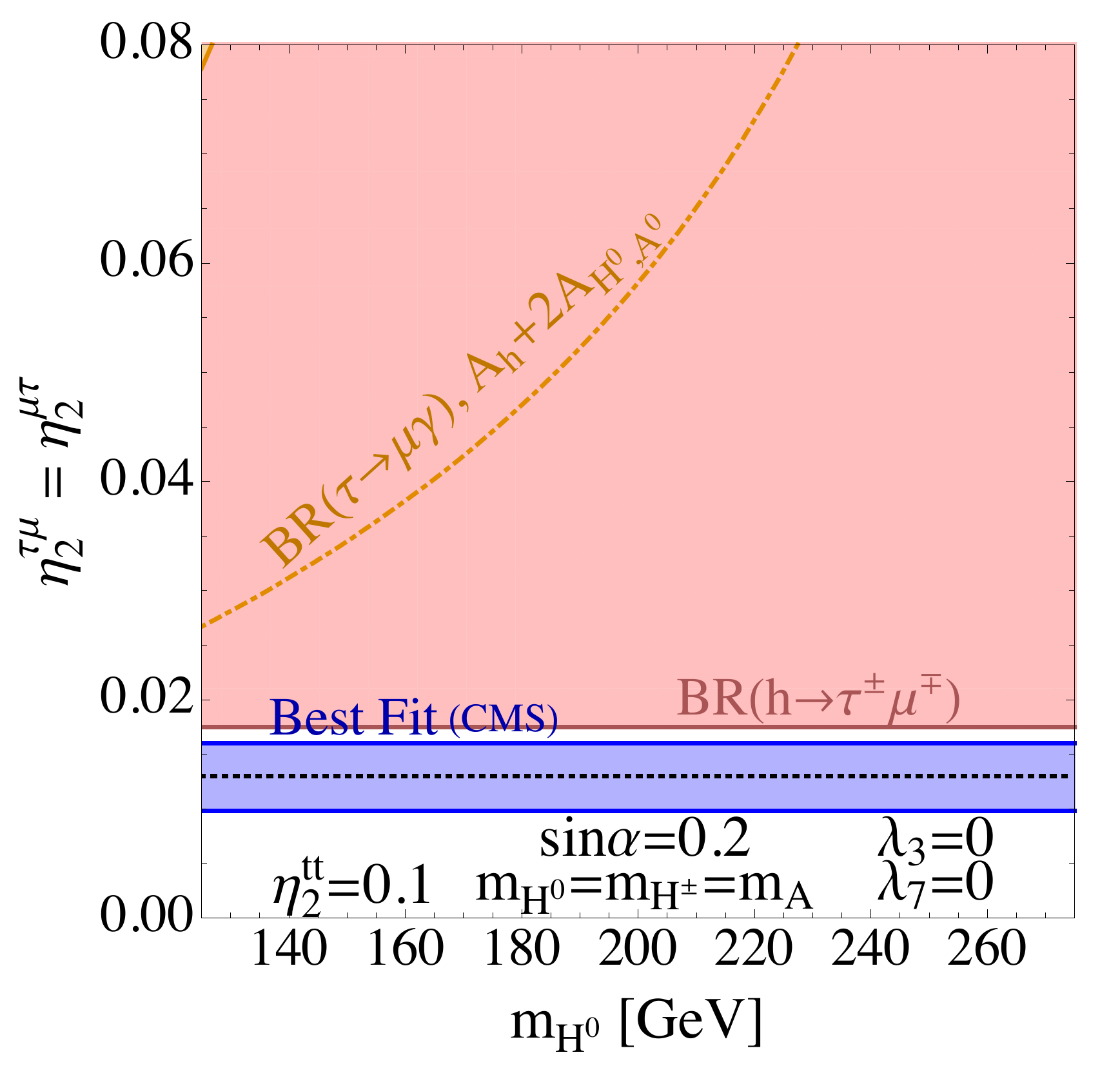} &
    \includegraphics[width=0.33\textwidth]{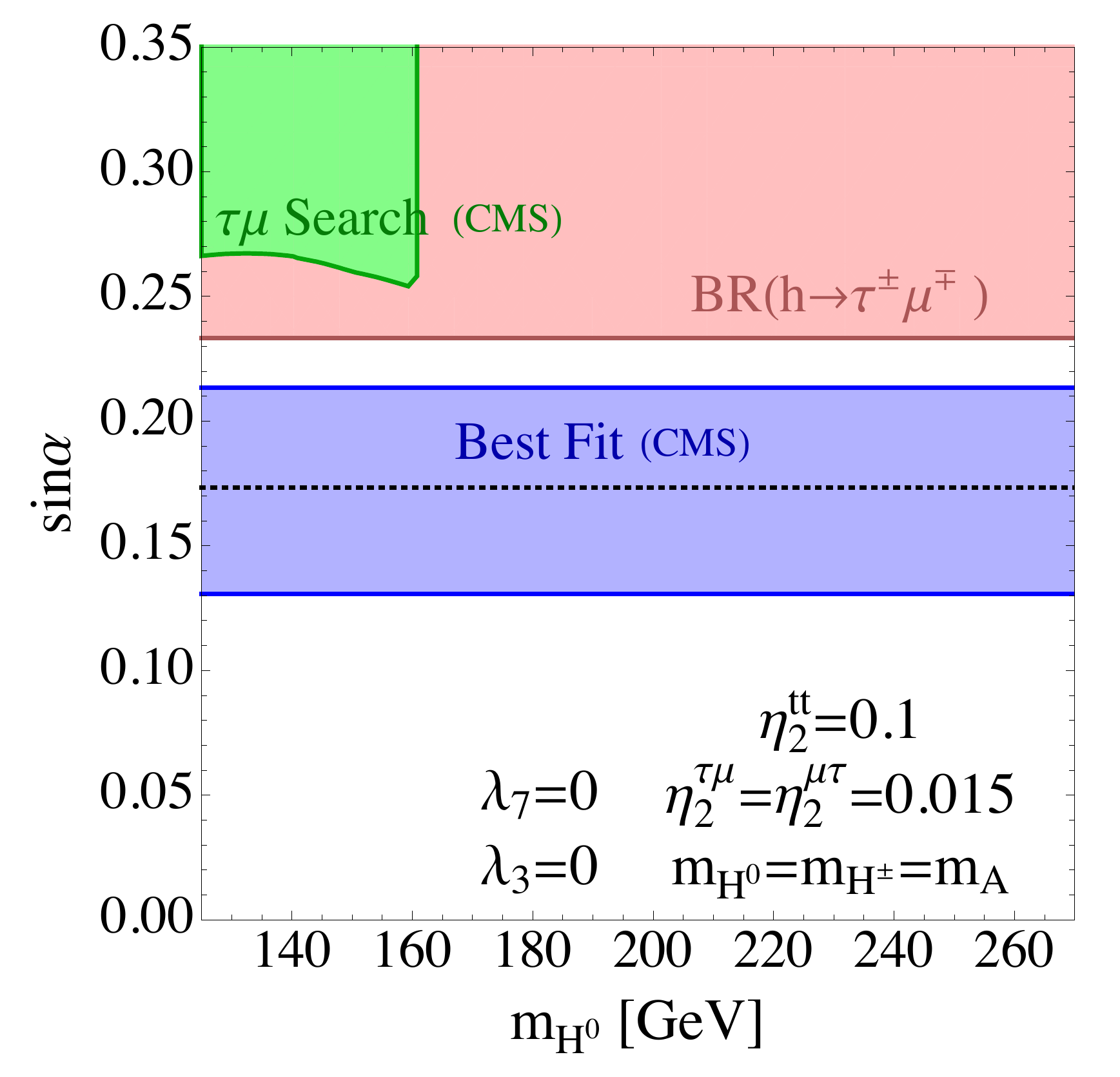} &
    \includegraphics[width=0.33\textwidth]{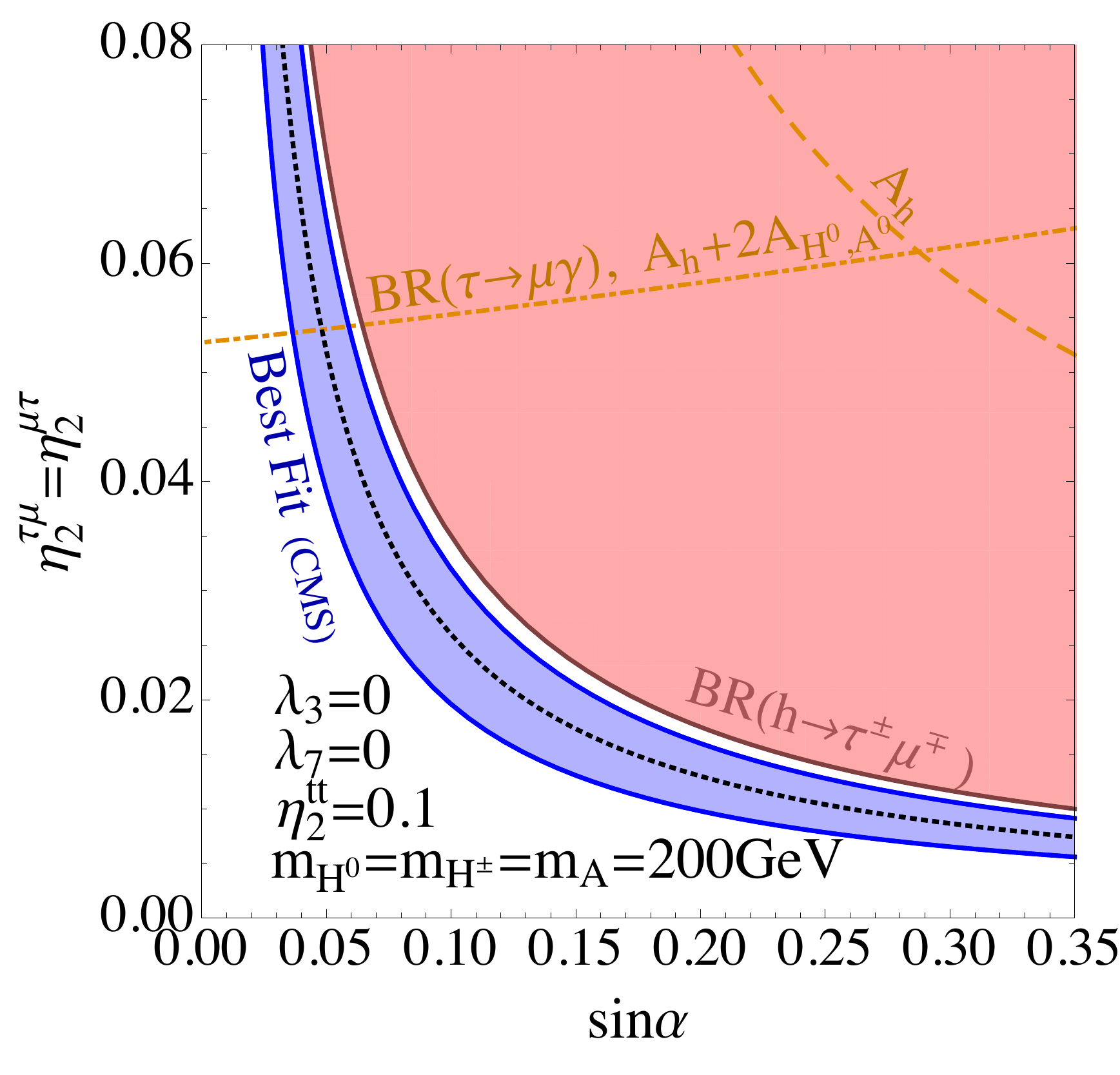} \\[-0.1cm]
    (g) & (h) & (i)
  \end{tabular}
  \caption{95\% CL constraints on the parameter space of the lepton flavor violating
    2HDM. We show results from our search for $H^0 \to \tau\mu$ based on
    recasting \cite{Khachatryan:2015kon} (green contours), from the CMS search for
    $h \to \tau\mu$ (red exclusion regions and blue $1\sigma$ preferred
    regions)~\cite{Khachatryan:2015kon}, and from
    $\tau \to \mu\gamma$ limits (brown/orange). For the $\tau \to \mu\gamma$ amplitude,
    we estimate that the contribution $A_{H^\pm}$ of diagrams involving $H^\pm$
    is of the same order as the contribution $A_{H^0,A^0}$ of diagrams
    involving $H^0$ or $A^0$ (solid curves, shaded
    regions). The uncertainty of this rough approximation is estimated by also
    showing the constraint in case the $H^\pm$ contribution is twice as large
    (dot-dashed curves) or cancels $A_{H^0,A^0}$ exactly (dashed curves). Note that typically
    not all of these curves are visible within the chosen plot ranges.
    The panels on the left show constraints on the heavy Higgs mass $m_{H^0}$ and
    the flavor violating Yukawa coupling $\eta_2^{\mu\tau} = \eta_2^{\tau\mu}$,
    the panels in the middle column show $m_{H^0}$ vs.\ the neutral Higgs mixing
    $\sin\alpha$, and the panels on the right display $\sin\alpha$ vs.\
    $\eta_2^{\mu\tau} = \eta_2^{\tau\mu}$. The three rows of plots correspond to
    different values of the top quark Yukawa coupling
    to the second Higgs doublet, $\eta_2^{tt}$. This coupling affects $H^0$ production
    through gluon fusion and the two-loop contributions to $\tau \to \mu\gamma$.}
  \label{fig:paramspace-H-tau-mu}
\end{figure*}

The main results of this section are presented in
\cref{fig:xsec-limit-H-tau-mu,fig:paramspace-H-tau-mu}.  In the former figure, we show
our limits on the cross section $\sigma(p p \to H^0 \to \tau\mu)$,
while in the latter we interpret these limits in the context of the
lepton flavor violating 2HDM,
translating them into constraints on the parameters $\sin\alpha$, $\eta_2^{\tau\mu} =
\eta_2^{\mu\tau}$, and $m_{H^0}$ of the model.  We also compare to
the limits from $\tau \to \mu\gamma$ and from flavor violating
decays of the light Higgs boson via $h \to \mu\tau$ (see \cref{sec:2HDM-tau-mu-xsecs}).
At low $m_{H^0}$, where $\BR(H^0 \to \tau\mu)$ is large, our limits are in
general stronger than these indirect constraints.  When $m_{H^0} > 2 m_W$,
the branching ratio to the $\mu$--$\tau$ final state drops rapidly, and
consequently also the sensitivity of our search is diminished.
Nevertheless, it can still provide the strongest constraints if $\sin\alpha$
is small and $\eta_2^{tt}$ is sizeable and negative. In this case, $H^0$ production
through gluon fusion is significant, while $\BR(h \to \tau\mu)$ and
$\BR(\tau \to \mu\gamma)$ are suppressed by $\sin\alpha$.  Note that our limits are
in general strongest for negative $\eta_2^{tt}$ because in this case, the
two terms on the right hand side of \cref{eq:xsec-ggH} interfere
constructively.

\section{Summary and Conclusion}
\label{sec:conclusions}

In summary, we have discussed several so far unexplored signatures related to flavor
violating Higgs couplings.  For the case of flavor violation in the quark
sector, we have studied the $t+hh$ (single top plus di-Higgs)
final state, working first in an effective field theory (EFT) framework with
operators of the form $\overline{Q_L^i} \tilde{H} u_R^j (H^\dag H)$ and
then in a Two Higgs Doublet Model (2HDM). In the EFT case, we find
that only the high-luminosity LHC may be sensitive to $t+hh$ production,
while in the 2HDM, discovery prospects are excellent already in Run II,
with $\mathcal{O}(300\ \text{fb}^{-1})$ of 13~TeV data, see \cref{fig:BR-htu-limit}.
In particular, the expected limits from our proposed search can surpass those
from the traditional search for $t \to h q$ decays by almost an
order of magnitude.
The reason for the enhanced discovery reach for $t+hh$ events in the 2HDM
is the contribution from the process $p p \to t + (H^0 \to h h)$, where
$H^0$ is the heavy CP even neutral Higgs boson.

We have considered also flavor violation in the lepton sector, as motivated
in particular by the recent CMS excess in the $h \to \tau\mu$ channel.
Perhaps the simplest explanations of this excess is provided by the 2HDM,
where it is related to the possibility of large flavor violating couplings
of the second (heavy) Higgs doublet, which mixes with $h$. Consequently, we have
studied direct production of heavy Higgs bosons and their flavor violating
decay in the process $p p \to H^0 \to \tau \mu$. We have used
existing CMS data to search for this process and to set new limits on the
lepton flavor violating 2HDM. Our limits are summarized in
\cref{fig:xsec-limit-H-tau-mu}.

To conclude, our study opens up new avenues to search for flavor changing
processes in the Higgs sector. It shows that searches in specific
models---in this case the 2HDM---can be orders of magnitude more sensitive
than searches based only on higher dimensional operators.  To illustrate this
point, we have derived limits on flavor violating couplings of heavy Higgs
bosons by recasting the CMS search for $h \to \tau\mu$.

\section*{Acknowledgments}

We would like to thank Matthew McCullough for very fruitful discussions that
inspired this project, and for collaboration in its early stages.
We are also grateful to Andreas Crivellin and Felix Yu for a very helpful discussion.
The work of MB, JK and JL is supported by the German Research Foundation (DFG)
in the framework of the Research Unit ``New Physics at the Large Hadron
Collider'' (FOR~2239) and of Grant No.\ \mbox{KO~4820/1--1}.  JK is moreover
supported by the European Research Council (ERC) under the European Union's
Horizon 2020 research and innovation programme (grant agreement No.\ 637506,
``$\nu$Directions'').  Additional support has been provided by the Cluster of
Excellence ``Precision Physics, Fundamental Interactions and Structure of
Matter'' (PRISMA -- EXC 1098), grant No.~05H12UME of the German Federal
Ministry for Education and Research (BMBF).


\bibliographystyle{JHEP}
\bibliography{referencelist}

\end{document}